\documentclass[12pt]{article}  

\usepackage[english]{babel}
\usepackage[utf8]{inputenc}
\usepackage{amsmath,amssymb, amsthm}
\usepackage{parskip}
\usepackage{graphicx}
\usepackage{setspace}
\usepackage[authoryear]{natbib} 
\usepackage{hyperref}
\usepackage{multirow}
\usepackage{booktabs} 
\usepackage{subfigure} 
\usepackage{dsfont}

\usepackage[top=2.5cm, left=2.5cm, right=2.5cm, bottom=4.0cm]{geometry}
\usepackage[table]{xcolor} 

\usepackage{booktabs}

\usepackage{authblk} 

\usepackage[titletoc,title]{appendix}  

\usepackage[normalem]{ulem}

\usepackage{amsfonts} 
\usepackage[T1]{fontenc}
\usepackage[linesnumbered,ruled,vlined]{algorithm2e}

\SetCommentSty{mycommfont}

\SetKwInput{KwInput}{Input}                
\SetKwInput{KwOutput}{Output}              

\renewcommand{\hat}{\widehat}
\renewcommand{\tilde}{\widetilde}

     \def\EE{\mathbb{E}}

     \def\PP{\mathbb{P}}
     
     \def\RR{\mathbb{R}}

 \def\cA{{\cal  A}}
 \def\cB{{\cal  B}}
\def\calC{{\cal  C}} 
\def\calD{{\cal  D}} 
\def\calE{{\cal  E}} \def\cE{{\cal  E}}
 
\def\calG{{\cal  G}} 
\def\calH{{\cal  H}} 
\def\calI{{\cal  I}}

\def\calN{{\cal  N}}

\def\calS{{\cal  S}} 
\def\calT{{\cal  T}} 
\def\calU{{\cal  U}}

\def\calX{{\cal  X}} 
 
 \def\cZ{{\cal  Z}}

\def\vec{\hbox{\rm vec}}

\def\one {\mathbbm{1}}

  \def\hbeta{\hat{\beta}}

\usepackage{mathtools}
\DeclarePairedDelimiterX{\infdivx}[2]{(}{)}{%
  #1 \; \delimsize\| \; #2%
}

\DeclareMathOperator{\argmin}{argmin}

\DeclareMathOperator{\diag}{diag}
\newcommand{\E}[1]{{\mathbb{E}} \left[ #1 \right]}

\DeclareMathOperator{\var}{var}

\newcommand{\vect}[1]{{\textsc{Vec}} \left( #1 \right)}

\newtheorem{assumption}{Assumption}[section]

\newtheorem{theorem}{Theorem}[section]
\newtheorem{corollary}{Corollary}[section]

\newtheorem{remark}{Remark}[section]

\definecolor{royalpurple}{rgb}{0.47, 0.32, 0.66}
\definecolor{greenfresh}{HTML}{00897B}
\definecolor{bluefresh}{HTML}{1E88E5}
\definecolor{redfresh}{HTML}{E53935}

\definecolor{royalpurple}{rgb}{0.47, 0.32, 0.66}

\def\beq{\begin{equation}}
\def\eeq{\end{equation}}

\def\bet{\begin{theorem}}
\def\eet{\end{theorem}}

\def\bel{\begin{lemma}}
\def\eel{\end{lemma}}

\def\vec{\mbox{vec}}

\newcommand{\R}{\mathbb{R}} 

\def\htf{\widehat{\tilde{f}}} 
\def\tf{\tilde f}

\def\tH{\tilde H} 
\def\tS{\tilde S}

\def\tU{\tilde U}
\def\Avar{\operatorname*{A\!var}} 
\def\one{\mathds{1}}

\def\Paik{P_{a_{ik}}} 
\def\Paikortho{P_{a_{ik}}^{\bot}} 

\newcommand{\tnorm}[1]{\left\| #1 \right\|_2} 
\newcommand{\tmean}{\frac{1}{T} \sum_t} 
\newcommand{\onorm}[1]{\left\| #1 \right\|_1}
\newcommand{\maxnorm}[1]{\left\| #1 \right\|_{\max}} 
\newcommand{\infnorm}[1]{\left\| #1 \right\|_{\infty}} 
\newcommand{\abs}[1]{\left| #1 \right|}

\newtheorem{lemma}{Lemma}[section]

\newcommand{\Xcal}{\mathcal{X}}

\begin{document}
\def\TITLE{Diffusion Index Forecasting with Tensor Data}
\newcommand{\blind}{0}

\if0\blind
{ \title{\textbf \TITLE\footnote{We thank the co-editor, Yanqin Fan, the associate editor and two referees for careful and constructive comments. We also thank Rong Chen, Frank Diebold, Tae-Hwy Lee, Kenwin Maung, Nese Yildiz and seminar participants at University of Connecticut, UC Riverside, University of Rochester, the ASSA 2025 Meeting, 2025 NBER-NSF Time Series Conference, Midwest Econometrics Group Conference 2025 for their useful comments and discussions. Any remaining errors are solely ours.}}
\author[1]{Bin Chen\thanks{ binchen@rochester.edu}}
\author[2]{Yuefeng Han \thanks{yuefeng.han@nd.edu}} 
\author[1]{Qiyang Yu\thanks{qyu13@ur.rochester.edu}} 

\affil[1]{University of Rochester}
\affil[2]{University of Notre Dame} 
 \date{}
	\maketitle
} \fi

\if1\blind
{
		\bigskip
	\title{\textbf{...}}
    \date{September, 2025}
	\maketitle
	\medskip
} \fi

\begin{abstract} 
     In this paper, we consider diffusion index forecasting with both tensor and non-tensor predictors, where the tensor structure is preserved with a Canonical Polyadic (CP) tensor factor model. When the number of non-tensor predictors is small, we study the asymptotic properties of the least squares estimator in this tensor factor-augmented regression, allowing for factors with different strengths. We derive an analytical formula for prediction intervals that accounts for the estimation uncertainty of the latent factors. In addition, we propose a novel thresholding estimator for the high-dimensional
covariance matrix that is robust to cross-sectional dependence. When the number of non-tensor predictors exceeds or diverges with the sample size, we introduce a multi-source factor-augmented sparse regression model and establish the consistency of the corresponding penalized estimator. Simulation studies validate our theoretical results and an empirical application to U.S. trade flows demonstrates the advantages of our approach over other popular methods in the literature.  

\bigskip

\smallskip

\bigskip

\end{abstract}


\section{Introduction} \label{sec:intro} 

Since the seminal work of \cite{StockWatson2002} and \cite{BaiNg2006}, diffusion index forecasting has been widely adopted by government agencies, policy institutes, and academic researchers around the world (see, e.g., \citealp{LudvigsonNg2007,LudvigsonNg2009,Kyle2015}). The classical diffusion index model predicts the target variable as a linear combination of factors extracted from a large panel of time series data, as well as other important predictors. Its strength lies in the ability to significantly reduce the dimensionality of the predictor space by summarizing it into a small number of factors, enabling effective use of large datasets while keeping the size of the forecasting model small.

However, as the availability and complexity of economic data have expanded, new challenges have emerged for forecasting models. In particular, multidimensional data, panel data with more than two dimensions, have attracted increasing attention in economics due to their ability to capture richer and more intricate relationships. For example, consider predicting U.S. import/export volumes with China using monthly time series data. While the traditional gravity model focuses on bilateral trade flows, it may overlook the influence of trade patterns between the U.S., China, and other countries due to substitution effects. Such data can be structured as a three-dimensional tensor, where the observed time series $\Xcal_t$ is of dimension $N \times N$ for each period $t$, with $N$ denoting the number of countries in the dataset. This type of multidimensional structure poses challenges to classical diffusion index forecasting, which is based on vector factor models. 

A natural approach to tackling this challenge is to flatten or vectorize the tensor time series (see, e.g., \citealp{LudvigsonNg2007}) to fit within the framework of vector factor models. However, this process changes the original data structure, potentially diminishing the interpretability of how information from different dimensions interacts. Furthermore, vectorizing tensors often leads to a significant increase in the number of parameters to estimate, which can result in high computational costs.

In this paper, we consider diffusion index forecasting with tensor and non-tensor predictors, where the tensor structure is preserved with a Canonical Polyadic (CP) tensor factor model\footnote{CP and Tucker structures are the two most commonly assumed low-rank structures for tensor factor models (see, e.g. \cite{KoldaBader2009}). We adopt the CP low-rank structure due to its parsimonious features.}. Common factors are extracted from tensor data using the contemporary covariance-based iterative simultaneous orthogonalization (CC-ISO) procedure proposed in \cite{infCP2024}. When the number of potential non-tensor predictors is small, we estimate the diffusion index model with ordinary least squares (OLS) and establish the consistency and asymptotic normality of the estimator. Unlike \cite{BaiNg2006}, we allow factors to exhibit different strengths. The convergence rate of the conditional mean prediction for the target variable depends on both the strength of the weakest factor and the sample size. To conduct valid inferences, we propose a thresholding-based covariance matrix estimator that is robust to cross-sectional correlation in the idiosyncratic component and demonstrate its consistency. 

When the number of potential non-tensor predictors is large, potentially comparable to or exceeding the sample size, we propose a two-step penalized regression approach, applying least absolute shrinkage and selection operator (LASSO) to select important non-tensor predictors. The combination of factor models and sparse regression has been explored in the literature. In a panel data context, \cite{Fanfactorandsparse} consider the factor-augmented sparse linear regression model, which includes a vector latent factor model and sparse regression as special cases. \cite{chenfanzhu2024} extend this framework to matrix-variate data and propose two new algorithms for estimation. However, both papers focus on a single type of predictor, either panel or matrix data. In economic forecasting, researchers often have access to mixed types of data. Consider the trade example again. While trade flows among various countries provide valuable information for predicting U.S. import/export volumes, other economic variables such as GDP, unemployment rates, exchange rates, and interest rates also play a critical role. These different types of data may reflect distinct sources of predictability. The tensor data on trade flows captures global factors while macroeconomic variables act as proxies for local predictability. Our model offers a novel framework for integrating these diverse data sources to improve forecast accuracy.   

Our work also relates to several recent developments in econometrics. Within tensor and matrix factor models, recent contributions include \cite{BabiiGhyselsPan2025}, \cite{BeyhumGautier2022}, \cite{bolivar2025threshold}, \cite{chen2021statistical}, \cite{infCP2024}, \cite{chen2024time}, \cite{han2022Rank}, \cite{han2024tensor}, \cite{yu2024dynamic}, among many others. Our framework differs by focusing on diffusion-index forecasting and integrating both tensor and non-tensor predictors within a unified structure. From the perspective of factor-augmented regressions, classical results often find factor estimation to be first-order neutral for OLS inference (\citealp{StockWatson2002}, \citealp{BaiNg2006} and \citealp{CaiKongWuZhao2025}), though it can matter in some important cases (\citealp{GONCALVES2014156}). We also contribute to the growing literature on high-dimensional covariance estimation (\citealp{Bickel2008},\citealp{Rothman2009},\citealp{Fan2013}) and on LASSO methods for dependent data (\citealp{KOCK2015325}, \citealp{MEDEIROS2016255}, \citealp{ChernozhukovHardleHuangWang2021}, \citealp{BabiiGhyselsStriaukas2022}, \citealp{BabiiGhyselsStriaukas2024} and \citealp{Beyhum2024}). Finally, our empirical application on forecasting international trade follows the standard macro-forecasting tradition, where autoregressive (AR), vector autoregressive (VAR), and diffusion-index models (\citealp{StockWatson2002}) serve as common benchmarks for evaluating forecast performance.

The rest of the paper is organized as follows. In Section 2, we introduce the diffusion index model based on the CP tensor factor model and develop the estimator when the number of non-tensor predictors is small. Section 3 derives the inferential theories for the diffusion index model and proposes a robust covariance matrix estimator. Section 4 introduces multi-source factor-augmented sparse regression to combine information from different sources and discusses the consistency of the proposed estimator. In Section 5, a simulation study is conducted to assess the reliability of the low- and high-dimensional
estimators in finite samples. In Section 6, an empirical example on U.S. export/import forecasting highlights the merits of our approach in comparison with some popular methods in the literature. All mathematical proofs and additional simulation results are contained in the Appendix.

\subsection{Notation and Preliminaries} \label{sec:notation} 

In this subsection, we introduce essential notations and basic tensor operations. For an in-depth review, readers may refer to \cite{KoldaBader2009}. 

Let $\|x\|_q = (x_1^q+\cdots+x_p^q)^{1/q}$, $q\ge 1$, for any vector $x=(x_1,\cdots,x_p)^\top$. In particular, $\| x \|_\infty = \max_{1 \leq j \leq p} | x_j |$. We employ the following matrix norms: matrix spectral norm $\|M\|_{2} = \underset{\|x\|_2=1,\|y\|_2=1}{\max}  |x^\top M y| = \sigma_1 (M)$, where $\sigma_1(M)$ is the largest singular value of $M$; max entry norm: $\maxnorm{M} = \max_{1 \leq i \leq p, 1 \leq j \leq q} | M_{ij} |$ for $M \in \R^{p \times q}$, where $M_{ij}$ denotes the $(i,j)$ entry of $M$. For two sequences of real numbers $\{a_n\}$ and $\{b_n\}$, we write $a_n\lesssim b_n$ (respectively, $a_n\gtrsim b_n$) if there exists a constant $C$ such that $|a_n|\leq C |b_n|$ (respectively, $|a_n|\geq C |b_n|$) holds for all sufficiently large $n$, and $a_n\asymp b_n$ if both $a_n\lesssim b_n$ and $a_n\gtrsim b_n$ hold. 

Consider two tensors $\cA\in\RR^{d_1\times d_2\times \cdots \times d_K}, \cB\in \RR^{p_1\times p_2\times \cdots \times p_N}$. The outer product $\otimes$ is defined as $\cA\otimes \cB\in \RR^{d_1\times \cdots \times d_K \times p_1\times \cdots \times p_N}$, where
$$(\cA\otimes\cB)_{i_1,...,i_K,j_1,...,j_N}=(\cA)_{i_1,...,i_K}(\cB)_{j_1,...,j_N} .$$
The mode-$k$ product of $\cA\in\RR^{d_1\times d_2\times \cdots \times d_K}$ with a matrix $U\in\RR^{d_k\times r_k}$ is an order $K$-tensor of size $d_1\times \cdots \times d_{k-1} \times r_k\times d_{k+1} \times \cdots \times d_K$, denoted as $\cA\times_k U^\top$, where
$$ (\cA\times_k U)_{i_1,...,i_{k-1},j,i_{k+1},...,i_K}=\sum_{i_k=1}^{d_k} \cA_{i_1,i_2,...,i_K} U_{j,i_k}.  $$

Given $\cA\in\RR^{d_1\times d_2\times \cdots \times d_K}$ and a sequence of $\{U_k\}_{k=1}^K$, where $U_k \in \R^{d_k \times r_k}$, the notation $\cA \times_{k=1}^K U_k^\top$ denotes a sequence of mode-$k$ product: 
$$
\cA \times_{k=1}^K U_k^\top = \cA \times_1 U_1^\top \times_2 U_2^\top \times \cdots \times_K U_K^\top \in \R^{r_1 \times r_2 \times \cdots \times r_K}. 
$$

The Khatri-Rao (or column-wise Kronecker) product of two matrices $A=(a_1,a_2,\cdots,a_r)$ and $B=(b_1,b_2,\cdots,b_r)$ is defined as $A * B = (a_1 \odot b_1, \cdots, a_r \odot b_r)$, where $\odot$ denotes the Kronecker product. Denote $d=d_1\times d_2\times \cdots \times d_K$, $d_{\min}=\min_{k\leq K} d_k$ and $d_{\max}=\max_{k\leq K} d_k$.

\section{Model and Estimation} \label{sec:model}

Assume that a decision maker is interested in predicting some univariate series $y_{t+h}$, conditional on $I_t$, the information
available at time $t$, which consists of a tensor-variate predictor $\calX_t\in \mathbb{R}^{d_1 \times d_2 \times \cdots \times d_K}$ and a set of other observable variables $w_t\in \R^{p}$, such as lags of $y_t$. We consider a diffusion index model as
\begin{equation}\label{eqn:di}
    y_{t+h} = \beta_0^\top w_t + \beta_1^\top f_t + \epsilon_{t+h}, \quad t=1,\cdots,T, 
\end{equation} 
where $h \geq 0$ is the lead time between information available and the target variable. The vector $f_t = (f_{1t}, \ldots, f_{rt})^\top$ consists of $r$ latent factors extracted from the observed tensor data $\calX_t$. Specifically, we model $\calX_t$ as a tensor factor model with a CP low-rank structure:
\begin{align}\label{eqn:cp}
    \calX_t &= \sum_{i=1}^r f_{it} ( \tilde a_{i1} \otimes \tilde a_{i2} \otimes \cdots \otimes \tilde a_{iK}) + \calE_t= \sum_{i=1}^r s_i f_{it} ( a_{i1} \otimes a_{i2} \otimes \cdots \otimes a_{iK}) + \calE_t, 
\end{align}
where $r$ denotes the fixed number of factors and $\tilde a_{ik}$ denotes the $d_k$-dimensional loading vector, which need not be orthogonal. Without loss of generality and to ensure identifiability, we assume that $\EE f_{it}^2 =1$ and normalize the factor loadings $\tilde a_{ik}$ so that $\| a_{ik}\|_2=1$, for all $1\le i\le r$ and $1\le k\le K$. Consequently, all factor strengths are captured by $s_i$. In the strong factor model case, $\|\tilde a_{ik}\|_2 \asymp \sqrt{d_k}$, which implies that $s_i \asymp \sqrt{d_1d_2\cdots d_K}$. The construction of $s_i$ is a matter of parametrization, which ensures the order of the estimated factor $\hat f_t$ to be $O_p(1)$ by convention\footnote{Incorporating $s_i$ into the loadings or factors does not improve the convergence speed for the asymptotic normality of the estimated factors discussed in Section 3.}. The noise tensor $\cal{E}_t$ is assumed to be uncorrelated with the latent factors but may exhibit weak correlations across different dimensions. Unlike classical vector factor models, which suffer from rotation ambiguity, the CP tensor factors are uniquely identified up to permutations and sign changes \citep{Kruskal1977,Kruskal1989,Bro2000}. Throughout this paper, we assume the sign of factors is known without loss of generality. 

To construct forecasts for $y_{T+h}$, the CP factor model \eqref{eqn:cp} needs to be estimated first. We adopt the CC-ISO method proposed by \cite{infCP2024} in our context. Specifically, we estimate $f_t$ via the following algorithm.

Step 1. Obtain the initial value $\widehat A_k^{(0)}=(\widehat a_{1k}^{(0)},\ldots,\widehat a_{rk}^{(0)})\in\R^{d_k\times r}$ via randomized composite PCA \citep{infCP2024} or tensor PCA \citep{BabiiGhyselsPan2025} and compute $\widehat B_k^{(0)} = \widehat A_k^{(0)}(\widehat A_k^{(0)\top}\widehat A_k^{(0)})^{-1} = (\widehat b_{1k}^{(0)},\cdots,\widehat b_{rk}^{(0)})$, where $1\le k\le K$.

Step 2. Given the previous estimates $\widehat a_{ik}^{(m-1)}$, where $m$ is the iteration number, calculate
\begin{align*}
&\cZ_{t,ik}^{(m)}=\calX_t \times_1 \widehat  b_{i1}^{(m)\top} \times_2 \cdots \times_{k-1} \widehat  b_{i,k-1}^{(m)\top} \times_{k+1} \widehat  b_{i,k+1}^{(m-1)\top} \times_{k+2}\cdots\times_K \widehat  b_{iK}^{(m-1)\top} ,
\end{align*}
for $t=1,\cdots,T$. Then the updated loading vectors $\widehat a_{ik}^{(m)}$ are obtained as the top eigenvector of the contemporary covariance $\widehat\Sigma( \cZ_{1:T,ik}^{(m)} )=\frac{1}{T}\sum_{t=1}^T \cZ_{t,ik}^{(m)} \cZ_{t,ik}^{(m)\top}$, where $1\le i\le r$ and $1\le k\le K$. 

Step 3. Update $\widehat B_k^{(m)} = \widehat A_k^{(m)}(\widehat A_k^{(m)\top}\widehat A_k^{(m)})^{-1} = (\widehat b_{1k}^{(m)},...,\widehat b_{rk}^{(m)})$ with $\widehat A_k^{(m)}=(\widehat a_{1k}^{(m)},\ldots,\widehat a_{rk}^{(m)})$.

Step 4. Repeat Steps 2 and 3 until the maximum number of iterations $M$\footnote{In our simulation, we set the maximum number of iterations to $M=100$, but convergence is typically achieved in fewer than 5 iterations.} is reached or $\max_{1\le i\le r}\max_{1\le k\le K}\| \widehat a_{ik}^{(m)} \widehat a_{ik}^{(m)\top} - \widehat a_{ik}^{(m-1)} \widehat a_{ik}^{(m-1)\top} \|_{2}\le \epsilon$, where the default accuracy is set to $\epsilon = 10^{-5}$.

Step 5. Obtain the estimated signal as $\hat s_i = \sqrt{ \frac{\sum_{t=1}^T \left(\calX_t\times_{k=1}^K \widehat  b_{ik}^\top \right)^2}{T}}$ and the estimated factors as $\hat{f}_{it}=\hat{s}_i^{-1}\left(\calX_t\times_{k=1}^K \widehat  b_{ik}^\top\right)$, for  $i=1,\cdots,r$ and $t=1,\cdots,T$.

The estimated factors, $\hat f_t$, along with $w_t$, are then used to estimate the coefficients in Equation \eqref{eqn:di}. When the dimension of the non-tensor predictors $w_t$ is small, we estimate \eqref{eqn:di} with OLS and the forecast for $y_{T+h}$ is obtained as  

$$
\hat y_{T+h} = \hat \beta_0^\top w_T + \hat \beta_1^\top \hat f_T, 
$$
where $\hat \beta_0^\top$ and $\hat \beta_1^\top$ are OLS estimates.

The above forecasting procedure assumes that the rank of $\calX_t$ is known. However, we need to estimate it in practice. We adopt the contemporary covariance-based unfolded eigenvalue ratio estimator considered in \cite{infCP2024}. Other estimators, such as the inner-product-based eigenvalue ratio estimator and autocovariance-based eigenvalue ratio estimator, work as well. More details can be found in \cite{han2024cp} and \cite{infCP2024}. 
\begin{remark}
    If $y_t$ in (\ref{eqn:di}) is a vector of $d$ series and $f_t$ is a vector of $r$ univariate factors obtained from $\calX$ via (\ref{eqn:cp}), a tensor CP factor-augmented vector autoregression (TFAVAR) of order $q$ can be constructed as 
\begin{align*}
    y_{t+1} = \sum_{k=0}^q \alpha_{11,k}y_{t-k}+\sum_{k=0}^q \alpha_{12,k}f_{t-k}+\epsilon_{1t+1},\\
    f_{t+1} = \sum_{k=0}^q \alpha_{21,k}y_{t-k}+\sum_{k=0}^q \alpha_{22,k}f_{t-k}+\epsilon_{2t+1}, \nonumber
\end{align*} 
where $\alpha_{11,k}$, $\alpha_{12,k}$, $\alpha_{21,k}$ and $\alpha_{22,k}$ are model parameters. The inference can be conducted following \cite{BaiNg2006}. To stay focused, we only consider the diffusion index forecasting and leave TFAVAR analysis for future research.
\end{remark}

\section{Asymptotic Properties} \label{sec:theory} 
In this section, we consider the asymptotic properties of our estimation when the number of non-tensor predictors is relatively small ($p\ll T$) and thus no regularization is required. \cite{infCP2024} propose the CC-ISO method and focus on the estimation and inference of loadings while the asymptotic properties of latent factors are unknown. Hence, we first fill in the gap by presenting the consistency and asymptotic normality of the estimated latent factors in Section \ref{sec:thm_factor}. Then we derive the inferential theories for the diffusion index model in Section \ref{sec:thm_di}. Section \ref{sec:threshold} introduces a robust covariance matrix estimator of the factor process for conducting inference on the conditional mean forecasts.

\subsection{Estimation of Factors}\label{sec:thm_factor}
 We start with some assumptions that are necessary for our theoretical development. 

\begin{assumption}\label{asmp:factor_error}\label{asmp:factor}\label{asmp:error}
Denote $e_t = \vec(\cE_t) \in \R^d$ where $d = \prod_{k=1}^K d_k$ and $f_t=(f_{1t},...,f_{rt})^\top$, 
\begin{enumerate}
    \item [(i)] For any $v\in\R^{r}$ with $\|v\|_2=1$ and any $u \in \R^d$ with $\|u\|_2 = 1$, 
    \begin{align}\label{cond2}
    &\max_t \PP\left(|u^\top e_t| \geq x\right) \leq c_1 \exp(-c_2 x^{\nu_1}), \\ 
    &\max_t\PP\left( \left| v^\top f_{t} \right| \geq x \right) \le c_1 \exp\left( -c_2x^{\nu_2} \right),
    \end{align}
    for some constants $c_1,c_2,\nu_1,\nu_2 > 0$. 
    \item [(ii)] Assume $\left(f_{t}, e_t \right)$ is stationary and $\alpha$-mixing. The mixing coefficient satisfies
    \begin{align}\label{cond1}
    \alpha(m) \le \exp\left( - c_0 m^{\gamma} \right)
    \end{align}
    for some constants $c_0>0$ and $\gamma\ge 0$, where
    \begin{align*} 
    \alpha(m) = \sup_t\Big\{&\Big|\PP(A\cap B) - \PP(A)\PP(B)\Big|: \\ 
    &A\in \sigma\left( \left(f_{s}, e_s\right), s\le t\right), B\in \sigma\left(\left(f_{s}, e_s\right), 1\le i\le r, s\ge t+m\right)\Big\}.
    \end{align*} 
    \item [(iii)] Denote $\Sigma_e = \EE(e_t e_t^\top)$ and $\Sigma_f = \EE(f_t f_t^\top)$. There exists a constant $C_0 > 0$ such that $\| \Sigma_e \|_2 \leq C_0$ and $C_0^{-1} \leq \lambda_r(\Sigma_f) \leq \cdots \leq \lambda_1(\Sigma_f) \leq C_0$, where $\lambda_i(\Sigma_f)$ denotes the $i^{\text{th}}$ largest eigenvalue of $\Sigma_f$. 
    \item [(iv)] The factor process $f_{t}$ is independent of the errors $e_t$.
\end{enumerate}
\end{assumption}  

\begin{assumption}\label{asmp:weakfactor} 
    Denote $d_{\max} = \max_{1 \leq k \leq K} d_k$, $\frac{1}{\eta_1} = \frac{2}{\nu_1} + \frac{1}{\gamma}$, $\frac{1}{\eta_2} = \frac{\nu_1 + \nu_2}{\nu_1 \nu_2} + \frac{1}{\gamma}$ and $\frac{1}{\eta_3} = \frac{2}{\nu_2} + \frac{1}{\gamma}$. Assume $\min\{\frac{1}{\eta_1}, \frac{1}{\eta_2}, \frac{1}{\eta_3}\} > 1$. The signal components satisfy $s_i^2 \asymp d^{\alpha_i}$ for some $0 < \alpha_r \leq \alpha_{r-1} \leq \cdots \leq \alpha_1 \leq 1$ such that: 
    $$
    \sqrt{\frac{d_{\max}}{d^{\alpha_r} T}} + \frac{d_{\max}^{1/\eta_1}}{d^{\alpha_r}T} + \frac{d_{\max}^{1/\eta_2}}{d^{\alpha_r/2}T} + \frac{1}{\sqrt{T}} = O(1).
    $$  
\end{assumption}


  Assumption \ref{asmp:error} (i) assumes that the tails of the error and factor processes exhibit exponential decay, which includes a sub-Gaussian distribution as an important example. This assumption could be extended to account for polynomial-type tails with bounded moment conditions. Unlike \cite{lam2012}, \cite{han2024cp} and \cite{infCP2024}, Assumption \ref{asmp:error}(ii) and (iii) allow both weak cross-sectional and serial correlations in the error term. Assumption \ref{asmp:factor}(ii) assumes the $\alpha$-mixing property on the factor process, a standard assumption assumed in the tensor factor literature to capture temporal dependence (e.g., \citealp{chen2021statistical} and \citealp{han2024cp}). We acknowledge that the $\alpha$-mixing condition might not be flexible enough to accommodate certain time series models (\citealp{Andrews1984}). Nevertheless, to maintain focus on the essential theoretical developments and ensure analytical tractability, we adopt the $\alpha$-mixing framework in the main analysis.
Possible relaxations of this assumption are discussed in Appendix E. 

A sufficient condition for Assumption \ref{asmp:factor} is $\max_{j} \sum_{l = 1}^d |\E{ e_{jt} e_{lt}}|<\infty$, which ensures that the aggregate dependence across all pairs of cross-sectional units remains bounded as $d$ increases. This condition is mild and commonly used in large-dimensional factor, panel, and matrix-valued time series models (e.g., \citealp{Bai2003}, \citealp{chen2021statistical}). If the
cross-sectional dimension has some natural ordering (e.g., spatial
or social network data), ${e_{jt}}$ may be assumed to be $\alpha$-mixing
in the cross-sectional dimension as well. Namely, for each $t=1,\cdots,T$, ${e_{jt}}$ is $\alpha$-mixing with mixing coefficients $\alpha_t(m)$ such that $\sup_t\alpha_t(m)\leq\alpha(m)$. Then it is straightforward to verify that Assumption \ref{asmp:factor} (iii) holds by the mixing inequality. Alternatively, if we take into account the tensor structure, we can consider an example as in Appendix C, which allows for exponentially decaying error correlation along both tensor modes. If there is no natural ordering for cross-sectional indices, one can follow
\cite{CHEN201271} by introducing a "distance
function" between cross-sectional units to define a weak dependence structure that also satisfies Assumption \ref{asmp:factor} (iii). 

Assumption \ref{asmp:factor} (iv) imposes independence between factors and errors, which simplifies the analysis of both the ISO algorithm and the forecast model. A more general assumption allowing for limited dependence, as suggested by \cite{Bai2003}, could also be considered, though it would introduce significantly greater theoretical complexity.

Unlike \cite{Bai2003} and \cite{Fanfactorandsparse}, Assumption \ref{asmp:weakfactor} allows for varying factor strengths by incorporating a mix of strong and weak factors, with certain conditions on the weakest signal strength, the dimensions of tensor data, and the sample size. Specifically, it ensures that, as $d, T \to \infty$,  $\max_{i \leq r, k \leq K} \| \hat a_{ik} \hat a_{ik}^\top - a_{ik} a_{ik}^\top \|_2 \to 0$, thereby guaranteeing the consistency of the factor estimation. 

Let $\psi_0$ denote the estimation error of the warm-start initial estimates for the factor loading vectors. Define 
\begin{equation}\label{eqn:rotation_matrix}
    H = \diag(s_1\hat{s}_1^{-1},\cdots,s_r\hat{s}_r^{-1}). 
\end{equation}
For ease of notation, we define 
\begin{equation}\label{eq:psi}
 \psi = \sqrt{\frac{d_{\max}}{d^{\alpha_r} T}} + \frac{d_{\max}^{1/\eta_1}}{d^{\alpha_r}T} + \frac{d_{\max}^{1/\eta_2}}{d^{\alpha_r/2}T} + \frac{1}{d^{\alpha_r}},
\end{equation}
which represents the final estimation error for the factor loading vectors. We first present the performance bounds of $\hat{f}_t$ below. 

\begin{theorem}\label{thm:factor1} 
    Suppose Assumptions \ref{asmp:error}-\ref{asmp:weakfactor} hold. Assume that 
    
    $\max_{k\le K}\| A_k^\top A_k - I_r\|_2<1$ and $T\le C\exp\left(d_{\max} \right)$ for some constant $C$. Suppose that the initial estimation error bounds satisfy the condition:  
    \begin{align}
    &C_{1,K}\left(\frac{s_1^2}{s_r^2} \right) \psi_0^{2K-3}
    + C_{1,K}\frac{s_1}{s_r}\left(\sqrt{\frac{\log T }{T}} + \frac{(\log T)^{1/\eta_3}}{T}  \right) \psi_0^{K-2} \le \rho <1 ,  \label{eqn:thm1_cond}
    \end{align} where $C_{1,K}$ is some constant depending on $K$ only. Then the estimated tensor factors satisfy 
    \begin{align}
        \begin{split}\label{eqn:fbound}
        &(i)\quad \|\hat f_{t} - H f_{t}\|_2 =  O_p\left(\psi + \frac{1}{d^{\alpha_r/2}}\right) ,
        \\ 
        &(ii)\quad \|\hat f_{t} - f_{t}\|_2 = O_p\left( \psi + \frac{1}{d^{\alpha_r/2}} + \sqrt{\frac{1}{T}} \right), 
        \end{split}
    \end{align} 
    where $H$ is defined in \eqref{eqn:rotation_matrix}. 
    \end{theorem}

Theorem \ref{thm:factor1} shows that $\hat{f}_t$ is a consistent estimator of the latent factor $f_t$. However, the convergence rate of $\hat{f}_t$ to $f_t$ may be slower compared to its convergence to $H f_t$. This discrepancy arises due to the non-negligible estimation error associated with the factor signal $s_i$. Nevertheless, this does not affect the prediction of $y_{t+h}$, as the impact is absorbed by the coefficient $\beta_1$. We will provide further discussion on this point in Section \ref{sec:emp}. When all factors are strong, i.e., $\alpha_i = 1$ for all $1 \leq i \leq r$, Theorem \ref*{thm:factor1} implies the following:   
\begin{align*}
    \begin{split}
        &\| \hat f_{t} - H f_{t}\|_2 = O_p\left(\sqrt{\frac{d_{\max}}{d T}} + \frac{d_{\max}^{1/\eta_1}}{dT} + \frac{d_{\max}^{1/\eta_2}}{d^{1/2} T} + \sqrt{\frac{1}{d}}\right) , \\ 
        &\| \hat f_{t} - f_{t}\|_2 = O_p\left(\sqrt{\frac{d_{\max}}{d T}} + \frac{d_{\max}^{1/\eta_1}}{dT} + \frac{d_{\max}^{1/\eta_2}}{d^{1/2} T} + \sqrt{\frac{1}{d}} + \sqrt{\frac{1}{T}}\right).  
        \end{split} 
\end{align*} 
If we further assume that the error term is serially uncorrelated and follows a sub-Gaussian distribution, then the rates simplify to:
\begin{align*}
    \begin{split}
    \| \hat f_{t} - H f_{t}\|_2 &= O_p\left(\sqrt{\frac{d_{\max}}{d T}} + \sqrt{\frac{1}{d}}\right) ,\\ 
    \| \hat f_{t} - f_{t}\|_2 &= O_p\left(\sqrt{\frac{1}{T}} + \sqrt{\frac{1}{d}}\right).
    \end{split}
\end{align*}
\begin{remark}
As $ \| a_{ik}\|_2^2=1$, $\| A_k^\top  A_k - I_{r}\|_{2}<1 $ is used to measure the correlation among columns of $ A_k$. If the loadings are orthogonal, this condition is automatically satisfied. If we define the maximum coherence level as $\varrho_k=\max_{i\neq j}|a_{ik}a_{jk}|$, one sufficient condition is $(r-1)\varrho_k<1$.
\end{remark}
\begin{remark}
The matrix $H$ is introduced to capture the estimation uncertainty of the factor strengths $s_i$, $i=1,\cdots,r$. 
Since the factor strengths must be estimated in order to recover $f_t$ (rather than the scaled version $Hf_t$), the presence of the $1 / \sqrt{T}$ term is inevitable. The use of $H$ effectively removes this source of uncertainty, and the resulting convergence rate with $H$ in Theorem \ref{thm:factor1} is indeed optimal in the time series setting, according to state-of-the-art technical tools \citep{merlevede2011}.
\end{remark}

\begin{remark}
    Under the assumption that the error $\calE_t$ is serially uncorrelated and both $d$ and $T$ go to infinity, the consistency results require $\sqrt{d_{\max} / (d^{\alpha_r}T)} \to 0$. Setting $d_{\max} = d^{\vartheta_d}$ and $T = d^{\vartheta_{T}}$, this condition simplifies to $\alpha_r + \vartheta_{T} > \vartheta_d$. If PCA is applied to the vectorized $\calX_t$, with some modifications to the proofs in \cite{Bai2023}, \cite{Huang2022} and \cite{Gao2024}, it can be shown that consistency requires $\alpha_r + \vartheta_{T} > 1$. Since $\vartheta_d \leq 1$, the CC-ISO algorithm imposes a weaker sample size requirement than PCA. Specifically, CC-ISO remains consistent in the range where $\vartheta_d < \alpha_r + \vartheta_{T} < 1$, whereas PCA does not. Appendix D provides numerical examples and simulations to illustrate this point.
\end{remark}

Denote $B = (b_1, b_2, \ldots, b_r) \in \R^{d \times r} $ where $b_i$ = $b_{iK} \odot b_{iK-1} \odot \cdots \odot b_{i1}$ with $b_{ik}$ defined as $B_k = A_k(A_k^{\top} A_k)^{-1} = (b_{1k},...,b_{rk}) \in\R^{d_k\times r}$, $A_k=(a_{1k},\ldots,a_{rk})\in \R^{d_k\times r}$. And denote $\hat S = \diag(\hat s_1,\cdots, \hat{s}_r)$. 


\begin{assumption}\label{asmp:normality} 
Assume $\sum_{j=1}^d B_{j.} e_{jt} \xrightarrow{d} N(0, \Sigma_{Be})$, where $ B_{j.}$ is the $j^{th}$ row of $B$ and $\Sigma_{Be} = \lim_{d \to \infty} \sum_{j=1}^d \sum_{l = 1}^d \E{B_{j.} B_{l.}^\top e_{jt} e_{lt}}$ is non-singular\footnote{Assumption \ref{asmp:factor}(iii) implies that $\| \Sigma_{Be} \|_2 \leq C_0$.}.

\end{assumption}

\begin{theorem}\label{thm:factor_normality}  
Under Assumptions \ref{asmp:error}- \ref{asmp:normality} and further assume $s_1 \psi=o(1)$, as $d, T\rightarrow \infty$, we have
\begin{equation}\label{eqn:fclt}
    \hat S \left(\hat f_t - H f_t\right)  \xrightarrow{d} N(0, \Sigma_{Be}).
\end{equation} 
\end{theorem}

Theorem \ref{thm:factor_normality} establishes the asymptotic normality of the estimated factors, confirming that the normal approximation is valid in this context. This result is consistent with the findings of \cite{Bai2003} for vector factor models. Additionally, Theorem \ref{thm:factor_normality} derives the asymptotic variance of $\hat{f}_T$, which provides a theoretical foundation for inference in the diffusion index model \eqref{eqn:di} (or \eqref{eqn:di2}) discussed below. The scaling matrix $H$ does not affect such inference, as it only involves the inner product $\beta_1^\top f_t$, and $\beta_1^\top f_t = \beta_1^\top H^{-1} H f_t$ for any invertible matrix $H$. Thus, the inference remains valid irrespective of $H$. Theorems \ref{thm:factor1} and \ref{thm:factor_normality} complement our earlier results in \cite{infCP2024}, which focus on estimation and inference of loadings. 
   
 \subsection{Inference for Diffusion Index Model}\label{sec:thm_di}
 We first consider the properties of the OLS estimates when the CC-ISO estimates of the latent factors are used as regressors, and then discuss how to construct a confidence interval for the conditional mean of (\ref{eqn:di}).
 
 To take advantage of the faster convergence rate of $\hat{f}_t$ to $Hf_t$, we rewrite the diffusion index model \eqref{eqn:di} as 
 \begin{align}\label{eqn:di2}
    y_{t+h} &= \beta_0^\top w_t + \tilde \beta_1^\top H f_t + \epsilon_{t+h}, \qquad t = 1, \ldots, T. 
 \end{align}
where $\tilde \beta_1 = H^{-1} \beta_1$. The conditional mean of $y_{T+h}$ given the information available at time $T$ is
\begin{align} \label{eqn:infeasible}
    y_{T+h|T} = \beta_0^\top w_{T} + \beta_1^\top f_{T} = \beta_0^\top w_{T} + \tilde \beta_1^\top H f_{T},
\end{align}
which is an infeasible predictor since it involves the unknown parameters $\beta_0$, $\tilde \beta_1$ and latent factors $f_T$. 

To obtain a feasible forecast, the factor process $f_t$ is first estimated using the CC-ISO algorithm discussed in Section \ref{sec:model}. Then the coefficients  $ \tilde \beta = (\beta_0^\top, \tilde \beta_1^\top)^\top$ are estimated via OLS: 
\begin{equation}\label{eqn:diest} 
    \hbeta = \left( \frac{1}{T} \sum_{t=1}^{T-h} \hat z_t \hat z_t^\top\right)^{-1} \left( \frac{1}{T} \sum_{t=1}^{T-h} \hat z_t y_{t+h}\right), 
\end{equation}
where $\hat z_t = (w_t^\top, \hat f_t^\top)^\top$ and the feasible prediction of $y_{T+h|T}$ is then given by
\begin{align} \label{eqn:feasible}
    \hat y_{T+h|T} = \hbeta^\top \hat z_T = \hbeta_0^\top w_{T} + \hbeta_1^\top \hat f_{T}.
\end{align}

Denote $z_t = (w_t^\top, f_t^\top)^\top$. To study the asymptotic normality of the OLS estimator $\hat \beta$, we impose the following assumptions. 

\begin{assumption}\label{asmp:di}
        \begin{itemize}
        \item[(i)] $z_t$ and $\epsilon_{t+h}$ are independent of $\cE_s$ for all $t$ and $s$. 
        \item[(ii)] For any $u_z \in \R^{p+r}$ with $\| u_z \|_2 = 1$, $z_t$ satisfies: 
        $$
        \max_t \PP\left( \left| u_z^\top z_t \right| \geq x \right) \leq c_1 \exp\left( -c_2 x^{\nu_3}\right),
        $$ 
        and $\epsilon_{t+h}$ satisfies 
        $$
        \max_t \PP\left( \left| \epsilon_{t+h} \right| \geq x \right) \leq c_1 \exp\left( -c_2 x^{\nu_4}\right),
        $$
        for some constants $c_1, c_2, \nu_3, \nu_4 > 0$.  
        \item[(iii)] $\left(z_t, e_t, \epsilon_t \right)$ is stationary and $\alpha$-mixing. The mixing coefficients satisfy 
        $$
        \alpha(m) \leq \exp\left(-c_0 m^{\gamma} \right)
        $$
        for some constant $c_0 > 0$, where $\gamma$ is defined in Assumption \ref{asmp:factor_error}.  
        \item[(iv)] $\E{\epsilon_{t+h}|y_t, z_t, y_{t-1}, z_{t-1}, \ldots} = 0$ for all $t$. 
        \item[(v)] Define $\Sigma_{zz} = \E{z_t z_t^\top}$ and $\Sigma_{zz,\epsilon} = \E{z_t z_t^\top \epsilon_{t+h}^2}$. Assume $\Sigma_{zz}$ and $\Sigma_{zz,\epsilon}$ are nonsingular.
        \item[(vi)] Let $1/\eta_4 = (\nu_1 + \nu_3)/(\nu_1 \nu_3) + 1/\gamma > 1$ and $1/\eta_5 = (\nu_1 + \nu_4)/(\nu_1 \nu_4) + 1/\gamma > 1$. Define $1/\eta^* = \max\{1/\eta_2, 1/\eta_4, 1/\eta_5\}$ and 
        $$
        \psi^* = \sqrt{\frac{d_{\max}}{d^{\alpha_r} T}} + \frac{d_{\max}^{1/\eta_1}}{d^{\alpha_r}T} + \frac{d_{\max}^{1/\eta^*}}{d^{\alpha_r/2}T} + \frac{1}{d^{\alpha_r}}.
        $$
        Assume $\left( d^{\alpha_1/2} + \sqrt{T} \right)\psi^* =o(1) $. 
    \end{itemize}
    \end{assumption} 

    These assumptions are standard in both factor and regression analysis. Assumption \ref{asmp:di} (ii) is weaker than the common assumption that regressors and errors are sub-Gaussian with $\nu_3 = \nu_4 = 2$ (see, for example, \citealp{Fanfactorandsparse}, \citealp{Huang2022}, \citealp{Gao2024}). Given this weaker condition, Assumption \ref{asmp:di} (vi) imposes additional conditions on the dimensionality and strength of the signals to ensure the consistency and asymptotic normality of $\hat f_t$, $\hat \beta$, and $\hat y_{T+h|T}$. In particular, it assumes that $d^{\alpha_r}$ grows faster than $d_{\max}$. In the case where $K=2$ and $d_1 \asymp d_2$ such that $d_{\max} \asymp d^{1/2}$, $\alpha_r$ is assumed to be larger than $1/2$, which is also imposed by \cite{Bai2023}. In the simulation section, however, we demonstrate that the results in the following theorem are robust to the setting where $\alpha_r < 1/2$ when $T$ is large enough. While Assumption \ref{asmp:di} (ii) could be further relaxed to require only bounded fourth moments for errors and regressors, as in \cite{BaiNg2006}, doing so would necessitate more complex restrictions on dimension and signal strengths. Assumption \ref{asmp:di} (iv) imposes a martingale difference condition on the errors, following \cite{BaiNg2006}. This assumption could be relaxed to allow for serial correlation at the cost of estimating the long-run variance. To simplify the analysis and maintain interpretability, we maintain the current assumption framework.


\begin{theorem}\label{thm:beta} 
    Under Assumptions \ref{asmp:error} to \ref{asmp:di} and conditions of Theorem \ref{thm:factor1}, and $\min\{\frac{2}{\nu_3}, \frac{2}{\nu_4} \} + \frac{1}{\gamma} > 1$, we have 
    $$
    \sqrt{T} (\hat \beta - \tilde \beta) \xrightarrow{d} N(0,\Sigma_{zz}^{-1} \Sigma_{zz, \epsilon} \Sigma_{zz}^{-1}).
    $$
\end{theorem} 

Theorem \ref{thm:beta} shows the asymptotic normality of $\hat \beta$, centered by $\tilde \beta$, the scaled true coefficient. This result does not hold for the unscaled true coefficient $\beta= (\beta_0^\top, \beta_1^\top)^\top$ because the estimation error of $\hat f_t$ with respect to $f_t$ is of order $\sqrt{T}$. Nonetheless, it does not affect the inference for the prediction $\hat y_{T+h|T}$ as shown below. A consistent estimator of the asymptotic variance of $\hbeta$ can be obtained by the sample covariance matrix of the residuals: 
\begin{equation}\label{eqn:avar_beta_hetero}
    \widehat{\Avar{\hbeta}} = \left( \frac{1}{T} \sum_{t=1}^{T-h} \hat z_t \hat z_t^\top\right)^{-1} \left( \frac{1}{T} \sum_{t=1}^{T-h} \hat z_t \hat z_t^\top \hat \epsilon_{t+h}^2 \right) \left( \frac{1}{T} \sum_{t=1}^{T-h} \hat z_t \hat z_t^\top\right)^{-1}.
\end{equation}
Under conditional homoskedasticity such that  $\E{\epsilon_{t+h}^2|z_t} = \sigma^2_\epsilon$, Equation (\ref{eqn:avar_beta_hetero}) can be simplified to 
\begin{equation}\label{eqn:sigma_zz_epsilon_homo}
     \widehat{\Avar{\hbeta}}= \hat \sigma^2_\epsilon \left( \frac{1}{T} \sum_{t=1}^{T-h} \hat z_t \hat z_t^\top\right)^{-1},
\end{equation}
where $\hat \sigma^2_\epsilon = \frac{1}{T} \sum_{t=1}^{T-h} \hat \epsilon_{t+h}^2$. 
\begin{theorem}\label{thm:di_pred}
    Under the assumptions of Theorem \ref{thm:beta}, we have
    $$
    \frac{\hat y_{T+h|T} - y_{T+h | T}}{\sigma_{y_{T+h|T}}}\xrightarrow{d} N(0, 1), 
    $$
    where $\sigma_{y_{T+h|T}} = \sqrt{\frac{1}{T} z_T^\top \Avar(\hat \beta) z_T + \beta_1^\top S^{-1} \Avar(\hat f_T) S^{-1} \beta_1}$ with $\Avar(\hat f_T) = \Sigma_{Be}$ defined in Assumption \ref{asmp:normality} and $S = \diag \left( s_1, \ldots, s_r \right)$. 
\end{theorem} 

The convergence is understood as conditional on $z_T$, which enters only the forecast evaluation but not the estimation of $\hat{\beta}$. Specifically, given data $\{y_t, z_t\}_{t=1}^T$, our goal is to forecast $y_{T+h|T}$ for a fixed $h$. The coefficient $\beta_0$ is estimated using $\{ y_{t+h}, z_t \}_{t=1}^{T-h}$, since the future observations ${y_{t}:t>T}$ are unavailable.

The two terms in the asymptotic variance of $\hat y_{T+h|T}$ decay at different rates, so the convergence rate of $\hat y_{T+h|T}$ is $d^{-\alpha_r/2}+T^{-1/2}$, which implies the efficiency improves with the increase of both the number of observations $T$ and the dimension of the tensor for factor estimation. 


Given consistent estimators of $\Avar(\hbeta)$ and $\Avar(\hat f_T)$, the prediction interval for $y_{T+h|T}$ with confidence level $\alpha$ can be constructed as 
\begin{equation}\label{eqn:PI}
  \left( \hat y_{T+h|T} - q_{1-\alpha/2}\hat \sigma_{y_{T+h|T}}, \quad \hat y_{T+h|T} + q_{1-\alpha/2}\hat \sigma_{y_{T+h|T}}\right),   
\end{equation}
where $q_{1-\alpha/2}$ is the $1-\alpha/2$ quantile of the standard normal distribution, and 
$$
\hat \sigma_{y_{T+h|T}}^2 = \frac{1}{T} \hat z_T^\top \widehat{\Avar(\hat \beta)} \hat z_T + \hat \beta_1^\top \hat S^{-1} \widehat{\Avar(\hat f_T)} \hat S^{-1} \hat \beta_1. 
$$
With $\widehat{\Avar(\hat \beta)}$ given in Equation (\ref{eqn:avar_beta_hetero}), a consistent estimator of $\Avar(\hat f_T)$ is still needed. Assuming the components of $\cE_t$ are cross-sectionally independent, such that $\Sigma_e = \diag(\sigma_1^2, \ldots, \sigma_d^2)$, $\Avar(\hat f_T)$ can be consistently estimated by 
\begin{equation}\label{eqn:avarf_est_iid}
    \hat \Gamma_1 = \sum_{j=1}^d \hat B_{j.} \hat B_{j.}^\top \frac{1}{T} \sum_{t=1}^{T-h} \hat e_{jt}^2,
\end{equation} where $\hat B$ is the estimated $B$ defined on page 10, with the CC-ISO estimator $\hat a_{jk}$ replacing the unknown $a_{jk}$, $\hat e_{t}= \vec(\hat \cE_t)$ and $\hat \cE_t=\calX_t- \sum_{i=1}^r \hat s_i \hat f_{it} ( \hat a_{i1} \otimes \hat a_{i2} \otimes \cdots \otimes \hat a_{iK})$. If cross-sectional dependence is allowed, a robust variance estimator will be introduced in the next section.

\subsection{Covariance matrix estimation of factor process by thresholding}\label{sec:threshold}

In the context of vector factor models, \cite{BaiNg2006} propose the cross-sectional HAC-type estimator of $\Avar(\hat f_T)$ robust to cross-sectional correlation as 
$$
\widehat{\Avar(\hat f_T)} = \frac{1}{n}\sum_{j=1}^n \sum_{l=1}^n \hat \Lambda_{j} \hat \Lambda_{l}^\top \frac{1}{T} \sum_{t=1}^{T-h} \hat e_{jt} \hat e_{lt}, 
$$
where $n$ diverges at a slower rate than $\min\{d, T\}$, and $\hat \Lambda_j$ denotes the estimated factor loading. This estimator could be extended to the CP factor model by replacing $\lambda_{j}$ with $\tilde B_{j.}$. However, it is well documented that HAC-type long-run variance estimators often exhibit poor finite-sample performance, particularly when the cross-sectional dimension is large relative to $T$ (see \citealp{DENHAAN1997}; \citealp{Kiefer2000}). 
Our simulation study in Appendix C confirms this finding in the tensor setting, where the HAC-type estimator tends to produce unreliable variance estimates.

To obtain a more reliable estimator in high-dimensional settings, we adopt a regularized covariance estimation approach that directly targets the structure of $\Sigma_e$. 
Specifically, we estimate $\Sigma_e$ via a thresholded sample covariance matrix, which shrinks small off-diagonal elements toward zero and yields a more stable and high-dimensionality-robust estimator. 
This regularization approach replaces the kernel-based smoothing of HAC estimators with an elementwise shrinkage scheme that adapts to approximate sparsity in the error covariance structure.




Recall from Theorem \ref{thm:factor_normality} that the asymptotic variance of $\hat{f}_t$ is given by  
$$
\Avar(\hat f_T)=\Sigma_{Be}  = B^\top \Sigma_e B, 
$$
where $B$ can be consistently estimated using the CC-ISO estimator $\hat B = (\hat b_1, \ldots, \hat b_r)$, as shown in \cite{infCP2024}. The primary challenge lies in estimating the high-dimensional covariance matrix $\Sigma_e$ in the presence of cross-sectional dependence. To address this, we propose a thresholding estimator $\hat \Sigma_e^{\calT}$: 
$$
\hat \Sigma_e^{\calT} = \calT_\lambda \left(\frac{1}{T} \sum_{t=1}^T \hat e_t \hat e_t^\top \right), 
$$ 
where $\left(\hat \Sigma_e^\calT\right)_{(j,l)} = \calT_\lambda \left(\frac{1}{T} \sum_{t=1}^T \hat e_{jt} \hat e_{lt} \right)$, $\calT_\lambda (\cdot)$ is a thresholding operator and $\hat e_t$ is the vectorized estimated error using the CC-ISO algorithm.  
Following \cite{Rothman2009}, the thresholding operator $\calT_\lambda(\cdot)$ is defined to satisfy the following conditions: 
\begin{itemize}\label{cond:threshold}
    \item[(i)] $|\calT_{\lambda}(z)| \leq |z|$;
    \item[(ii)] $\calT_{\lambda}(z) = 0$ for $|z| \leq \lambda$; 
    \item[(iii)] $| \calT_{\lambda}(z) - z | \leq \lambda$ for all $z$. 
\end{itemize} Examples of generalized thresholding include the LASSO penalty rule:
$$
\calT_{\lambda}(z)=\operatorname{sgn}(z)\left(|z|-\lambda\right)_{+}
$$
and the SCAD thresholding rule proposed by \cite{Fan2001scad}:
$$
\calT_\lambda(z) = \begin{cases}
    \operatorname{sgn}(z)(|z| - \lambda)_+ & \text{if } |z| \leq 2\lambda\\ 
    \left[ (a-1)z - \operatorname{sgn}(z)a\lambda\right] / (a-2) & \text{if } 2\lambda < |z| \leq a\lambda\\ 
    z & \text{if } |z| > a\lambda.
\end{cases}
$$
The bound for the estimation error of $\hat \Sigma_e^{\calT}$ is established uniformly over a class of covariance matrices, as introduced by \cite{Bickel2008} and \cite{Rothman2009}: 
\begin{equation}\label{eqn:covclass}
    \calU(q,c_0(d),M) = \left\{ \Sigma: \sigma_{ii} < M, \ \max_{i} \sum_{j = 1}^d |\sigma_{ij}|^q \leq c_0(d) \right\}, 
\end{equation}
for $0 \leq q < 1$. When $q = 0$,  this class represents exact sparse covariance matrices, where the number of non-zero entries per column is bounded by $c_0(d)$. For $q > 0$, this class defines approximately sparse covariance matrices, where most of the entries in each column are small. Additional assumptions are imposed to derive the bound for the estimation error of $\hat \Sigma_e^{\calT}$. 

Let $\tilde a_{ik,j}$ be the $j^{th}$ entry of $\tilde a_{ik}$ where $\tilde a_{ik} = d_k^{\alpha_i/2} a_{ik}$.

\begin{assumption}\label{asmp:threshold}
    \begin{enumerate}
        \item[(i)] For all $i$ and $k$, $\max_{1 \leq j \leq d_k} |\tilde a_{ik,j}| \leq C$ for some constant $C > 0$. 
        \item[(ii)] $\log(d)^{2/\mu - 1} = o(T)$ where $\mu = \min\{\eta_1, \eta_2 \}$. 
        \item[(iii)] $\frac{d_{\max}}{d^{\alpha_r}} + \frac{d_{\max}^{2/\eta_1}}{d^{2\alpha_r}T} + \frac{d_{\max}^{2/\eta_2}}{d^{\alpha_r}T} = O\left(\log(d)\right)$.  
    \end{enumerate}
\end{assumption}

Assumption \ref{asmp:threshold} (i) bounds the maximum entry of the factor loadings in model \eqref{eqn:cp}. Similar conditions are used in the strong factor model literature such as \cite{Bai2003} and \cite{Fan2013}. In strong factor models, Assumption \ref{asmp:threshold} (i) ensures that the factor loadings for each mode are ``dense'', i.e., the number of zero entries in each column of $\tilde A=(\tilde a_1,...,\tilde a_r)$, $\tilde a_i$ = $\vec(\tilde a_{iK} \odot \tilde a_{iK-1} \odot \cdots \odot \tilde a_{i1})$, does not increase with $d$. In weaker factor models, however, this number is allowed to increase in $d$ with the rate depending on the factor strength $s_i$. Assumption \ref{asmp:threshold} (ii) is imposed to ensure that the bound of  $\left|e_{it}e_{jt} - \E{e_{it}e_{jt}}\right|$ is the same as in \cite{Bickel2008} and \cite{Rothman2009} to accommodate stationary and ergodic errors. This assumption is also imposed in \cite{Fan2011} and \cite{Fan2013}. 

The following theorem provides the rate of convergence for $\hat \Sigma_e^{\calT}$ over the class $\calU(q,c_0(d),M)$. 
\begin{theorem}\label{thm:coverate}
    Suppose Assumptions \ref{asmp:error}-\ref{asmp:weakfactor} and \ref{asmp:threshold} hold. Assume the true covariance matrix $\Sigma_e$ lies in the set $\calU(q,c_0(d),M)$ defined in Equation (\ref{eqn:covclass}) with parameter $q$, $c_0(d)$ and $M$, and the threshold $\lambda = C'\left( \sqrt{\frac{\log(d)}{T}} + \frac{1}{d^{\alpha_r/2}}\right)$, where $C' > 0$ is a sufficiently large constant. Then we have 
    $$
    \| \hat \Sigma_e^\calT - \Sigma_e \|_2 = O_p\left( c_0(d) \left( \sqrt{\frac{\log(d)}{T}} + \frac{1}{d^{\alpha_r/2}}\right)^{1-q} \right). 
    $$
\end{theorem} 

\begin{remark}
    If Assumption \ref{asmp:threshold} (i) is replaced with a ``dense'' factor loading assumption, that is, there exists a constant $C > 0$ such that $\max_{j} |a_{ik,j}| \leq \frac{C}{\sqrt{d_k}}$ for all $i$ and $k$, where $a_{ik,j}$ denotes the $j^{th}$ entry of $a_{ik}$, Theorem \ref{thm:coverate} could be strengthened by replacing $d^{\alpha_r}$ with $d$ in both the threshold and rate. In particular, letting $\lambda =  C'\left( \sqrt{\frac{\log(d)}{T}} + \sqrt{\frac{1}{d}}\right)$, we can obtain 
    $$
        \| \hat \Sigma_e^\calT - \Sigma_e \|_2 = O_p\left( c_0(d) \left( \sqrt{\frac{\log(d)}{T}} + \sqrt{\frac{1}{d}}\right)^{1-q} \right). 
    $$
\end{remark}

\cite{Fan2013} show that the thresholding estimator $\hat \Sigma_e^\calT$ with the adaptive thresholding method developed by \cite{Cai2011} achieves the same rate as in Theorem \ref{thm:coverate} within the strong vector factor model framework. While these results could, in principle, be extended to the CP factor model, the adaptive threshold method presents significant computational challenges when applied to tensor data. Specifically, it requires estimating $\var(e_{jt}e_{lt})$ for all $j$ and $l$, which substantially increases the computational cost due to the high dimensionality of the tensor data. In addition, the adaptive thresholding approach allows $\Sigma_e$ to have diverging diagonal entries, whereas in the CP factor model, the spectral norm of $\Sigma_e$ is typically assumed to be bounded \citep{infCP2024,han2024cp}. This boundedness assumption aligns with both the theoretical framework and practical considerations of the CP factor model, making the results in Theorem \ref{thm:coverate} sufficient for inference in the diffusion index model. 

Define $\hat \Gamma_2 = \hat B^\top \hat \Sigma_e^\calT \hat B$. Theorem \ref{thm:coverate} implies the consistency of $\hat \Gamma_2$, as summarized below. 

\begin{corollary}\label{cor:threshold}
    Under the Assumptions of Theorem \ref{thm:coverate}, suppose $c_0(d) \left( \sqrt{\frac{\log(d)}{T}} + \sqrt{\frac{1}{d^{\alpha_r}}}\right)^{1-q} = o(1)$, then $ \| \hat \Gamma_2 - \Sigma_{Be} \|_2 = o_p(1)$.
\end{corollary}
Corollary \ref{cor:threshold} guarantees a valid prediction interval for $y_{T+h|T}$ that remains robust in the presence of potential cross-sectional error correlations.

\section{Multi-Source Factor-Augmented Sparse Regression}\label{sec:thm_hd}


While diffusion index forecasting with OLS is effective when the number of predictors is relatively small, some real-world applications might involve a large number of potential predictors, sometimes exceeding the sample size. This high-dimensional setting arises in macroeconomic forecasting, financial modeling and trade analysis, where policymakers and researchers need to integrate information from multiple sources. In such contexts, OLS estimation might become unreliable. Moreover, some predictors may be irrelevant, introducing noise rather than improving forecast accuracy. Therefore, it is important to employ variable selection techniques that identify the most relevant predictors while preserving the predictive power of the model. In this section, we extend diffusion index forecasting to accommodate high-dimensional predictors by incorporating regularization---specifically, Multi-Source Factor-Augmented Sparse Regression (MS-FASR)---to ensure robust estimation and improved out-of-sample performance.    

Let $w_t \in \R^p$ denote the set of high-dimensional predictors, alongside the tensor time series $\calX_t$. We consider the diffusion index forecast model:   
\begin{align}
    y_{t+h} &= \beta_0^\top w_t + \beta_1^\top f_t + \epsilon_{t+h}, \label{eqn:hd_y} \\ 
    w_t &= \Lambda f_t + V_t, \label{eqn:hd_w} \\
    \calX_t &= \sum_{i=1}^r s_i f_{it} ( a_{i1} \otimes a_{i2} \otimes \cdots \otimes a_{iK}) + \calE_t \label{eqn:hd_x}, 
\end{align} 
where $p$ is allowed to diverge with the sample size $T$. 

Substituting Equation (\ref{eqn:hd_w}) into Equation (\ref{eqn:hd_y}), we obtain: 
\begin{equation*}
    y_{t+h} = \beta_0^\top V_t + \beta_1^{*\top} f_t + \epsilon_{t+h}, 
\end{equation*}
where $\beta_1^* = \Lambda^\top \beta_0 + \beta_1$. After estimating the factors $f_t$ and $V_t$ from Equation \eqref{eqn:hd_x} and \eqref{eqn:hd_w}, we obtain the estimators of the unknown parameters $\beta_0$ and $\beta_1^*$ via the following penalized regression: 

\begin{equation}\label{eqn:hd_lasso}
    \left(\hat \beta_0, \hat \beta_1^*\right) = \argmin_{\beta_0, \beta_1^*} \frac{1}{2T} \sum_{t=1}^{T-h} \left(y_{t+h} - \beta_0^\top \hat V_t - \beta_1^{*\top} \hat f_t\right)^2 + \lambda \|\beta_0\|_1, 
\end{equation}

where $\lambda > 0$ is a tuning parameter. Since $\hat V_t$ is orthogonal to $\hat f_t$ by construction, the solution to the penalized regression can be obtained via the following steps:

Step 1.   Obtain $\hat f_t$ using the CC-ISO algorithm described in Section \ref{sec:model}.  

Step 2. Estimate $\Lambda$ and $V_t$ via OLS: 
\begin{align*}
    \hat \Lambda &= \sum_{t=1}^T w_t \hat f_t^\top \left(\sum_{t=1}^T \hat f_t \hat f_t^\top\right)^{-1}, \\ 
    \hat V_t &= w_t - \hat \Lambda \hat f_t.   
\end{align*}

Step 3. Obtain the projection residuals $\tilde y_{t+h}$ by regressing $y_{t+h}$ on $\hat f_t$: 
\begin{align*} 
    \hat \beta_1^{*} &= \left(\sum_{t=1}^{T-h} \hat f_t \hat f_t^\top\right)^{-1} \left(\sum_{t=1}^{T-h} \hat f_t y_{t+h}\right), \\ 
    \tilde y_{t+h} &= y_{t+h} - \hat \beta_1^{*\top} \hat f_t. 
\end{align*}

Step 4. Estimate $\beta_0$ by regressing $\tilde y_{t+h}$ on $\hat V_t$ using LASSO: 
$$
\hat \beta_0 = \argmin_{\beta_0} \frac{1}{2T} \|\tilde Y - \hat V \beta_0\|_2^2 + \lambda \|\beta_0\|_1, 
$$

where $\hat V = (\hat V_1, \ldots, \hat V_{T-h})^\top \in \R^{(T-h) \times p}$ and $\tilde Y = (\tilde y_{1+h}, \ldots, \tilde y_{T}) \in \R^{T-h}$. 

Step 5. Estimate $\beta_1$ by 
$$
\hat \beta_1 = \hat \beta_1^* - \hat \Lambda \hat \beta_0, 
$$
and forecast the conditional mean $y_{T+h | T}$ by 
$$
\hat y_{T+h|T} := \hat \beta_0^{\top} \hat V_T + \hat \beta_1^{*\top} \hat f_T.
$$

The algorithm is based on residual-on-residual regression, so $V_t$ in Equation (\ref{eqn:hd_w}) should be interpreted as a projection error, rather than the true error from a structural equation. That is, Equation (\ref{eqn:hd_w}) does not necessarily represent the true data generating process (DGP); $w_t$ may have a nonlinear relationship or no relationship with $f_t$.  This formulation simplifies theoretical analysis. 

For $\varsigma \geq 0$, define the sparsity index set $\calS_\varsigma := \left\{j: | \beta_{0,j} | > \varsigma \right\}$. Let $p_0 := \left| \calS_0 \right|$ denote the cardinality of the support set of $\beta_0$. The following additional assumptions are imposed. 
\begin{assumption}\label{asmp:hd_di}
    \begin{itemize}
        \item[(i)] For any $u \in \R^p$ with $\| u \|_2 = 1$, $V_t$ satisfies: 
        $$
        \max_t \PP\left( \left| u^\top V_t \right| \geq x \right) \leq c_1 \exp\left( -c_2 x^{\nu_5}\right),
        $$ 
        and $\epsilon_{t+h}$ satisfies 
        $$
        \max_t \PP\left( \left| \epsilon_{t+h} \right| \geq x \right) \leq c_1 \exp\left( -c_2 x^{\nu_6}\right),
        $$
        for some constants $c_1, c_2, \nu_5, \nu_6 > 0$.  
        \item[(ii)] $\left( f_t, e_t, V_t, \epsilon_t \right)$ is stationary and $\alpha$-mixing. The mixing coefficients satisfy 
        $$
        \alpha (m) \leq \exp\left(-c_0 m^{\gamma} \right)
        $$
        for some constant $c_0> 0$, where $\gamma$ is defined in Assumption \ref{asmp:factor_error}. 
        \item[(iii)] For a general index set $\calS$, define  the compatibility constant 
        $$
        \phi_{\Sigma_V}(\calS) = \min_{\beta \in \calC(\calS,3)} \frac{ | \calS | \beta^\top \Sigma_V \beta}{\| \beta_S \|_1^2}, 
        $$
        where $\Sigma_V = \E{V_t V_t^\top}$, $\calC(\calS,3) = \left\{ \beta \in \R^p: \| \beta_{\calS^C} \|_1 \leq \onorm{\beta_\calS} \right\}$ and $\beta_\calS = (\beta_j)_{j \in \calS}$. Assume that $\phi^2_{\Sigma_V}(\calS_\lambda) \geq 1 / C$ for some constant $C > 0$. 
        \item[(iv)] $\E{V_t f_t} = \E{V_t \epsilon_{t+h}} = \E{f_t \epsilon_{t+h}} = \E{V_t e_t} = 0$. 
        \item[(v)] Let $\Lambda_j$ denote the $j^{th}$ row of $\Lambda$. $\max_{j=1,\ldots,p} \|\Lambda_j\|_2 \leq C$ for some constant $C$. 
        \item[(vi)] $\beta_0$ satisfies  $\| \beta_0 \|_1 = O(p_0)$. 
        \item[(vii)] Assume $1/\eta_{\min} = \min\{2/\nu_1,2/\nu_2,2/\nu_5,2/\nu_6\} + 1/\gamma > 1$. And assume $\log(p)^{2/\eta_{\min} - 1} = o(T)$. 
    \end{itemize}
\end{assumption} 

These assumptions are standard in the analysis of high-dimensional regressions. Assumption \ref{asmp:hd_di}(i) is weaker than the common assumption that regressors and errors are sub-Gaussian, as seen in the high-dimensional regression literature (e.g., \citealp{Loh2012} and \citealp{Fanfactorandsparse}). Assumption \ref{asmp:hd_di}(iii) imposes a compatibility condition, which is less restrictive than directly assuming the positive definiteness of the sample or population covariance matrix. Since $V_t$ is not directly observable in the data, it is more natural to impose the compatibility condition on the population covariance matrix rather than its sample counterpart, as is often done in the high-dimensional regression literature. This approach is also adopted in \cite{Adamek2023}. 

\begin{theorem}\label{thm:hd_di} 
    Under Assumption \ref{asmp:factor_error}, \ref{asmp:weakfactor}, \ref{asmp:hd_di} and conditions on Theorem \ref{thm:factor1} and $p = O\left(\exp\left(  d^{\alpha_r \nu_5/2}\right) + \exp\left(d_{\max}\right)\right)$, if the tuning parameter $\lambda = C \left( \psi^2 + \frac{1}{s_r^2} + \sqrt{\log(p) / T}\right)$ for some constant $C$ that is large enough, we have 
    \begin{align*}
        &\| \hat \beta_0 - \beta_0 \|_1 = O_p\left( p_0 \left(\sqrt{\frac{\log(p)}{T}} + \frac{1}{d^{\alpha_r}} + \psi^2 \right)\right), \\ 
        & \| \hat \beta_1 - \beta_1 \|_2 = O_p\left( p_0 \left( \psi + \frac{1}{d^{\alpha_r/2}} + \sqrt{\frac{\log(p)}{T}}\right) \right), \\ 
        & | \hat y_{T+h|T} - y_{T+h|T} | = O_p \left(p_0 \left( (\log p)^{1/\nu_5} \sqrt{\frac{\log p}{T}} + \psi + \frac{1}{d^{\alpha_r/2}}\right)\right), 
    \end{align*}
    where $\psi$ is defined in \eqref{eq:psi}. 
\end{theorem} 
Theorem \ref{thm:hd_di} shows that diffusion index forecasting remains consistent even in the presence of a large number of potential predictors. The convergence rate of $\widehat\beta_0$ equals the usual LASSO rate plus an additional component associated with factor estimation, while the rate of $\widehat\beta_1$ depends on the estimation error of $\widehat\beta_0$.\footnote{In a standard linear regression estimated by OLS, the Frisch--Waugh--Lovell (FWL) theorem implies that the estimation of  $\beta_1$ is unaffected by the estimation of $\beta_0$. However, under the $\ell_1$-penalized framework, the orthogonality doesn't hold. Because the LASSO penalty applies to $\beta_0$, the shrinkage changes the fitted residuals that determine $\widehat{\beta}_1$, and therefore the numerical value and convergence rate of $\widehat{\beta}_1$ depend on the estimation error of $\widehat{\beta}_0$. 
This feature has been well documented in the literature (see, e.g., \citealp{chernozhukov2018}; \citealp{Fanfactorandsparse}).} The rate condition on $p$ is imposed to simplify the consistency result. Furthermore, by assuming $d^{\alpha_r} \psi^2 = o(1)$, the result can be improved by eliminating the $\psi$ term in the rates. While selection consistency of the penalized regression could be established with much more involved theoretical derivations and additional assumptions, our primary focus is on prediction. Therefore, we leave this extension for future research to maintain clarity and focus.     

\begin{remark}
    If we further let the restricted eigenvalue condition in Assumption \ref{asmp:hd_di}(iii) hold with $\phi_{\Sigma_V}^*(\calS) := \min_{\beta \in \calC(\calS,3)} \frac{ \beta^\top \Sigma_V \beta}{\| \beta \|_2^2}$, we can bound the estimation error of $\beta_0$ with $\ell_2$ norm:
    $$
    \| \hat \beta_0 - \beta_0 \|_2 = O_p\left( \sqrt{p_0} \left(\sqrt{\frac{\log(p)}{T}} + \frac{1}{d^{\alpha_r}} + \psi^2 \right)\right). 
    $$
\end{remark}

\begin{remark}
    Suppose there exist low-dimensional predictors $g_t$ that are strong predictors for $y_{t+h}$ and should be selected for sure. The proposed model can be extended to incorporate $g_t$ by including $g_t$ in the regression equations \eqref{eqn:hd_y} and \eqref{eqn:hd_w}. The theoretical results in Theorem \ref{thm:hd_di} remain valid in this extended setting, provided that $g_t$ satisfies additional tail conditions, mixing properties, and moment conditions, corresponding to Assumption \ref{asmp:hd_di}(i), (ii) and (iv).
\end{remark}

\begin{remark}\label{cp-pca}
    Compared to the regression with low-dimensional predictors studied in Section \ref{sec:theory}, the magnitude of the forecast error $\hat y_{T+h|T}-y_{T+h|T}$ resulting from the estimation uncertainty of $\hat\beta$ differs. For comparison, assume that $\psi=O(T^{-1/2}+d^{-\alpha_r})$, which typically holds for factor loading estimations \citep{han2024cp,lam2012,Bai2003}, and let $p_0=O(1)$. In the low-dimensional case, the error is of order $T^{-1/2}+d^{-\alpha_r/2}$, whereas in the high-dimensional setting it increases to order $(\log p)^{1/\nu_5+1/2}/\sqrt{T}+\psi+d^{-\alpha_r/2}$ as the number of predictors grows. The first term $(\log p)^{1/\nu_5+1/2}/\sqrt{T}$, present in both the MS-FASR model based on the CP factor structure and the one with vector factors, stems from regularization in high-dimensional settings. Consequently, as $p$ and $d$ increase---making this term increasingly dominant---the forecast performance of MS-FASR-CP and MS-FASR-PCA converge. This theoretical insight is consistent with our simulation results in Section \ref{sec:sim_high} and the empirical findings in Section \ref{sec:emp_out_sample}.
\end{remark}
    
\begin{remark}\label{inference}
Theoretical inference for diffusion-index forecasts with a high-dimensional set of non-tensor predictors $w_t$ is substantially more involved than in the low-dimensional OLS case. The presence of model selection and regularization complicates the limiting distribution of the forecast mean, as the LASSO estimator introduces bias that is typically of the same order as the usual dominating term that determines the limiting distribution in the absence of bias. Although recent progress has been made on debiased or post-selection inference in
high-dimensional regressions (e.g., \citealp{lee2016,liu2018}), extending these results to time-series settings with estimated factors remains analytically challenging and warrants separate investigation. To provide practical guidance, Appendix F outlines a post-selection debiased LASSO (PD-LASSO) approach for constructing prediction intervals around the conditional mean $\hat y_{T+h|T}$. This procedure applies the debiasing step only to the selected coefficients to balance interval validity and efficiency. Simulation evidence shows that the PD-LASSO intervals achieve coverage rates close to the nominal level while remaining substantially tighter than those from the fully debiased estimator.
\end{remark}
    

\section{Simulation} \label{sec:sim} 

In this section, we examine the finite-sample properties of the proposed estimators through a simulation study. We consider the following DGP for $\calX_t$ with $r=3$ and $K=2$: 
\begin{align*}
    \calX_t &= \sum_{i=1}^r f_{it} s_i a_{i1}a_{i2}^{\top} + \cE_t, \\
    f_{it} &=  \rho_i f_{it-1} + \sqrt{1 - \rho_i^2} u_{it}, \qquad\left(\rho_1,\rho_2,\rho_3 \right) = (0.6,0.5,0.4)\\
    \cE_t &= \Sigma_{\cE,1}^{1/2} Z_t \Sigma_{\cE,2}^{1/2}, 
  \end{align*}
where $u_{it}$ and entries of $Z_t$ are generated independently from $\calN(0, 1)$. Throughout the section, we let $d_1 = d_2$ and let $\Sigma_{\calE,k} = \text{Toeplitz}(0.5, d_k)$, $k=1,2$, such that the $(j,l)^{th}$ entry of $\Sigma_{\calE,k}$ is equal to $0.5^{|j-l|}$. Factor loadings $A_k = \left(a_{1k},\ldots,a_{rk}\right)$ are generated as follows: let $\tilde A_k^{(\calN)} \in \R^{d_k \times r}$ whose elements are generated independently from $\calN(0,1)$. We first generate $\tilde A_k$ by orthonormalizing $\tilde A_k^{(\calN)}$ through QR decomposition, i.e., $\tilde A_k = \left(\tilde a_{1k}, \ldots, \tilde a_{rk} \right) = \operatorname{QR}(\tilde A_k^{(\calN)})$. Then $A_k = (a_{1k}, \ldots, a_{rk})$ is generated by $a_{ik} = \Sigma_{\calE,k}^{1/2} \tilde a_{ik} / \sqrt{\tilde a_{ik}^\top \Sigma_{\calE,k} \tilde a_{ik}}$. We set the factor strength $s_i = (r-i+1) \sqrt{d^{\alpha}}$ with $\alpha \in \left\{0.6,0.4 \right\}$. 

In Section \ref{sec:sim_factor_constcy} and \ref{sec:sim_factor_norm}, we evaluate the consistency and asymptotic distribution of factor estimators. Section \ref{sec:sim_PI} compares the coverage rates of the prediction intervals by CC-ISO and by PCA. Section \ref{sec:sim_high} illustrates the convergence rates of the LASSO estimators and associated predictions studied in Section \ref{sec:thm_hd}. Additional simulation results, including settings with correlated and persistent factors, stronger error dependence, and heavy-tailed (Student-t) disturbances, are provided in Appendix G. Across all designs, the proposed method maintains strong predictive performance and estimation accuracy,
confirming its robustness. 

\subsection{Factor Estimator Consistency}\label{sec:sim_factor_constcy}
In this section, we evaluate the finite-sample performance of the factor estimator $\hat f_{t}$. Estimation errors are measured as $\|\hat f_t - Hf_t \|_2$ at $t = T$ where $H$ is defined in Equation \eqref{eqn:rotation_matrix}\footnote{Since our primary interest is in forecasting, we report results for $t=T$. Figures for $t=\frac{T}{2}$ show a similar pattern.}. We vary $d_k$ in $\{20,40,60,80\}$ and $T$ in $\{300,400,500\}$. 

\begin{figure}[t]
    \centering
    \includegraphics[width = 0.9\textwidth]{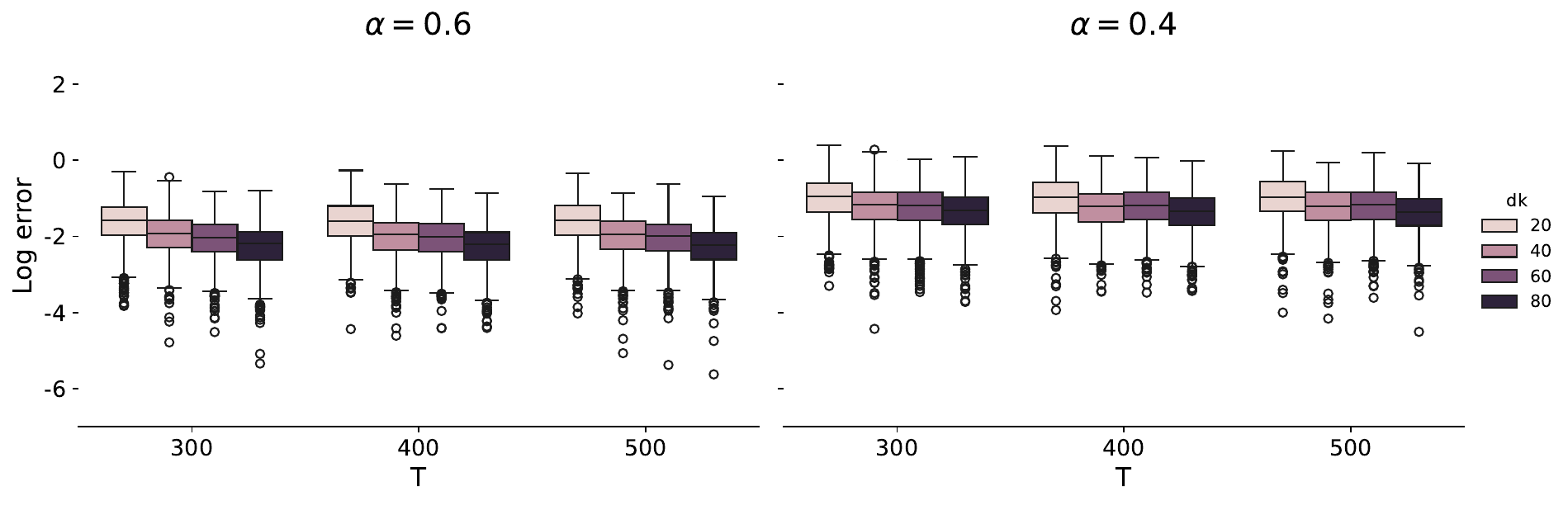}
    \caption{Boxplots of log estimation errors of $\hat f_T$.}
    \label{fig:sim_fTerr}
\end{figure}

Figure \ref{fig:sim_fTerr} presents boxplots of log estimation errors over 1000 repetitions. In all settings, estimation errors decrease as $d_k$ increases. Additionally, estimation errors decrease as factors are stronger. These findings align with Theorem \ref{thm:factor1}. 

\subsection{Factor Estimator Distribution}\label{sec:sim_factor_norm}
Next, we conduct simulations to assess the asymptotic normality of $\hat f_{t}$, as stated in Theorem \ref{thm:factor1}, and to evaluate the proposed covariance matrix estimator in Theorem \ref{thm:coverate}. We vary $d_k$ in $\{40,60,80\}$ and let $T = 800 + \lceil d^{3/4} \rceil$. 

Specifically, we use the SCAD thresholding function developed by \cite{Fan2001scad}, defined as  
$$
\calT_\lambda(z) = \begin{cases}
    \operatorname{sgn}(z)(|z| - \lambda)_+ & \text{if } |z| > a\lambda\\ 
    \left[ (a-1)z - \operatorname{sgn}(z)a\lambda\right] / (a-2) & \text{if } 2\lambda < |z| \leq a\lambda\\ 
    z & \text{if } |z| \leq 2\lambda,
\end{cases}
$$
where we set $a = 3.7$ as suggested in \cite{Fan2001scad}.\footnote{
The other three thresholding functions (hard thresholding, soft thresholding and adaptive LASSO) considered in \cite{Rothman2009} are also evaluated, yielding similar simulation results.} The threshold $\lambda$ is set to $\sqrt{\log(d) / T} + \sqrt{1/d}$. 

\begin{figure}[t]
    \centering
    \includegraphics[width = \textwidth]{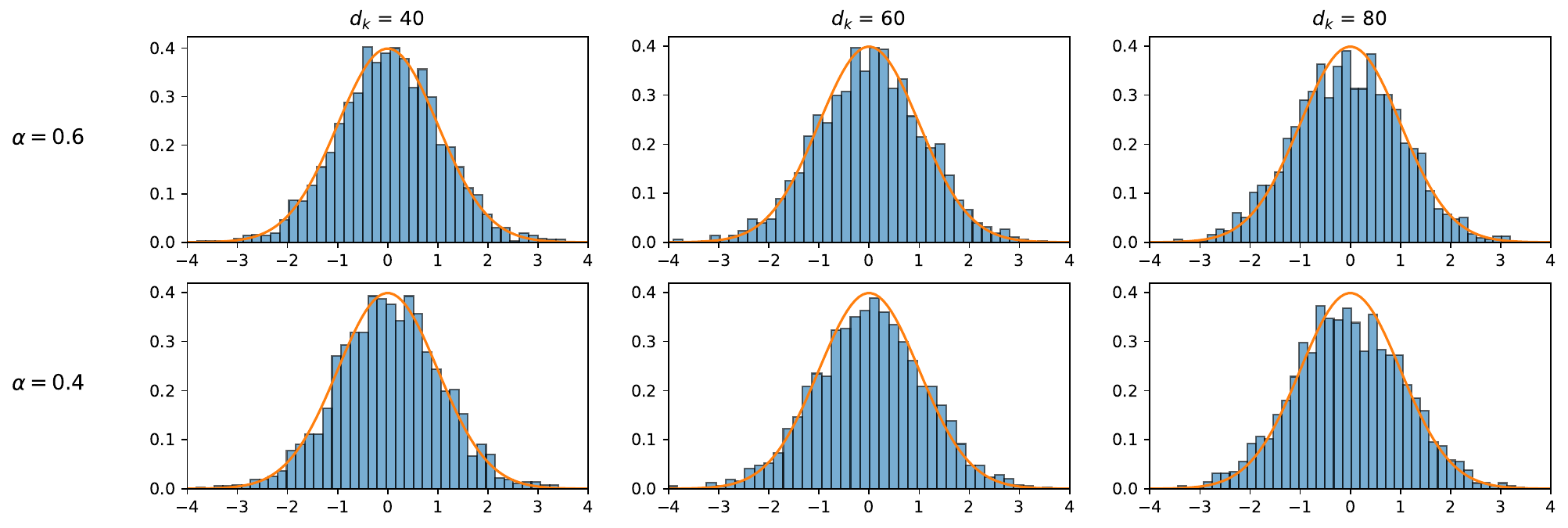}
    \caption{Sample distribution of $\hat \Sigma_{Be}^{-1/2}(\htf_t - \tf_t)$ over 2000 repetitions. The orange line is the pdf of standard normal.}
    \label{fig:sim_fclt}
\end{figure} 

Figure \ref{fig:sim_fclt} shows the distribution of $\hat \Sigma_{Be}^{-1/2}\hat S(\hat f_T - H f_T)$ over 2000 repetitions under two factor strengths. We note that the distribution of $\hat f_T$ approximates the standard normal distribution, which validates Theorem \ref{thm:factor_normality}. Furthermore, the result remains robust to cross-sectional dependence, supporting the effectiveness of the proposed thresholding covariance matrix estimation. 

\subsection{Prediction Interval} \label{sec:sim_PI}
In this section, we examine the prediction intervals for $y_{T+1|T}$ constructed based on Theorem \ref{thm:di_pred}. The target variable $y_{t+1}$ is generated as
$$
y_{t+1} = \beta_0 + \beta_1^\top f_t + \epsilon_{t+1},
$$
where $\beta_0 = 0.5$ and $\beta_1 = \left(0.5,0.5,0.5\right)$. The idiosyncratic error $\epsilon_{t+1}$ is drawn independently from $\calN(0,\nu_t)$ with $\nu_t$ drawn independently from $U[0.5,1.5]$. 

Set $d_1 = d_2 \in \{20,40,60,80,120,160\}$ and $T = 800 + \lceil d^{3/4}\rceil$, with a confidence level of 0.95.
We assess the finite-sample performance of the prediction interval for $\hat y_{T+1|T}$ proposed in Equation (\ref{eqn:PI}) and compare it with the vector PCA method of \cite{BaiNg2006}. For this comparison, we apply the classical PCA method to the vectorized tensor $x_t := \Vec(\calX_t) \in \R^{d}$ and construct the confidence interval following \cite{BaiNg2006} and \cite{Bai2023}: 
\begin{equation}\label{eqn:PCAPI}
    \quad \left( \hat y_{T+h|T} - q_{1-\alpha/2}\hat \sigma_{y_{T+h|T},pca}, \quad \hat y_{T+h|T} + q_{1-\alpha/2}\hat \sigma_{y_{T+h|T},pca}\right),
\end{equation}
where $\hat \sigma^2_{y_{T+h|T},pca} = \frac{1}{T} \hat z_T^{(pca)\top} \ \hat \Avar(\hat \beta^{(pca)}) \hat z_T^{(pca)} + \frac{1}{d} \hat \beta_1^{(pca)\top} \hat \Avar(\hat f_T^{(pca)}) \hat \beta_1^{(pca)}$. The variance estimator for $\hat f_T^{(pca)}$ is given by
$$
\hat \Avar(\hat f_T^{(pca)}) = \Tilde V^{-1} \hat \Gamma_t \Tilde V^{-1}, 
$$
where $\Tilde V$ is a diagonal matrix with diagonal elements equal to the top $r$ eigenvalues of $\frac{1}{dT} \sum_{t=1}^T x_t x_t^\top$, and $\hat z_T^{(pca)}$, $\hat \beta^{(pca)}$, and $\hat f_T^{(pca)}$ are the corresponding PCA estimators. We consider two types of $\hat \Gamma_t$. The first one is the $\hat A^{(PCA)^\top}\hat \Sigma^{(\calT)}_{e,pca} \hat A^{(PCA)}$ where $\hat A^{(PCA)}$ are factor loadings estimated via PCA and $\hat \Sigma_{e,pca}^{(\calT)}$ is the proposed thresholding estimator of the covariance matrix of error terms for PCA. The second one is the HAC-type estimator proposed by \cite{BaiNg2006} and \cite{Bai2023}: 
$$
\hat \Gamma_t^{(HAC)} = \frac{1}{n} \sum_{j = 1}^n \sum_{l = 1}^n \hat A_{j:}^{(\text{PCA})} \hat A_{l:}^{\top(\text{PCA})} \frac{1}{T} \sum_{t=1}^T \hat e_{jt}^{(\text{PCA})} \hat e_{lt}^{(\text{PCA})}, 
$$

where $\hat A_{j:}^{(\text{PCA})}$ and $\hat e_{jt}^{(\text{PCA})}$ are factor loadings and errors estimated via PCA, respectively. The tuning parameter is set as $n = \min\{ \sqrt{d}, \sqrt{T}\}$ as suggested by \cite{BaiNg2006}. For both CP and PCA approaches, $\Avar(\hbeta)$ is estimated using Equation \eqref{eqn:avar_beta_hetero}.

Table \ref{tab:sim_ci} shows the coverage rates of three estimated prediction intervals under two different values of $\alpha$, with a confidence level $95\%$. For $\alpha=0.6$, the coverage rates for the CP-based approach are close to the nominal level. For $\alpha = 0.4$, the coverage rate is slightly lower when $d_k = 20$ but converges to the nominal level as $d_k$ increases. In contrast, the PCA-based approach fails to produce reliable prediction intervals: its coverage rates deviate significantly from the nominal level and show no improvement with increasing $d_k$. 

\begin{table}[htbp]
  \centering
  \caption{Coverage rate of CP and PCA prediction intervals}
  \fontsize{10}{15}\selectfont{
    \begin{tabular}{ccccccc}
    \toprule
          & \multicolumn{3}{c}{$\alpha=0.6$} & \multicolumn{3}{c}{$\alpha=0.4$} \\
    \midrule
    $d_k$    & CP    & PCA(T) & PCA(H) & CP    & PCA(T) & PCA(H) \\
    \midrule
    20    & 0.925 & 0.783 & 0.731 & 0.880  & 0.456 & 0.426 \\
    40    & 0.923 & 0.602 & 0.654 & 0.896 & 0.361 & 0.393 \\
    60    & 0.935 & 0.716 & 0.723 & 0.921 & 0.420  & 0.421 \\
    80    & 0.939 & 0.696 & 0.722 & 0.924 & 0.367 & 0.381 \\
    120   & 0.939 & 0.727 & 0.739 & 0.932 & 0.391 & 0.396 \\
    160   & 0.960  & 0.774 & 0.789 & 0.954 & 0.398 & 0.406 \\
    \bottomrule
    \end{tabular}%
    }
  \label{tab:sim_ci}%
  \parbox{0.65\textwidth}{ 
    \small \textit{Notes:} (1) PCA(T) and PCA(H) refer to the prediction interval constructed using the PCA approach, where the covariance matrix of the factors is estimated via the proposed thresholding covariance estimator and the HAC-type estimator proposed by \cite{BaiNg2006} and \cite{Bai2023}, respectively. (2) The nominal confidence level is $95\%$. 
}
\end{table}%

\subsection{Multi-Source Factor-Augmented Sparse Regression}\label{sec:sim_high}

In this section, we evaluate the convergence rate of $\hat \beta_0$ and $\hat y_{T+1|T}$ in Theorem \ref{thm:hd_di}. Consider the following DGP for $y_{t+h}$ and $w_t \in \R^p$: 
\begin{align*}
    y_{t+1} &= \beta_0^\top w_t + \beta_1^\top z_t + \epsilon_{t+1} ,\\ 
    w_t &= \Lambda z_t + V_t, 
\end{align*}
where $z_t = \left(1, f_t^\top\right) \in \R^{r+1}$. We set the predictor dimension to $p=200$, with the first three elements of $\beta_0$ equal to $0.5$ and the remaining elements set to 0. Each entry of $\Lambda$ is drawn from the uniform distribution $U[-1,1]$, and the entries of $V_t$ are generated independently from $N(0,1)$. The idiosyncratic errors $\epsilon_{t+h}$ follow the same setting as in Section \ref{sec:sim_PI}. We fix $d_1 = d_2 = 40$ and vary $T$. 

In this setting, the rate of $\| \hat \beta_0 - \beta_0 \|_1$ is bounded above by $p_0 \sqrt{\log(p) / T}$, while the forecast error $\left|\hat y_{T+h|T} - y_{T+h|T}\right|$ is bounded above by $p_0 \log(p) / \sqrt{T}$, given $\nu_5=2$ for Gaussian $V_T$. We choose $T$ such that $p_0 \sqrt{\log(p) / T}$ takes values on a uniform grid in $[0.15,0.5]$, which implies that $p_0 \log(p) / \sqrt{T}$ ranges in $(0.34,1.15)$. The tuning parameter for the LASSO regression is fixed at $\sqrt{\log(d)/T} + 1/\hat{s_r}$, where $\hat{s_r}$ is the estimated weakest factor signal, $s_r$.

\begin{figure}
    \centering 
    \includegraphics[width = \columnwidth]{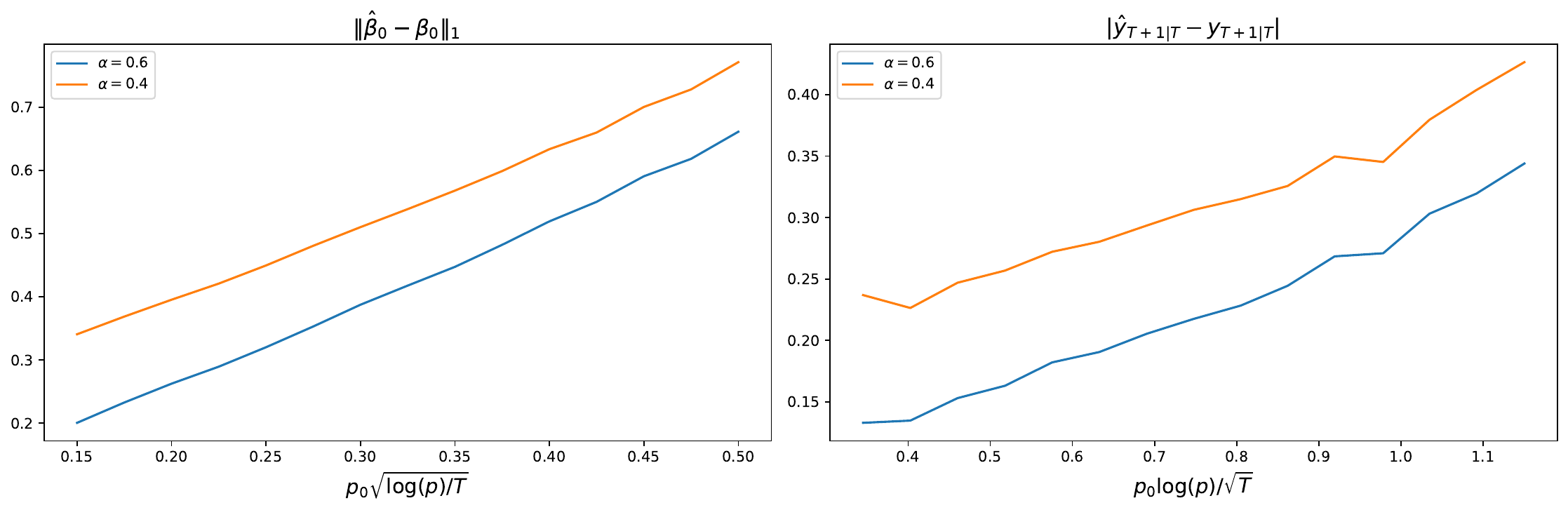} 
    \caption{Estimation error of $\beta_0$ and prediction error of $y_{T+h|T}$ over 1000 repetitions under strong and weak factor setting.} 
    \label{fig:sim_hd}
\end{figure} 

Figure \ref{fig:sim_hd} reports the estimation and prediction errors. The results provide further support for the theoretical findings established in Section \ref{sec:thm_hd}.

\section{Empirical Application} \label{sec:emp} 
Understanding trade flow patterns and forecasting their dynamics are essential for policymaking, firm optimization, and risk management. Trade data inherently form a dynamic sequence of tensor variates, which can capture network-like structures, underlie common dynamics, and reveal intricate interaction patterns.  In this section, we consider diffusion index forecasting based on the CP tensor factor model for international trade data, providing a unified framework to estimate global trade factors and predict future variations in U.S. trade.

\subsection{Data and sample}

We analyze monthly bilateral import and export volumes of commodity goods among 24 countries and regions from January 1999 to December 2018, using data from the International Monetary Fund Direction of Trade Statistics (IMF-DOTS). The countries and regions included in the dataset are: Australia (AU), Canada (CA), China Mainland (CN), Denmark (DK), Finland (FI), France (FR), Germany (DE), Hong Kong (HK), Indonesia (ID), Ireland (IE), Italy (IT), Japan (JP), Korea (KR), Malaysia (MY), Mexico (MX), Netherlands (NL), New Zealand (NZ), Singapore (SG), Spain (ES), Sweden (SE), Taiwan (TW), Thailand (TH), United Kingdom (GB), and the United States (U.S.).

In our study, we employ the diffusion index model with a CP low rank structure, as defined in (\ref{eqn:di}) and (\ref{eqn:cp}). Specifically, we represent the trade data as a $24 \times 24$ two-dimensional tensor, where each element $x_{i,j,t}$ denotes the monthly variation of exports from country $i$ to country $j$ at month $t$. For simplicity, self-exports are set to zero, i.e., $x_{i,i,t} = 0$ for all $i$ and $t$. The target variables for our analysis are the monthly variation of U.S. aggregate export and import to/from countries in the sample, denoted by $y_t^{ex}$ and $y_t^{im}$, respectively. 

The number of common factors is determined using the eigenvalue-ratio-based method proposed by \cite{Ahn2013} and \cite{infCP2024}, which identifies four common factors explaining 51.1\% of the total variance.  Let $f_t$ denote the common factors extracted from the growth rate of bilateral trade. We then construct one-month-ahead forecasts for monthly variations in U.S. aggregate exports and imports using the following regression: 
\begin{equation} \label{eq:emp_model} 
      \begin{split}
            y_{t+1}^{(ex)} &= \beta_{00}^{(ex)} + \beta_{01}^{(ex)} y_{t}^{(ex)} + \beta_{02}^{(ex)} y_t^{(im)} + \beta_1^{(ex)\top} f_t + \epsilon_{t+1}^{(ex)}, \\ 
            y_{t+1}^{(im)} &= \beta_{00}^{(im)} + \beta_{01}^{(im)} y_{t}^{(ex)} + \beta_{02}^{(im)} y_t^{(im)} + \beta_1^{(im)\top} f_t + \epsilon_{t+1}^{(im)}. 
      \end{split}
\end{equation}

\subsection{In-sample analysis} \label{sec:emp_in_sample} 

\begin{table}[t]
  \centering
  \caption{In-sample estimation results for monthly variations in U.S. exports and imports}
  \fontsize{10}{15}\selectfont{
    \begin{tabular}{ccccccccc}
    \toprule
      & Const & $y_{t}^{(ex)}$ & $y_{t}^{(im)}$ & $\hat f_{1t}$ & $\hat f_{2t}$ & $\hat f_{3t}$ & $\hat f_{4t}$ & $R^2$ \\
    \midrule
    \multicolumn{9}{l}{$y_{t+1}^{(ex)}$} \\
    \multirow{2}[0]{*}{(a)} & 294.066 & \textbf{-0.325} & -0.069 &       &       &       &       & 0.163 \\
          & (0.97) & (-3.48) & (-1.06) &       &       &       &       &  \\
    \multirow{2}[0]{*}{(b)} & 384.766 &       &       & \textbf{0.726} & \textbf{-0.234} & \textbf{0.422} & -0.125 & 0.289 \\
          & (1.363) &       &       & (6.73) & (-2.13) & (2.99) & (-0.74) &  \\
    \multirow{2}[1]{*}{(c)} & 365.062 & \textbf{-0.989} & \textbf{0.337} & \textbf{1.118} & \textbf{0.347} & \textbf{0.6} & \textbf{1.179} & 0.402 \\
          & (1.4) & (-5.91) & (3.57) & (7.13) & (2.55) & (4.39) & (3.0) &  \\
    \midrule
    \multicolumn{9}{l}{$y_{t+1}^{(im)}$} \\
    \multirow{2}[0]{*}{(a)} & 586.208 & -0.056 & \textbf{-0.307} &       &       &       &       & 0.114 \\
          & (1.31) & (-0.41) & (-3.21) &       &       &       &       &  \\
    \multirow{2}[0]{*}{(b)} & 613.43 &       &       & \textbf{0.906} & 0.243 & \textbf{0.86} & -0.457 & 0.211 \\
          & (1.45) &       &       & (5.61) & (1.47) & (4.07) & (-1.81) &  \\
    \multirow{2}[1]{*}{(c)} & 597.329 & \textbf{-1.376} & -0.069 & \textbf{0.759} & \textbf{0.941} & \textbf{1.21} & \textbf{2.4} & 0.301 \\
          & (1.49) & (-5.35) & (-0.47) & (3.15) & (4.5) & (5.76) & (3.97) &  \\
    \bottomrule
    \end{tabular}%
    }
\label{tab:in_sample_tab}%
\vspace{0.7em} 
\parbox{0.8\textwidth}{ 
 \small \textit{Note:} The table reports results from the in-sample diffusion index model of monthly variation in U.S. total export and import to countries in the dataset on lagged variables named in the first row. $\hat f_{it}$ is the $i$-th common factor extracted from the bilateral trade flow tensor data. The t-values are reported in parentheses. Coefficients that are statistically significant at the 5\% level are in bold.
}
\end{table}%

Table \ref{tab:in_sample_tab} reports the in-sample forecasting results based on Equation \eqref{eq:emp_model}. As a benchmark, regression (a) predicts each target variable using only the first lag of changes in U.S. exports and imports. In contrast, regression (b) demonstrates that incorporating common factors extracted from the tensor data significantly increases predictive power compared to using lagged values alone. Specifically, the common factors explain $29\%$ of the variation in monthly export changes and 21\% of the variation in import changes. Regression (c) integrates both the lagged target variables and the common factors, leading to a substantial improvement in explanatory power, with R-squared values increasing to 40\% for U.S. exports and 30\% for imports. Moreover, all four common factors are statistically significant predictors for both trade flows. These findings highlight the crucial role of common factors derived from the tensor data in enhancing the accuracy of monthly U.S. export and import forecasts.

\begin{figure}[t]
      \centering 
      \includegraphics[width=\textwidth]{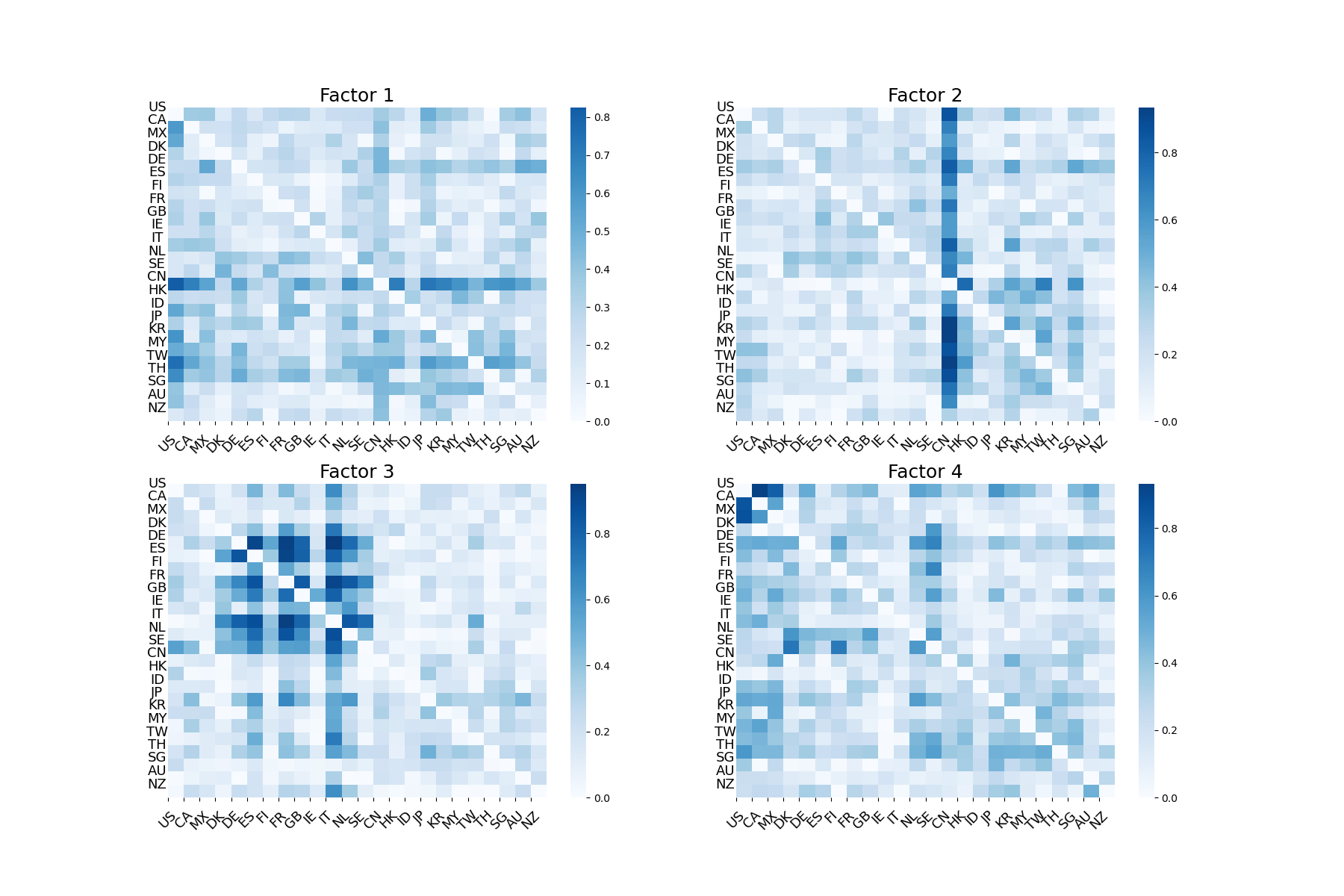}
      \caption{\small Heatmap of the absolute value of correlation between the common factors and the monthly variation in bilateral tradeflow among selected countries. The saturation represents the correlation strength, with high saturation indicating a stronger correlation.} 
      \label{fig:factor_heatmap}
\end{figure}

Since the factors in the CP tensor model are identified only up to sign changes, it is meaningful to explore their economic interpretation. To characterize these factors, we examine their correlations with monthly variations in bilateral trade flows among the selected countries. These correlations are visualized in the heatmap shown in Figure \ref{fig:factor_heatmap}, where stronger correlations between a factor and bilateral trade flows are indicated by deeper blue shades. 

The heatmap reveals distinct regional patterns for each factor. Factor 1 is closely associated with exports from Asian countries to the rest of the dataset, with the highest correlation observed in exports from CN. Factor 2 is highly correlated with China's imports from most countries in the dataset and also shows notable correlations with trade flows among key Asian economies, including CN, KR, JP, and SG. Factor 3 is mainly correlated with bilateral trade flows among European countries, while Factor 4 predominantly captures trade flows within North America, specifically among U.S., CA and MX. In summary, Factors 1 and 2 contain information on trade flows within Asia, particularly involving China. Factor 3 relates to trade dynamics within Europe, and Factor 4 captures variations in trade among North American countries. The in-sample analysis in Section \ref{sec:emp_in_sample} demonstrates that these factors are not only economically interpretable but also provide significant predictive power for monthly variations in U.S. exports and imports.

\subsection{Out-of-sample analysis} \label{sec:emp_out_sample} 

In this section, we evaluate the out-of-sample performance of the diffusion index model \eqref{eq:emp_model} based on the CP low-rank structure and compare it with alternative methods, in particular the vector factor model studied by \cite{BaiNg2006}. In addition to Model \eqref{eq:emp_model}, we incorporate 126 macroeconomic variables from FRED-MD \citep{McCrackenNg2016} and up to 12 lags of monthly variations of U.S. aggregate exports and imports. This allows us to assess the performance of MS-FASR, introduced in Section \ref{sec:thm_hd} and investigate whether including U.S. macroeconomic variables and additional lags of the target variables improves the out-of-sample forecast of U.S. aggregate export and import variations. The MS-FASR model is specified as follows: 
\begin{equation} \label{eq:emp_model_hd} 
      \begin{split}
            y_{t+1}^{(ex)} &= \beta_{00}^{(ex)} + \beta_{01}^{(ex)} y_{t}^{(ex)} + \beta_{02}^{(ex)} y_t^{(im)} + \beta_1^{(ex)\top} f_t + \beta_{2}^{(ex)\top}w_t + \epsilon_{t+1}^{(ex)}, \\ 
            y_{t+1}^{(im)} &= \beta_{00}^{(im)} + \beta_{01}^{(im)} y_{t}^{(ex)} + \beta_{02}^{(im)} y_t^{(im)} + \beta_1^{(im)\top} f_t + \beta_{2}^{(im)\top}w_t + \epsilon_{t+1}^{(im)}, 
      \end{split}
\end{equation} 

where $w_t \in \R^{148}$ includes 126 macroeconomic variables and lagged U.S. aggregate exports and imports from lag 2 to lag 12\footnote{$y_{t}^{(ex)}$ and $y_{t}^{(im)}$ represent lag 1 export and import variations and are already included in the model.}. 

The out-of-sample analysis follows an expanding-window approach, where model parameters are re-estimated as new data become available. The process begins with an initial five-year sample from December 1999 to December 2004. Factors and parameters are estimated using data from December 1999 to November 2004,\footnote{The predictor variables span December 1999 to October 2004, while the target variables cover January 2000 to November 2004, forming a five-year training sample.} and the model is then used to forecast monthly variations in U.S. aggregate exports and imports for December 2004. This procedure is repeated iteratively until the end of the sample, resulting in a total of 169 monthly forecasts from December 2004 to December 2018.

The tuning parameter $\lambda$ is selected via an expanding forecast validation scheme following  \cite{song2011large} and \cite{han2015direct}, which is appropriate for time-series settings. Specifically, we divide the sample into an initial training subsample $t = 1, \ldots, \lceil \gamma T\rceil$ and a validation sample $t = \lceil \gamma T\rceil + 1, \ldots, T$, with $\gamma=0.8$. For each candidate penalty $\lambda_k$, the model is recursively re-estimated and used to generate one-step-ahead forecasts over the validation period. 
The value of $\lambda_k$ that minimizes the mean squared prediction error is selected.

We compare the performance of Model \eqref{eq:emp_model} and Model \eqref{eq:emp_model_hd} against various alternative methods\footnote{ To our knowledge, there are no well-established forecasting benchmarks for international trade flows. Some research, such as \cite{Bussiere2009} and \cite{Greenwood2012}, employ Global VAR (GVAR) to capture international linkages. However, GVAR relies on pre-specified weighting matrices and requires a consistent set of macroeconomic indicators across countries at the same frequency, which is infeasible for monthly data.}: 
\begin{itemize}
    \item Benchmark: Predicts the target variable using only the first lag of U.S. exports and imports along with a constant; 
    \item DI(CP): Model \eqref{eq:emp_model} with factors estimated by CC-ISO; 
    \item DI(PCA): Model \eqref{eq:emp_model} with factors estimated via PCA on $\Vec\left(\calX_t\right)$; 
    \item MS-FASR(CP): Model \eqref{eq:emp_model_hd} with factors estimated by CC-ISO; 
    \item MS-FASR(PCA): Model \eqref{eq:emp_model_hd} with factors estimated via PCA on $\Vec\left(\calX_t\right)$; 
    \item DI(CP) + DI(w): $y_{t+1}^{(\cdot)} = \beta_{00}^{(\cdot)} + \beta_{01}^{(\cdot)} y_{t}^{(ex)} + \beta_{02}^{(\cdot)} y_t^{(im)} + \beta_1^{(\cdot)\top} f_t + \beta_2^{(\cdot)\top} f_t^{(w)} + \epsilon_{t+1}^{(\cdot)}$, where $f_t^{(w)} \in \R^{r_w}$ consists of factors extracted from $w_t$ via PCA and $f_t$ is estimated using CC-ISO; 
    \item DI(PCA) + DI(w): Same as DI(CP) + DI(w) but with $f_t$ estimated via PCA on $\Vec\left( \calX_t \right)$; 
    \item LASSO(w): $y_{t+1}^{(\cdot)} = \beta_{00}^{(\cdot)} + \beta_{01}^{(\cdot)} y_{t}^{(ex)} + \beta_{02}^{(\cdot)} y_t^{(im)} + \beta_2^{(\cdot)\top} w_t + \epsilon_{t+1}^{(ex)}$, where $\beta_2$ is estimated with an $\ell_1$-norm constraint. 
\end{itemize}

Table \ref{tab:mse_all_r2} presents the mean squared error (MSE) ratios of the one-month-ahead out-of-sample forecast for each model relative to the benchmark. It also presents p-values from the forecast comparison tests of \cite{Diebold1995} (DM). These tests are one-sided, with the following alternatives:
\begin{itemize}
    \item DM(Benchmark): Competing methods outperform the benchmark model.
    \item DM(I): DI(CP) is more accurate than DI(PCA).
    \item DM(II): MS-FASR(CP) is more accurate than competing methods.
\end{itemize}

Our findings indicate that, for both exports and imports, MS-FASR(CP) achieves the lowest MSE among all the methods considered. First, MS-FASR(CP) significantly outperforms LASSO(w), reinforcing our in-sample results that common factors extracted from tensor data are valuable for predicting U.S. export and import variations. Furthermore, MS-FASR(CP) outperforms both DI(CP) and DI(PCA) with DM test p-values smaller than any conventional significance level, suggesting that macroeconomic variables provide additional predictive power. Additionally, MS-FASR(CP) also outperforms DI(CP) + DI(w) and DI(PCA) + DI(w) models, which attempt to incorporate factors from multiple sources. This result provides strong empirical support for combining the CP tensor model with sparse regression. Notably, between CP and PCA, DI(CP) significantly outperforms DI(PCA); MS-FASR(CP) modestly improves upon MS-FASR(PCA), consistent with the theoretical argument in Remark~\ref{cp-pca}.

The superior empirical performance of the MS-FASR method can be attributed to its ability to integrate multiple sources of information in a statistically coherent and efficient way. Specifically, the method jointly exploits (i) low-dimensional factors extracted from the tensor predictor, which capture common cross-country dynamics, and (ii)  a high-dimensional set of macroeconomic predictors $w_t$, which provide complementary, country-specific signals. 
Unlike conventional diffusion-index regressions, MS-FASR selectively penalizes only the coefficients on $w_t$ while keeping factor components unpenalized. 
This structure preserves systematic global information from the tensor factors while preventing overfitting from noisy or redundant local predictors. Moreover, the residual-on-residual estimation step ensures that the penalized regression operates on information orthogonal to the factor space, mitigating multicollinearity and enhancing out-of-sample stability. 
Competing models either rely solely on factor information or treat all predictors symmetrically, which can reduce forecasting efficiency when predictive sources are heterogeneous. 
MS-FASR’s hybrid structure thus allows it to combine global coherence with local adaptability, yielding substantial gains in predictive accuracy.



To quantify the relative contributions of global and local information, we conduct a two-component Shapley attribution (\citealp{Shapley}) that decomposes the total gain in forecast accuracy 
relative to the benchmark into contributions from local predictors ($w_t$) and global factors ($f_t$)\footnote{ The Shapley decomposition provides an order-invariant and symmetric measure of how much each information source contributes to the overall reduction in MSE relative to the benchmark model. Let $\text{MSE}_{\text{Benchmark}}$, $\text{MSE}_{\text{Lags+f}}$, $\text{MSE}_{\text{Lags+w}}$, and $\text{MSE}_{\text{MS-FASR}}$ denote the MSEs of the benchmark, lags and global factor only, lags and local predictors only, and MS-FASR models, respectively. 
The Shapley contributions are 
\[
\phi_f = \tfrac{1}{2}\big[\text{MSE}_{\text{Benchmark}} - \text{MSE}_{\text{Lags+f}}\big]
       + \tfrac{1}{2}\big[\text{MSE}_{\text{Lags+w}} - \text{MSE}_{\text{MS-FASR}}\big],
\]
\[
\phi_w = \tfrac{1}{2}\big[\text{MSE}_{\text{Benchmark}} - \text{MSE}_{\text{Lags+w}}\big]
       + \tfrac{1}{2}\big[\text{MSE}_{\text{Lags+f}} - \text{MSE}_{\text{MS-FASR}}\big],
\]
with $\phi_f+\phi_w = \text{MSE}_{\text{Benchmark}} - \text{MSE}_{\text{MS-FASR}}$. 
Each $\phi$ represents the average marginal MSE reduction attributable to that component.}. The result shows that both local and global information contribute meaningfully to MS-FASR’s forecasting gains, with local predictors accounting for a slightly larger share. For exports, $54.5\%$ of the total MSE reduction is attributed to local information and $45.5\%$ to global factors; for imports, the shares are $61.3\%$ and $38.7\%$, respectively. This suggests that while local variation remains somewhat more influential, global factors also provide substantial complementary information, particularly for export forecasting.

\begin{table}[htbp]
  \centering
  \caption{MSE ratios of out-of-sample forecasts} 
    \fontsize{9}{12}\selectfont{
  \resizebox{\textwidth}{!}{
    \begin{tabular}{cccccccc}
    \toprule
          & DI(CP) & DI(PCA) & MS-FASR(CP) & MS-FASR(PCA) & DI(CP) + DI(w) & DI(PCA) + DI(w) & LASSO(w) \\
    \midrule
    \multicolumn{8}{l}{Export} \\
    MSE ratio & 0.7733 & 0.8207 & 0.4354 & 0.449 & 0.8574 & 0.9205 & 0.6951 \\
    DM(Benchmark) & 0.0108 & 0.0371 & <0.0001 & <0.0001 & 0.1182 & 0.2574 & 0.0034 \\
    DM(I) & -     & 0.0599 & -     & -     & -     & -     & - \\
    DM(II) & <0.0001 & <0.0001 & -     & 0.2356 & <0.0001 & <0.0001 & 0.0001 \\
    \midrule
    \multicolumn{8}{l}{Import} \\
    MSE ratio & 0.8159 & 0.8965 & 0.5156 & 0.5116 & 1.0079 & 1.0526 & 0.6441 \\
    DM(Benchmark) & 0.0012 & 0.028 & <0.0001 & <0.0001 & 0.4751 & 0.286 & 0.0001 \\
    DM(I) & -     & 0.0015 & -     & -     & -     & -     & - \\
    DM(II) & <0.0001 & <0.0001 & -     & 0.4407 & <0.0001 & <0.0001 & 0.0083 \\
    \bottomrule
    \end{tabular}%
    }
    }
    \makebox[\textwidth]{
    \parbox{0.95\textwidth}{ 
        \textit{Notes:} (1) The table reports the out-of-sample MSE ratios relative to the benchmark model, which only includes $y_t=(y_t^{(ex)},y_t^{(im)})^\top$ and a constant. (2) The MSE Ratio row shows the ratio of each method's MSE to that of the benchmark model; DM(Benchmark) reports DM test p-values with the alternative being that the competing method is more accurate than the benchmark model; DM(I) reports DM test p-values with the alternative being that DI(CP) outperforms DI(PCA); DM(II) reports DM test p-values with the alternative being that MS-FASR(CP) outperforms the competing method. (3) The number of factors is selected by the unfolded eigenvalue ratio method by \cite{infCP2024}. (4) The tuning parameter $\lambda$ for LASSO and MS-FASR is selected by the EV scheme.
    }
    }
  \label{tab:mse_all_r2}%
\end{table}%

\section{Conclusion}

Factor models are powerful tools for extracting meaningful information from high-dimensional data, which can then be used for prediction. This paper studies the case where the data naturally take the form of a tensor and can be represented by CP decomposition. We develop inferential theories for factor estimation and predictive intervals in the diffusion index forecasting model. We establish that the least squares estimates from predictive regressions are $\sqrt{T}$-consistent and asymptotically normal, even in the presence of weaker factors. Furthermore, we show that the conditional mean remains consistent and asymptotically normal, with its convergence rate determined by $T$ and the strength of the weakest factor. For predictive inference, we propose a consistent estimator for the high-dimensional covariance matrix of cross-sectionally correlated and heteroskedastic errors. 

Additionally, we consider settings where multiple data sources with different structures are available and introduce the MS-FASR model, which effectively integrates information across datasets. Simulation studies confirm our theoretical results, and an empirical application demonstrates that leveraging the tensor structure enhances predictive performance. Our findings suggest that incorporating tensor-based factor extraction can lead to substantial improvements over existing forecasting methods.

\bibliographystyle{apalike} 
\bibliography{ref}

\clearpage 
\appendix
\begin{center}
\LARGE
Supplementary Material of \\ ``Diffusion Index Forecasting with Tensor Data''
\end{center}
\setcounter{page}{1}

\appendix
\section{Proofs of Theorems} 

\begin{proof}[Proof of Theorem \ref{thm:factor1}] 

Let $\htf_{it}=\hat s_i \hat f_{it} $ and $ \tf_t = s_i f_{it}$.
For $(iii)$, since 
$$
\| \htf_t - \tf_t \|_2 = \left\| \sum_{i=1}^r \htf_{it} - \tf_{it} \right\|_2 \leq \sum_{i=1}^r \| \htf_{it} - \tf_{it} \|_2,
$$
it suffices to show that $| \htf_{it} - \tf_{it} | = O_p(s_i \psi)$. The same logic can be applied to $(iv)$ and $(v)$.
\begin{align*}
    \htf_{it} - \tilde f_{it} &= \calX_t \times_{k=1}^K \hat b_{ik}^\top - \tilde f_{it} \\ 
    &= \tf_{it}(\prod_{k=1}^K a_{ik}^\top \hat b_{ik} - 1) + \sum_{j \neq i}^r \tf_{jt} (\prod_{k=1}^K a_{jk}^\top \hat b_{ik}) + \calE_t \times_{k=1}^K \hat b_{ik}^\top \\ 
    &:= \Pi_1 + \Pi_2 + \Pi_3. 
\end{align*}
For $\Pi_1$, by construction of $\hat b_{ik}$,  $\hat a_{ik}^\top \hat b_{ik} = 1$. 
\begin{equation}\label{eqn:aikbikhat}
a_{ik}^\top \hat b_{ik} = a_{ik}^\top \hat b_{ik} - 1 + 1 = (\hat a_{ik} - a_{ik})^\top \hat b_{ik} + 1 \leq \|\hat a_{ik} - a_{ik}\|_2 \| \hat b_{ik} \|_2 + 1. 
\end{equation}

To bound $\|\hat a_{ik} - a_{ik}\|_2$, 
\begin{align*}
    \|\hat a_{ik} - a_{ik}\|_2^2 &= (\hat a_{ik} - a_{ik})^\top (\hat a_{ik} - a_{ik}) \nonumber \\ 
    &= 2(1 - a_{ik}^\top \hat a_{ik}) \nonumber \\ 
    &\leq 2\left(1 - (a_{ik}^\top \hat a_{ik})^2 \right) \nonumber \\ 
    &= 2 \| \hat a_{ik} \hat a_{ik}^\top - a_{ik} a_{ik}^\top \|_2^2, 
\end{align*} 
which yields 
\begin{equation}\label{eqn:ahatminua}
\|\hat a_{ik} - a_{ik}\|_2 \leq \sqrt{2} \psi. 
\end{equation}
To bound $\| \hat b_{ik} \|_2$, denote $A_k^\top A_k = \Sigma_k$ and $\hat A_k^\top \hat A_k = \hat \Sigma_k$. Observe that 
\begin{align}\label{eqn:bikhat}
    \| \hat b_{ik} \|_2 &= \|\hat A_k (\hat A_k^\top \hat A_k)^{-1} e_1\|_2 \nonumber \\ 
    &= e_1^\top (\hat A_k^\top \hat A_k)^{-1} \hat A_k^\top \hat A_k (\hat A_k^\top \hat A_k)^{-1} e_1 \nonumber \\ 
    &= e_1^\top (\hat A_k^\top \hat A_k)^{-1} e_1 \nonumber\\ 
    &\leq \| \hat \Sigma_k^{-1} \|_2 \nonumber \\ 
    &= \left\| \left(\Sigma_k - (\Sigma_k - \hat \Sigma_k)\right)^{-1} \right\|_2 \nonumber \\ 
    &\leq \frac{1}{\lambda_{min}\left(\Sigma_k - (\Sigma_k - \hat \Sigma_k)\right)} \nonumber \\ 
    &\leq \frac{1}{\lambda_{min}(\Sigma_k) - \lambda_{max}(\Sigma_k - \hat \Sigma_k)} \nonumber \\ 
    &\leq \frac{1}{\lambda_{min}(\Sigma_k) - \|\hat \Sigma_k - \Sigma_k\|_2}, 
\end{align} 
where the second last inequality is by Weyl's inequality. 
For $\Sigma_k$, for all $1 \leq i \leq r$, 
$$
\left| \lambda_i(\Sigma_k) - 1 \right| \leq \delta, 
$$
which implies $\lambda_{min}(\Sigma_k) \geq 1 - \delta$ and $\lambda_{max}(\Sigma_k) \leq 1 + \delta_k$. From the bound of $\lambda_{max}(\Sigma_k)$, we have $\|A_k\|_2 = \sqrt{\|A_k^\top A_k\|_2} \leq \sqrt{1 + \delta_k}$. 
For $\| \hat \Sigma_k - \Sigma_k \|_2$, 
\begin{align} \label{eqn:sigmahatdiff}
    \| \hat \Sigma_k - \Sigma_k \|_2 &= \| \hat A_k^\top \hat A_k - A_k^\top A_k \|_2 \nonumber \\ 
    &= \| (\hat A_k - A_k)^\top (\hat A_k^\top - A_k) + \hat A_k^\top A_k + A_k^\top \hat A_k - 2 A_k^\top A_k  \|_2 \nonumber \\ 
    &\leq \| \hat A_k - A_k \|_2^2 + 2\|(\hat A_k - A_k)^\top A_k\|_2 \nonumber \\ 
    &\leq \|\hat A_k - A_k \|_2^2 + 2 \| \hat A_k - A_k \|_2 \| A_k \|_2.
\end{align} 

Note 
\begin{align*}
    \|\hat A_k - A_k \|_2 &= \max_{\|x\| = 1} \| (\hat A_k - A_k) x\|_2 \\ 
    &= \max_{\|x\| = 1} \| \sum_{i=1}^r (\hat a_{ik} - a_{ik}) x_i\|_2 \\ 
    &\leq \max_{\|x\| = 1} \left(\sum_{i=1}^r \| \hat a_{ik} - a_{ik} \|_2^2\right)^{\frac{1}{2}} \left(\sum_{i=1}^r x_i\right)^{\frac{1}{2}} \\ 
    &= \left(\sum_{i=1}^r \| \hat a_{ik} - a_{ik} \|_2^2\right)^{\frac{1}{2}} \\ 
    &\leq \sqrt{2r} \psi. 
\end{align*}
Plug it to \eqref{eqn:sigmahatdiff}, we have 
$$
\| \hat \Sigma_k - \Sigma_k \|_2 \leq 2r \psi^2 + 2\sqrt{2r} \psi \sqrt{1 + \delta_k}. 
$$
Plug it to \eqref{eqn:bikhat}, we have 
$$
\| \hat b_{ik} \|_2 \leq \frac{1}{1 - \delta_k - 2r \psi^2 - 2\sqrt{2r} \psi \sqrt{1 + \delta_k}}.
$$

As $r$ is fixed, we have 
\begin{equation}\label{eqn:bikhat_bound}
    \| \hat b_{ik} \|_2 = O_p(1). 
\end{equation}

By \eqref{eqn:aikbikhat} and \eqref{eqn:ahatminua}, we have 
\begin{equation}\label{eqn:aikbikhat_bound}
a_{ik}^\top \hat b_{ik} \leq \sqrt{2}\psi + 1. 
\end{equation}
Therefore, 
\begin{equation}\label{eqn:Omega1} 
    \Pi_1 = \tf_{it}\left( \prod_{k=1}^K a_{ik}^\top \hat b_{ik} - 1 \right) \leq \tf_{it}\left( \prod_{k=1}^K (\sqrt{2}\psi + 1) - 1\right) = O_p(s_i \psi). 
\end{equation} 

For $\Pi_2$, similarly to $\Pi_1$, for $i \neq j$, 
$$
\hat a_{jk}^\top \hat b_{ik} = 0. 
$$
So 
$$
a_{jk}^\top \hat b_{ik} = (a_{jk} - \hat a_{jk})^\top \hat b_{ik} \leq \|\hat a_{jk} - a_{jk}\|_2 \| \hat b_{ik} \|_2 \lesssim \psi. 
$$ 
Therefore, 
\begin{equation}\label{eqn:Omega2}
    \sum_{j\neq i} f_{jt} (\prod_{k=1}^K a_{jk}^\top \hat b_{ik}) = O_p(s_1 \psi^K). 
\end{equation}
For $\Pi_3$, denote $\hat g_{ik} = \frac{\hat b_{ik}}{\| \hat b_{ik} \|_2}$ and $g_{ik} = \frac{b_{ik}}{\| b_{ik} \|_2}$. 
\begin{align}\label{eqn:ebhat} 
    \calE_t \times_{k=1}^K \hat b_{ik} &= \prod_{k=1}^K \|\hat b_{ik}\|_2 \cdot \calE_t \times \hat g_{ik}^\top \nonumber \\ 
    &\leq \prod_{k=1}^K \| \hat b_{ik}\|_2 \cdot \max_{\|u_k\|_2 = 1} \calE_t \times_{k=1}^K u_k^\top \nonumber \\ 
    &\leq \prod_{k=1}^K \| \hat b_{ik}\|_2 \cdot \max_{\|u\|_2} u^\top \vec(\calE_t) \nonumber \\ 
    &= \prod_{k=1}^K \| \hat b_{ik}\|_2 \cdot \max_{\|u\|_2} u^\top e_t.
\end{align} 
By Assumption \ref{asmp:error}, 
\begin{align}\label{eqn:ue_ord}
    P \left(u^\top e_t > x\right) &< \frac{u^\top\E{e_t e_t^\top}u}{x} \nonumber \\ 
    &\leq \frac{\lambda_1(\Sigma_e)}{x} \leq \frac{C_0}{x}. 
\end{align}
Therefore, 
$$
\max_{\|u\|_2 = 1} u^\top e_t = O_p(1).
$$
By \eqref{eqn:bikhat_bound} and \eqref{eqn:ue_ord}, we have 
$$
\Pi_3 = O_p(1). 
$$
Therefore, putting them all together: 
$$
\htf_{it} - \tilde f_{it} = O_p(s_i \psi + s_1 \psi^K + 1). 
$$
Since $r$ is fixed, by Assumption \ref{asmp:weakfactor},
$$
\| \htf_{t} - \tilde f_{t}\|_2 = O_p(s_1\psi + 1) = O_p(d^{1/2 - \alpha_r} \sqrt{\frac{d_{\max}}{T}} + \sqrt{\frac{d^{1-\alpha_r}}{T}} + d^{1/2 - \alpha_r} + 1), 
$$
 \\~\\
Notice that 
\begin{align}\label{eqn:fhatfexpand}
    \hat f_{it} - f_{it} &= \hat s_i^{-1} \htf_{it} - f_{it} \nonumber \\ 
    &= \hat s_i^{-1} \left(\htf_{it} - s_i f_{it} \right) + \left(\hat s_i^{-1} - s_i^{-1}\right) s_i f_{it} \nonumber \\ 
    &= \left(\hat f_{it} - h_i f_{it}\right) + \left(\hat s_i^{-1} - s_i^{-1}\right) \tf_{it}. 
\end{align} 

So $\hat f_{it} - f_{it} $ has one additional term involving $\hat s_i$, compared with $\hat f_{it} - h_i f_{it}$. 
For $\hat f_{it} - h_i f_{it}$, 
\begin{align}\label{eqn:fhathfexpand}
    \hat f_{it} - h_i f_{it} &= (\hat s_i^{-1} - s_i^{-1}) (\htf_{it} - \tf_{it}) + s_i^{-1} (\htf_{it} - \tf_{it}) \nonumber \\
    &= (\hat s_i^{-1} - s_i^{-1}) (\htf_{it} - \tf_{it}) + O_p(\psi + \frac{1}{s_i}).
\end{align}

By Taylor expansion, 
\begin{equation}\label{eqn:sitaylor}
    \hat s_i^{-1} - s_i^{-1} = -\frac{1}{2}s_i^{-3} (\hat s_i^2 - s_i^2) + \frac{3}{8} s_i^{-5} (\hat s_i^2 - s_i^2)^2 + O(s_i^{-7} (\hat s_i^2 - s_i^2)^3). 
\end{equation}

To bound $\hat s_i^2 - s_i^2$, observe that 
\begin{align*}
    \hat s_i^2 - s_i^2 &= \frac{1}{T} \sum_{t=1}^T \htf_{it}^2 - \E{\tf_{it}^2} \\ 
    &= \frac{1}{T}\sum_{t=1}^T(\htf_{it}^2 - \tf_{it}^2) + \frac{1}{T} \sum_{t=1}^T \left(\tf_{it}^2 - \E{\tf_{it}^2}\right) \\ 
    &:= \Gamma_2 + \Gamma_1.
\end{align*} 

For $\Gamma_1$, by Bernstein inequality for $\alpha$-mixing processes by \cite{merlevede2011} and by assumption \ref{asmp:factor}, for $\frac{1}{\gamma} = \frac{2}{\gamma_1} + \frac{1}{\gamma_2}$, 
\begin{equation*}
    \begin{split}
    P \left[T s_i^{-2} \left(\frac{1}{T} \sum_{t=1}^T (\tf_{it}^2 - s_i^2) \geq x \right) \right] \leq T \exp\left(-\frac{x^\gamma}{c_1}\right) + &\exp\left(-\frac{x^2}{c_2 T}\right)\\ 
    &+ \exp\left(-\frac{x^2}{c_3 T} \exp\left( \frac{x^{\gamma(1-\gamma)}}{c_4 (\log(x))^\gamma} \right)\right) .
\end{split}
\end{equation*} 

Let $x \asymp \sqrt{T \log(T)} + (\log(T))^{1/\gamma}$. Then with probability at  $\frac{1}{2} T^{-c_2}$, 
$$
s_i^{-2} \frac{1}{T} \sum_{t=1}^T (\tf_{it}^2 - s_i^2) \geq \sqrt{\frac{\log(T)}{T}} + \frac{1}{T} \left(\log(T)^{1/\gamma}\right), 
$$ 
which implies 
\begin{equation}\label{eqn:ft2bound}
    \Gamma_1 = \frac{1}{T} \sum_{t=1}^T (\tf_{it}^2 - s_i^2) = s_i^2 O_p(\sqrt{\frac{1}{T}}). 
\end{equation} 

For $\Gamma_2$, we consider the following general form: 
$$
\frac{1}{T} \sum_{t=1}^T \left(\htf_{it}\htf_{jt} - \tf_{it}\tf_{jt}\right).
$$
Expanding it, we have 
\begin{align*}
    &\frac{1}{T} \sum_{t=1}^T \left(\htf_{it}\htf_{jt} - \tf_{it}\tf_{jt}\right) \\ 
    &= \frac{1}{T} \sum_{t=1}^T \left(\calX_t \otimes \calX_t\right) \times_{k=1}^K \hat b_{ik}^\top \times_{k=K+1}^{2K} \hat b_{jk}^\top - \tf_{it}\tf_{jt}\\ 
    \begin{split}
        &= \frac{1}{T} \sum_{t=1}^T \left(\calE_t \otimes \calE_t\right) \times_{k=1}^K \hat b_{ik}^\top \times_{K+1}^{2K} \hat b_{jk}^\top \\ 
        &\quad + \frac{1}{T} \sum_{t=1}^T \tf_{it}\tf_{jt}\left(\prod_{k=1}^K a_{ik}^\top \hat b_{ik} \prod_{k=1}^K a_{jk}^\top \hat b_{jk} - 1\right) \\ 
        &\quad +\frac{1}{T} \sum_{t=1}^T \sum_{l_1 \neq i} \sum_{l_2 \neq j} \tf_{l_1t} \tf_{l_2t} \prod_{k=1}^K a_{l_1k}^\top \hat b_{ik} \prod_{k=1}^K a_{l_2k}^\top \hat b_{jk} \\ 
        &\quad + \frac{1}{T} \sum_{t=1}^T \sum_{l \neq i} \tf_{lt} \tf_{jt} \prod_{k=1}^K a_{lk}^\top \hat b_{ik} \prod_{k=1}^K a_{jk}^\top \hat b_{jk} + \frac{1}{T} \sum_{t=1}^T \sum_{l \neq j} \tf_{it} \tf_{lt} \prod_{k=1}^K a_{ik}^\top \hat b_{ik} \prod_{k=1}^K a_{lk}^\top \hat b_{jk} \\
        &\quad + \frac{1}{T} \sum_{t=1}^T \tf_{it}\left(\prod_{k=1}^K a_{ik}^\top \hat b_{ik}\right)\left(\calE_t \times_{k=1}^K \hat b_{jk}^\top\right) + \frac{1}{T} \sum_{t=1}^T \tf_{jt}\left(\prod_{k=1}^K a_{jk}^\top \hat b_{jk}\right)\left(\calE_t \times_{k=1}^K \hat b_{ik}^\top\right) \\ 
        &\quad + \frac{1}{T} \sum_{t=1}^T \sum_{l \neq i} \tf_{lt} \left(\prod_{k=1}^K a_{lk}^\top \hat b_{ik}\right)\left(\calE_t \times_{k=1}^K \hat b_{jk}^\top\right) + \frac{1}{T} \sum_{t=1}^T \sum_{l \neq j} \tf_{lt} \left(\prod_{k=1}^K a_{lk}^\top \hat b_{jk}\right)\left(\calE_t \times_{k=1}^K \hat b_{ik}^\top\right) \\ 
    \end{split}\\
    &= \sum_{i=1}^9 \Delta_i. 
\end{align*} 

For $\Delta_1$, 
\begin{align}
\Delta_1&=\frac{1}{T} \sum_{t=1}^T \calE_t \otimes \calE_t \times_{k=1}^K \hat b_{ik}^\top \times_{K+1}^{2K} \hat b_{jk}^\top \nonumber \\ 
    &= \left(\prod_{k=1}^K \|\hat b_{ik}\|_2 \prod_{k=1}^K \| \hat b_{jk}\|_2\right) \frac{1}{T} \sum_{t=1}^T \calE_t \otimes \calE_t \times_{k=1}^K \hat g_{ik}^\top \times_{k=K+1}^{2K} \hat g_{jk}^\top \nonumber  .
\end{align} 

By \eqref{eqn:bikhat_bound}, 
$$
\prod_{k=1}^K \|\hat b_{ik}\|_2 \prod_{k=1}^K \| \hat b_{jk}\|_2 = O_p(1). 
$$ 

Expand the outer product, we have 
\begin{align*}
    &\frac{1}{T} \sum_{t=1}^T \calE_t \otimes \calE_t \times_{k=1}^K \hat g_{ik}^\top \times_{k=K+1}^{2K} \hat g_{jk}^\top \\ 
    &=\frac{1}{T} \sum_{t=1}^T \calE_t \otimes \calE_t \times_1 g_{i1} \times_{k=2}^K \hat g_{ik}^\top \times_{k=K+1}^{2K} \hat g_{jk}^\top + \frac{1}{T} \sum_{t=1}^T \calE_t \otimes \calE_t \times_1 (\hat g_{i1} - g_{i1}) \times_{k=1}^K \hat g_{ik}^\top \times_{k=K+1}^{2K} \hat g_{jk}^\top \\ 
    \begin{split}
        &\leq \frac{1}{T} \sum_{t=1}^T \calE_t \otimes \calE_t \times_1 g_{i1} \times_{k=2}^K \hat g_{ik}^\top \times_{k=K+1}^{2K} \hat g_{jk}^\top \\ 
        &\quad + \| \hat g_{i1} - g_{i1} \|_2 \max_{\|u_{ik}\| = \|u_{jk}\|_2 = 1} \frac{1}{T} \sum_{t=1}^T \calE_t \otimes \calE_t \times_1 u_{i1} \times_{k=1}^K u_{ik}^\top \times_{k=K+1}^{2K} u_{jk}^\top
    \end{split} \\ 
    &\leq \cdots \\ 
    \begin{split}
        &\leq \frac{1}{T} \sum_{t=1}^T \calE_t \otimes \calE_t \times_{k=1}^K g_{ik}^\top \times_{k=K+1}^{2K} g_{jk}^\top \\ 
        &\quad + (\sum_{k=1}^K \|\hat g_{ik} - g_{ik}\|_2 + \sum_{k=1}^K \| \hat g_{jk} - g_{jk}\|_2 ) \max_{\|u_{ik}\| = \|u_{jk}\|_2 = 1} \frac{1}{T} \sum_{t=1}^T \calE_t \otimes \calE_t \times_1 u_{i1} \times_{k=1}^K u_{ik}^\top \times_{k=K+1}^{2K} u_{jk}^\top .
    \end{split}
\end{align*} 

By Assumption \ref{asmp:error}, 
$$
\frac{1}{T} \sum_{t=1}^T \calE_t \otimes \calE_t \times_{k=1}^K g_{ik}^\top \times_{k=K+1}^{2K} g_{jk}^\top = g_i^\top \left( \frac{1}{T} \sum_{t=1}^T e_t e_t^\top\right) g_j = O_p(1).
$$

By \cite{infCP2024}, $\| \hat g_{ik} - g_{ik} \|_2 \lesssim \psi$ if $r$ is fixed. Therefore, since $r$ and $K$ are fixed, 
$$
\sum_{k=1}^K \|\hat g_{ik} - g_{ik}\|_2 + \sum_{k=1}^K \| \hat g_{jk} - g_{jk}\|_2 \lesssim \psi. 
$$ 

Denote $\epsilon$-net for $S^{d_k-1}$ with $\calN_k(\epsilon)$ for $1 \leq k \leq K$. Then the cartesian product of $\epsilon$-net for $S^{d_k-1}$, $1 \leq k \leq 2K$ form a $\sqrt{2K}\epsilon$-net for $S^{d_1-1} \times \cdots S^{d_K-1} \times S^{d_1 - 1} \times \cdots S^{d_K - 1}$. Denote it with $\calN(\sqrt{2K}\epsilon)$ and denote $u_i = \odot_{k=1}^K u_{ik}$ and $u_j = \odot_{k=1}^K u_{jk}$. By Corollary 4.2.13 and Lemma 4.4.1 in \cite{HDP}, take $\epsilon = \frac{1}{3}$, we have $|\calN(\sqrt{2K}\epsilon)| = 49^{d_1 + \cdots + d_K} \lesssim 7^{d_{\max}}$ and 
\begin{align*}
    \max_{\|u_{ik}\| = \|u_{jk}\|_2 = 1} \frac{1}{T} \sum_{t=1}^T \calE_t \otimes \calE_t \times_1 u_{i1} \times_{k=1}^K u_{ik}^\top \times_{k=K+1}^{2K} u_{jk}^\top \lesssim \max_{u_{ik}, u_{jk} \in \calN_k(\frac{1}{3\sqrt{2K}})} u_i^\top \left(\frac{1}{T} \sum_{t=1}^T e_t e_t^\top \right) u_j  .
\end{align*}

By Assumption \ref{asmp:error} and Lemma \ref{lemma:expprod}, for any conformable unit-norm vector $u_i$ and $u_j$, $u_i^\top e_t e_t^\top u_j$ is a general sub-exponential random variable with parameter $2/\nu_1$. By Theorem 1 in \cite{merlevede2011} , for $\frac{1}{\eta_1} = \frac{2}{\nu_1} + \frac{1}{\gamma}$,
$$
    \PP\left(u_i \left(\sum_{t=1}^T (e_t e_t^\top - \E{e_t e_t^\top}) \right) u_j > x\right) \leq T \exp\left(-\frac{x^{\eta_1}}{c_1}\right) + \exp\left(-\frac{x^2}{c_2T}\right) + \exp \left(-\frac{x^2}{c_3T} \exp\left(\frac{x^{{\eta_1}(1 - {\eta_1})}}{c_4 (\log x)^{\eta_1}}\right) \right).
$$ 
Let $x \asymp \sqrt{T( d_{\max} + \log T)} + (d_{\max} + \log T)^{1/{\eta_1}}$, by union bound and condition of Theorem \ref{thm:factor1}, we have, with probability at least $1 - \frac{1}{2}T^{-c_2}$, 
$$
\max_{\|u_{ik}\| = \|u_{jk}\|_2 = 1} \frac{1}{T} \sum_{t=1}^T \calE_t \otimes \calE_t \times_1 u_{i1} \times_{k=1}^K u_{ik}^\top \times_{k=K+1}^{2K} u_{jk}^\top \lesssim \sqrt{\frac{d_{\max} + \log(T)}{T}} + \frac{(d_{\max} + \log T)^{1/{\eta_1}}}{T} + 1, 
$$

which implies that 
$$
\max_{\|u_{ik}\| = \|u_{jk}\|_2 = 1} \frac{1}{T} \sum_{t=1}^T \calE_t \otimes \calE_t \times_1 u_{i1} \times_{k=1}^K u_{ik}^\top \times_{k=K+1}^{2K} u_{jk}^\top = O_p\left(\sqrt{\frac{d_{\max}}{T}} + \frac{d_{\max}^{1/{\eta_1}}}{T} + 1\right) .
$$

Therefore, we have 
\begin{equation}\label{eqn:D1_bound} 
    \Delta_1 = O_p\left(1 + \psi + \psi\left( \sqrt{\frac{d_{\max}}{T}} + \frac{d_{\max}^{1/{\eta_1}}}{T}\right)\right).
\end{equation} 

For $\Delta_2$, by \eqref{eqn:aikbikhat_bound}, 
$$
\prod_{k=1}^K a_{ik}^\top \hat b_{ik} \prod_{k=1}^K a_{jk}^\top \hat b_{jk} = \left(1 + O_p(\psi)\right)^{2K} = 1 + O_p(\psi). 
$$

Therefore 
\begin{align}\label{eqn:D2_bound} 
    \Delta_2 &= \frac{1}{T} \sum_{t=1}^T \tf_{it}\tf_{jt} O_p(\psi) \nonumber \\ 
    &\leq \left(\frac{1}{T} \sum_{t=1}^T \tf_{it}^2 \right)^{1/2} \left(\frac{1}{T} \sum_{t=1}^T \tf_{jt}^2\right)^{1/2} O_p(\psi) \nonumber \\ 
    &= \left(\frac{1}{T} \sum_{t=1}^T (\tf_{it}^2 - s_i^2) + s_i^2 \right)^{1/2} \left(\frac{1}{T} \sum_{t=1}^T (\tf_{jt}^2 - s_j^2) + s_j^2 \right)^{1/2} O_p(\psi) \nonumber \\ 
    &= O_p(s_i s_j \psi) .
\end{align} 
The last step is based on \eqref{eqn:ft2bound}.

For $\Delta_3$, 
\begin{align*}
\Delta_3  &=\frac{1}{T} \sum_{t=1}^T \sum_{l_1 \neq i} \sum_{l_2 \neq j} \tf_{l_1t} \tf_{l_2t} \prod_{k=1}^K a_{l_1k}^\top \hat b_{ik} \prod_{k=1}^K a_{l_2k}^\top \hat b_{jk} \nonumber \\ 
    &= \frac{1}{T}\sum_{t=1}^T\left(\sum_{l_1 \neq i} f_{l_1t}\prod_{k=1}^K a_{l_1k}^\top \hat b_{ik}\right) \left(\sum_{l_2 \neq j} f_{l_2t} \prod_{k=1}^K a_{l_2k}^\top \hat b_{jk}\right)\\ 
    &\leq \left(\frac{1}{T}\sum_{t=1}^T \left(\sum_{l_1 \neq i} f_{l_1t} \prod_{k=1}^K a_{l_1k}^\top \hat b_{ik}\right)^2\right)^{1/2} \left(\frac{1}{T}\sum_{t=1}^T \left(\sum_{l_2 \neq j} f_{l_2t} \prod_{k=1}^K a_{l_2k}^\top \hat b_{jk}\right)^2\right)^{1/2} \\ 
    &\leq \left(\sum_{l_1 \neq i} \sum_{l_2 \neq i} \frac{1}{T} \sum_{t=1}^T f_{l_1t} f_{l_2t}\right)^{1/2}\left(\sum_{l_1 \neq j} \sum_{l_2 \neq j} \frac{1}{T} \sum_{t=1}^T f_{l_1t} f_{l_2t}\right)^{1/2} O_p(\psi^{2K}). 
\end{align*}

By a similar argument with \eqref{eqn:ft2bound} and \eqref{eqn:D2_bound}, we have 
$$
\frac{1}{T} \sum_{t=1}^T f_{l_1t} f_{l_2t} = s_{l_1}s_{l_2} o_p(1) + s_{l_1} s_{l_2}.
$$
So for a fixed $r$, 
\begin{align}\label{eqn:D3_bound}
    \Delta_3 &= O_p(s_1^2 \psi^{2K}). 
\end{align}
By condition \eqref{eqn:thm1_cond}, $s_1 / s_r \psi^{K-1} \leq s_1 / s_r \psi_0^{K-1} \lesssim 1$, which implies 
\begin{equation}\label{eqn:s_1_s_r_psi}
    s_1 \psi^K \lesssim s_r \psi. 
\end{equation} 
Therefore, $\Delta_3 = O_p(s_r^2 \psi^2)$. 

Note that $\Delta_4$ and $\Delta_5$ have the similar bound. For $\Delta_4$, 
\begin{align}\label{eqn:D4_bound}
   \Delta_4 &=\frac{1}{T}\sum_{t=1}^T \sum_{l_2 \neq j} f_{it} f_{l_2t} \prod_{k=1}^K a_{ik}^\top \hat b_{ik} \prod_{k=1}^K a_{l_2k}^\top \hat b_{jk} \nonumber \\
    &= \frac{1}{T} \sum_{t=1}^T \sum_{l_2 \neq j} f_{it} f_{l_2t} O_p(1 + \psi) O_p(\psi^K) \nonumber \\ 
    &= O_p(s_i s_1 \psi^K) = O_p(s_i s_r \psi). \\
\end{align}
Similarly, $\Delta_5=O_p(s_j s_r \psi)$.

The bounds for $\Delta_6$ and $\Delta_7$ are similar. For $\Delta_6$, by \eqref{eqn:bikhat_bound}, \eqref{eqn:aikbikhat_bound} and bound for $\| \hat g_{ik} - g_{ik}\|_2$, 
\begin{align}\label{eqn:D6_expand} 
\Delta_6&= \frac{1}{T} \sum_{t=1}^T \tf_{it} \left(\prod_{k=1}^K a_{ik}^\top \hat b_{ik}\right) \left(\calE_t \times_{k=1}^K \hat b_{jk}^\top\right) \nonumber \\ 
&= \left(\frac{1}{T} \sum_{t=1}^T \tf_{it} \calE_t\right) \times_{k=1}^K \hat g_{jk}^\top \cdot O_p(1) \nonumber \\ 
&= \left(\left(\frac{1}{T} \sum_{t=1}^T \tf_{it} \calE_t\right) \times_1 (\hat g_{j1} - g_{j1}) \times_{k=2}^K \hat g_{jk}^\top + \left(\frac{1}{T} \sum_{t=1}^T \tf_{it} \calE_t\right) \times_1 g_{j1} \times_{k=2}^K \hat g_{j2}\right) O_p(1)  \nonumber \\ 
&\lesssim \left(\psi \max_{\|u_{jk}\|_2 = 1} \left(\frac{1}{T} \sum_{t=1}^T \tf_{it} \calE_t\right) \times_{k=1}^K u_{jk}^\top + \left(\frac{1}{T} \sum_{t=1}^T \tf_{it} \calE_t\right) \times_1 g_{j1} \times_{k=2}^K \hat g_{j2}\right) O_p(1) \nonumber \\
&\leq \cdots \notag \\ 
&\lesssim \left( \left(\frac{1}{T} \sum_{t=1}^T \tf_{it} \calE_t\right)\times_{k=1}^K g_{jk}^\top +  K\psi \max_{\|u_{jk}\|_2 = 1} \left(\frac{1}{T} \sum_{t=1}^T \tf_{it} \calE_t\right) \times_{k=1}^K u_{jk}^\top \right) O_p(1). 
\end{align}  

By assumption \ref*{asmp:error} and \ref*{asmp:factor}(iii), denote $g_i = \odot_{k=1}^K g_{ik}$,
\begin{align*}
    P(\frac{1}{T} \sum_{t=1}^T \tf_{it} e_t^\top g_i > x) &< \frac{g_i ^\top \E{\frac{1}{T^2} \left(\sum_{t=1}^T \tf_{it} e_t\right) \left(\sum_{t=1}^T \tf_{it} e_t^\top\right)} g_i}{x^2} \nonumber \\ 
    &= \frac{ s_i^2 g_i^\top \E{\frac{1}{T^2} \sum_{t=1}^T \sum_{s=1}^T f_{is}f_{it} e_s e_t^\top} g_i}{x^2} \\
    &= \frac{ s_i^2 g_i^\top \frac{1}{T^2} \sum_{s=1}^T \sum_{t=1}^T \E{f_{is}f_{it}} \E{e_s e_t^\top} g_i}{x^2} \\
    &\lesssim \frac{s_i^2}{Tx^2} .
\end{align*} 
Choosing $x \asymp s_i / \sqrt{T}$ yields
$$
\left(\frac{1}{T} \sum_{t=1}^T \tf_{it} \calE_t\right)\times_{k=1}^K g_{jk}^\top = O_p\left(\frac{s_i}{\sqrt{T}}\right). 
$$

Next, by Lemma \ref{lemma:expprod}, $f_{it}e_{t}$ is general sub-exponential with parameter $\nu_1\nu_2/(\nu_1 + \nu_2)$. Apply the same argument as in $\Delta_1$. With probability at least $1 - cT^{-c_2}$, 
$$
\max_{\|u_{jk}\|_2 = 1} \left(\frac{1}{T} \sum_{t=1}^T \tf_{it} \calE_t\right) \times_{k=1}^K u_{jk}^\top \lesssim s_i \left( \sqrt{\frac{d_{\max} + \log(T)}{T}} + \frac{(d_{\max} + \log(T))^{1/\eta_2}}{T}\right), 
$$
where $\eta_2 = \frac{\nu_1 + \nu_2}{\nu_1\nu_2} + \frac{1}{\gamma}$. So we have 
$$
\max_{\|u_{jk}\|_2 = 1} \left(\frac{1}{T} \sum_{t=1}^T \tf_{it} \calE_t\right) \times_{k=1}^K u_{jk}^\top = O_p\left(s_i\left(\sqrt{\frac{d_{\max}}{T}} + \frac{d_{\max}^{1/\eta_2}}{T}\right)\right).
$$

Therefore, 
\begin{align}\label{eqn:D6_bound}
    &\Delta_6 = O_p\left( \frac{s_i}{\sqrt{T}} + s_i \left(\sqrt{\frac{d_{\max}}{T}} + \frac{d_{\max}^{1/\eta_2}}{T}\right)\right).
\end{align} 

Similarly, 
\begin{align}\label{eqn:D7_bound}
&\Delta_7 = O_p\left( \frac{s_j}{\sqrt{T}} + s_j \left(\sqrt{\frac{d_{\max}}{T}} + \frac{d_{\max}^{1/\eta_2}}{T}\right)\right).
\end{align}

Note that $\Delta_8$ and $\Delta_9$ have the same bound. For $\Delta_8$, by results from $\Delta_7$, 
\begin{align}\label{eqn:D8_bound} 
    &\frac{1}{T} \sum_{t=1}^T \sum_{l \neq i} \tf_{lt} \left(\prod_{k=1}^K a_{lk}^\top \hat b_{ik}\right)\left(\calE_t \times_{k=1}^K \hat b_{jk}^\top\right) \nonumber \\ 
    &= \prod_{k=1}^K \| \hat b_{jk} \|_2 \sum_{l \neq i} \left(\frac{1}{T} \sum_{t=1}^T  \tf_{lt} \calE_t\right) \times_{k=1}^K \hat g_{jk}^\top \cdot O_p(\psi^K) \nonumber \\ 
    &= O_p\left(s_1 \frac{\psi^K}{\sqrt{T}} + s_1\psi^K\left( \sqrt{\frac{d_{\max}}{T}} + \frac{d_{\max}^{1/\eta_2}}{T} \right) \right) \nonumber \\ 
    &=O_p\left(s_r \frac{\psi}{\sqrt{T}} + s_r\psi\left( \sqrt{\frac{d_{\max}}{T}} + \frac{d_{\max}^{1/\eta_2}}{T} \right) \right). 
\end{align}
 
Putting \eqref{eqn:D1_bound} to \eqref{eqn:D8_bound} together, and as $1/\sqrt{T} \leq \sqrt{d_{\max} / T}\lesssim s_r \psi$, we have 
\begin{align}\label{eqn:fitfjt_bound}
\frac{1}{T} \sum_{t=1}^T \left(\htf_{it}\htf_{jt} - \tf_{it}\tf_{jt}\right)= O_p(1+ s_i s_j \psi) = O_p\left(1 + s_i s_j\sqrt{\frac{d_{\max}}{s_r^2 T}} + \frac{s_i s_j d_{\max}^{1/\eta_1}}{s_r^2T} + \frac{s_i s_j d_{\max}^{1/\eta_2}}{s_rT}\right). 
\end{align}

By \eqref{eqn:fitfjt_bound} and \eqref{eqn:ft2bound} and by Assumption \ref*{asmp:weakfactor}, denote $\Omega_i = \frac{1}{s_i^2} (\hat s_i^2 - s_i^2)$: 
\begin{align}\label{eqn:omega_bound}
\Omega_i := \frac{1}{s_i^2} (\hat s_i^2 - s_i^2) = O_p\left(\frac{1}{s_i^2} +\sqrt{\frac{d_{\max}}{s_r^2 T}} + \frac{ d_{\max}^{1/\eta_1}}{s_r^2 T} + \frac{ d_{\max}^{1/\eta_2}}{s_r T} + \sqrt{\frac{1}{T}}\right) = O_p\left(\psi + \sqrt{\frac{1}{T}}\right)= o_p(1). 
\end{align} 

Therefore, 
$$
\frac{1}{s_i^3} (\hat s_i^2 - s_i^2) = s_i^{-1} \Omega_i = s_i^{-1}o_p(1), 
$$
and 
$$
\hat s_i^{-1} - s_i^{-1} = s_i^{-1}\Omega_i + O(s_i^{-1} \Omega_i^2) = s_i^{-1} o_p(1). 
$$ 

By \eqref{eqn:fhathfexpand}, we have 
$$
\hat f_{it} - h_i f_{it} =O_p (\psi+ s_i^{-1})= O_p\left(\sqrt{\frac{d_{\max}}{d^{\alpha_r} T}} + \frac{d_{\max}^{1/\eta_1}}{d^{\alpha_r} T} + \frac{d_{\max}^{1/\eta_2}}{d^{\alpha_r/2} T} + \frac{1}{d^{\alpha_r/2}}\right). 
$$ 

For $\hat f_{it} - f_{it}$, observe that 
\begin{align*}
    (\hat s_i^{-1} - s_i^{-1}) \tf_{it} = s_i^{-1} \Omega_i \tf_{it} + O (s_i^{-1} \Omega_i^2 \tf_{it} )= O_p(\Omega_i)
\end{align*}

So, by \eqref{eqn:omega_bound}, 
\begin{align*}
\hat f_{it} - f_{it} = O_p\left(\psi + s_i^{-1} + \Omega_i \right) = O_p\left( \sqrt{\frac{d_{\max}}{s_r^2 T}} + \frac{ d_{\max}^{1/\eta_1}}{s_r^2 T} + \frac{ d_{\max}^{1/\eta_2}}{s_r T} + \frac{1}{s_i} + \sqrt{\frac{1}{T}} \right).
\end{align*}

For the central limit theorem \eqref{eqn:fclt}, denote $S = \diag(\{s_1, \ldots, s_r\}) \in \R^{r \times r}$. By \eqref{eqn:fhathfexpand} and the bounds proved, 
\begin{align*}
    \htf_{t} - \tf_t &= \left[  \calX_t \times_{k=1}^K \hat b_{ik}^\top - \tf_{it},\quad  i=1,...,r  \right]^\top  \\ 
    &= \left[  \calE_t \times_{k=1}^K \hat b_{ik}^\top + s_i f_{it} \prod_{k=1}^K a_{ik}^\top \hat b_{ik} -  s_i f_{it} + \sum_{j\neq i} s_j f_{jt} \prod_{k=1}^K a_{jk}^\top \hat b_{ik},\quad  i=1,...,r  \right]^\top  \\
    &=  \left[  \calE_t \times_{k=1}^K \hat b_{ik}^\top + O_p(s_i \psi) , \quad  i=1,...,r  \right]^\top  \\
    &= \left[  e_t^\top \hat b_{i} + O_p(s_i \psi) , \quad  i=1,...,r  \right]^\top \\ 
    &= \left[  \|\hat b_i\|_2 \, e_t^\top (\hat g_i - g_i) + \left(\| \hat b_i \|_2 - \| b_i \|_2 \right) e_t^\top g_i + e_t^\top b_i + O_p(s_i \psi) , \quad  i=1,...,r  \right]^\top \\ 
    &\leq \left[  \|\hat b_i\|_2 \|\hat g_i - g_i\|_2 \, \max_{\|u\|_2 = 1} e_t^\top u + \left(\| \hat b_i \|_2 - \| b_i \|_2 \right) e_t^\top g_i + e_t^\top b_i + O_p(s_i \psi) , \quad  i=1,...,r  \right]^\top .
\end{align*}

By \eqref{eqn:bikhat_bound}, 
$$
\| \hat b_i \|_2 = \| \odot_{k=K}^1 \hat b_{ik} \|_2 = \prod_{k=1}^K \|\hat b_{ik}\|_2 = O_p(1). 
$$ 

For $\| \hat g_i - g_i \|_2$, observe that 
\begin{align}\label{eqn:gihatgi_bound}
    \| \hat g_i - g_i \|_2 &= \| \hat g_{iK} \odot \hat g_{iK-1} \odot \ldots \odot \hat g_{i1} - g_{iK} \odot g_{iK-1} \odot \ldots \odot g_{i1} \|_2 \nonumber \\
    &= \| (g_{iK} + \hat g_{iK} - g_{ik}) \odot \hat g_{iK-1} \odot \cdots \odot \hat g_{i1} - g_{iK} \odot g_{iK-1} \odot \ldots \odot g_{i1} \|_2 \nonumber \\ 
    &= \| (\hat g_{iK} - g_{iK}) \odot \hat g_{iK-1} \odot \cdots \odot \hat g_{i1} + g_{iK} \odot \hat g_{iK-1} \odot \cdots \odot \hat g_{i1} - g_{iK} \odot g_{iK-1} \odot \cdots \odot g_{i1}\|_2 \nonumber\\ 
    &\leq \| \hat g_{iK} - g_{iK} \|_2 + \|g_{iK} \odot \hat g_{iK-1} \odot \cdots \odot \hat g_{i1} - g_{iK} \odot g_{iK-1} \odot \cdots \odot g_{i1} \|_2 \nonumber \\ 
    &= \cdots \nonumber \\ 
    &\leq \sum_{k=1}^K \| \hat g_{ik} - g_{ik} \|_2 + \| g_{iK} \odot g_{iK-1} \odot \cdots \odot g_{i1} - g_{iK} \odot g_{iK-1} \odot \cdots \odot g_{i1} \|_2 \nonumber \\ 
    &= \sum_{k=1}^K \| \hat g_{ik} - g_{ik} \|_2. 
\end{align} 

By \cite{infCP2024}, $\| \hat g_{ik} - g_{ik} \|_2 \leq C \psi$, so 
$$
\| \hat g_i - g_i \|_2 = O_p(K \psi) = O_p(\psi).
$$ 

By \eqref{eqn:ue_ord}, $\max_{\|u\|_2 = 1} e_t^\top u = O_p(1)$. So 
$$
\| \hat b_i \|_2 \|\hat g_i - g_i\|_2 \max_{\|u\|_2 = 1} e_t^\top u = O_p(\psi). 
$$

Note that 
$$
\| \hat b_{ik} - b_{ik} \|_2 \leq \max \{ \| \hat b_{ik}\|_2, \| b_{ik}\|_2 \} \| \hat g_{ik} - g_{ik} \|_2 = O_p(\psi). 
$$

By a similar argument with \eqref{eqn:gihatgi_bound}, 
\begin{align*}
    &\left(\| \hat b_i \|_2 - \| b_i \|_2 \right) e_t^\top g_i \\ 
    &\leq \| \hat b_i - b_i \|_2 \cdot e_t^\top g_i \\ 
    &\leq \sum_{k=1}^K \| \hat b_{ik} - b_{ik} \|_2 \cdot e_t^\top g_i \\  
    &= O_p(K \psi) = O_p(\psi).
\end{align*}

So 
$$
\htf_{t} - \tf_t = B^\top e_t + O_p(s_1 \psi) \xrightarrow{d} N(0, \Sigma_{Be}).
$$

\end{proof}  

\begin{proof}[Proof of Theorem \ref*{thm:beta}] 
    \begin{align}\label{eqn:betahatexpand}
        \sqrt{T} (\hat \beta - \tilde \beta ) &= \left(\frac{1}{T} \sum_{t=1}^{T-h} \hat z_t \hat z_t^\top \right)^{-1} \left(\frac{1}{\sqrt{T}} \sum_{t=1}^{T-h} \hat z_t \epsilon_{t+h} + \frac{1}{\sqrt{T}} \sum_{t=1}^{T-h} \hat z_t (\hat f_t - H f_t)^\top \tilde \beta\right). 
    \end{align} 
    Observe that 
    \begin{equation}\label{eqn:zzhat}
        \begin{split}
            \frac{1}{T} \sum_{t=1}^{T-h} \hat z_t \hat z_t^\top = \frac{1}{T} \sum_{t=1}^{T-h} z_t z_t^\top + \frac{1}{T} \sum_{t=1}^{T-h} (\hat z_t - z_t) z_t^\top + & \frac{1}{T} \sum_{t=1}^{T-h} z_t (\hat z_t - z_t)^\top \\ 
            & + \frac{1}{T} \sum_{t=1}^{T-h} (\hat z_t - z_t) (\hat z_t - z_t)^\top
        \end{split}
    \end{equation} 

    Recall $\Omega_i$ is defined in \eqref{eqn:omega_bound}. By Lemma \ref{lemma:samplemeanfitfjt} and Assumption \ref{asmp:weakfactor}, 
    \begin{align}\label{eqn:fhatfz_bound}
        \left\| \frac{1}{T} \sum_{t=1}^{T-h} (\hat z_t - z_t) z_t^\top \right\|_2 &= \left\|\frac{1}{T} \sum_{t=1}^{T-h} (\hat f_t - f_t)^\top z_t\right\| \nonumber \\ 
        &\leq \left(\frac{1}{T}\sum_{t=1}^{T-h} \|\hat f_t - f_t\|_2^2 \right)^{\frac{1}{2}} \left(\frac{1}{T} \sum_{t=1}^{T-h} \|z_t\|_2^2\right)^{\frac{1}{2}} \nonumber \\ 
        &= O_p\left(\max_i \Omega_i\right) = o_p(1),  
    \end{align} 
    and 
    \begin{align}\label{eqn: fhatfsq_bound}
        \left\| \frac{1}{T} \sum_{t=1}^{T-h} (\hat z_t - z_t) (\hat z_t - z_t)^\top \right\|_2 &= \left\| \frac{1}{T} \sum_{t=1}^{T-h} (\hat f_t - f_t) (\hat f_t - f_t)^\top \right\| \nonumber \\ 
        &\leq \frac{1}{T} \sum_{t=1}^{T-h} \|\hat f_t - f_t\|_2^2\\ 
        &= O_p(\max_i \Omega_i^2) = o_p(1).
    \end{align} 

    Then by \eqref{eqn:zzhat} and assumption \ref*{asmp:di}, 
    $$
    \frac{1}{T} \sum_{t=1}^{T-h} \hat z_t \hat z_t^\top = \frac{1}{T} \sum_{t=1}^{T-h} z_t z_t^\top + o_p(1) \xrightarrow{p} \Sigma_{zz}. 
    $$
For the second part of \eqref{eqn:betahatexpand}. Let $\tH = \diag(I_p, H)$ and $\tS = diag(I_p, S)$ and $\hat \tS = diag(I_p, \hat S)$. 
\begin{align*}
    \frac{1}{\sqrt{T}} \sum_{t=1}^{T-h} \hat z_t \epsilon_{t+h} &= \frac{1}{\sqrt{T}} \sum_{t=1}^{T-h}\tH z_t \epsilon_{t+h} + \frac{1}{\sqrt{T}} \sum_{t=1}^{T-h} (\hat z_t - \tH z_t) \epsilon_{t+h} \nonumber \\ 
\end{align*} 
For the first term, 
\begin{align*}
   \tilde H \frac{1}{\sqrt{T}} \sum_{t=1}^T z_t \epsilon_{t+h} &= \hat \tS^{-1} \tS \frac{1}{\sqrt{T}} \sum_{t=1}^T z_t \epsilon_{t+h} \nonumber \\ 
    &= (I_{p + r} + O(\max_i \Omega_i) ) \frac{1}{\sqrt{T}} \sum_{t=1}^T z_t \epsilon_{t+h} \nonumber \\ 
    &= \frac{1}{\sqrt{T}} \sum_{t=1}^T z_t \epsilon_{t+h} + o_p(1) \xrightarrow{d} N(0, \Sigma_{zz, \epsilon}). 
\end{align*}
For the second term, by Lemma \ref{lemma:samplemeanfitfjt} and the condition on Theorem \ref{thm:beta}, 
\begin{align*}
    \left\| \frac{1}{\sqrt{T}} \sum_{t=1}^{T-h} (\hat z_t - \tH z_t) \epsilon_{t+h} \right\|_2 &= \left\| \frac{1}{\sqrt{T}} \sum_{t=1}^{T-h} (\hat f_t - H f_t) \epsilon_{t+h} \right\|_2 \\ 
    &= O_p\left(\psi \left( \frac{\sqrt{d_{\max}}}{s_r} + \frac{d_{\max}^{1/\eta_7}}{s_r \sqrt{T}} + 1\right) + \frac{1}{s_r}\right) \\ 
    &= o_p(1). 
\end{align*} 

Therefore, by assumption \ref*{asmp:di} and conditions on Theorem \ref{thm:beta}, 
\begin{equation}\label{eqn:zhate}
    \frac{1}{\sqrt{T}} \sum_{t=1}^{T-h} \hat z_t \epsilon_{t+h} = \frac{1}{\sqrt{T}} \sum_{t=1}^{T-h}  z_t \epsilon_{t+h} + o_p(1) \xrightarrow{d} N(0, \Sigma_{zz, \epsilon}) . 
\end{equation} 

For the second term in the second part, by Lemma \ref{lemma:samplemeanfitfjt} and the condition on Theorem \ref{thm:beta},
\begin{align*}
    \left\|\frac{1}{\sqrt{T}} \sum_{t=1}^{T-h} \hat z_t (\hat f_t - H f_t)^\top \beta\right\|_2 &= \left\| \frac{1}{\sqrt{T}} \sum_{t=1}^{T-h} z_t (\hat f_t - H f_t)^\top \beta + \frac{1}{\sqrt{T}} \sum_{t=1}^{T-h} (\hat z_t - \tH z_t) (\hat f_t - H f_t)^\top \beta \right\|_2 \\ 
    &\leq \left\| \frac{1}{\sqrt{T}} \sum_{t=1}^{T-h} z_t (\hat f_t - H f_t)^\top  \right\|_2 \|\beta\|_2  + \left\| \frac{1}{\sqrt{T}} \sum_{t=1}^{T-h} (\hat f_t - H f_t) (\hat f_t - H f_t)^\top \right\|_2 \|\beta\|_2 \\ 
    &\leq  \left\| \frac{1}{\sqrt{T}} \sum_{t=1}^{T-h} z_t (\hat f_t - H f_t)^\top  \right\|_2 O_p(1) +  \frac{1}{\sqrt{T}} \sum_{t=1}^{T-h} \left\| \hat f_t - H f_t\right\|_2^2 O_p(1) \\ 
    &= O_p\left(\sqrt{T} \psi \left( \sqrt{\frac{d_{\max}}{s_rT}} + \frac{d_{\max}^{1/\eta_6}}{s_r T}+1\right) + \frac{1}{s_r}\right) + O_p\left(\sqrt{T} \psi^2 + \frac{\sqrt{T}}{s_r^2} \right) \\ 
    &= o_p(1), 
\end{align*} 

Putting them all together, we have 
$$
\sqrt{T} (\hat \beta - \beta) = \left(\frac{1}{T} \sum_{t=1}^{T-h} z_t z_t^\top\right) \left(\frac{1}{\sqrt{T}} \sum_{t=1}^{T-h} z_t \epsilon_{t+h}\right) + o_p(1) \xrightarrow{d} N(0, \Sigma_{zz}^{-1} \Sigma_{zz, \epsilon} \Sigma_{zz}^{-1}).
$$

\end{proof} 

\begin{proof}[Proof of Theorem \ref*{thm:di_pred}] 
    Observe that
    \begin{align*}
        \hat y_{T+h | T} - y_{T+h | T} &= w_T^\top \hat \beta_0 - w_T^\top \beta_0 + \hat f_T^\top \hat \beta_1 - f_T^\top H H^{-1} \beta_1 \\ 
        &= w_T^\top (\hat \beta_0 - \beta_0)  + \hat f_T^\top (\hat \beta_1 - H^{-1}\beta_1) + \beta_1^\top H^{-1} ( \hat f_T - H f_T)  \\ 
        &= \frac{1}{\sqrt{T}} \hat z_T^\top \sqrt{T}(\hat \beta - \tilde \beta) + \tilde \beta_1^\top \hat S^{-1} (\htf_T - \tf_T) \\ 
        \begin{split}
            &= \frac{1}{\sqrt{T}} z_T^\top \sqrt{T}(\hat \beta - \tilde \beta) + \frac{1}{\sqrt{T}} (\hat z_T - z_T)^\top \sqrt{T}(\hat \beta - \tilde \beta) + \tilde \beta_1^\top S^{-1} (\htf_T - \tf_T) \\ 
            &\quad + \tilde \beta_1^\top (\hat S^{-1} - S^{-1}) (\htf_T - \tf_T) 
        \end{split}\\
        &= \frac{1}{\sqrt{T}} z_T^\top \sqrt{T}(\hat \beta - \tilde \beta) + \frac{1}{\sqrt{T}} o_p(1) \sqrt{T}(\hat \beta - \tilde \beta) + \tilde \beta_1^\top S^{-1} (\htf_T - \tf_T) + \tilde \beta_1^\top o_p(1) (\htf_T - \tf_T) 
    \end{align*} 

    By Theorem \ref{thm:beta} and Theorem \ref{thm:factor1}, $\sqrt{T} (\hat \beta - \tilde \beta) \xrightarrow{d} N(0, \Sigma_{zz}^{-1} \Sigma_{zz, \epsilon} \Sigma_{zz}^{-1})$ and $\htf_T - \tf_T \xrightarrow{d} N(0, \Sigma_{Be})$. These two distributions are asymptotically independent since $\cE_t$ and $\epsilon_t$ are independent. Then the result follows. 

    
\end{proof} 

\begin{proof}[Proof of Theorem \ref*{thm:coverate}] ~\\ 
    By Lemma \ref{lemma:thresholding_bound}, it is sufficient to show that 
    $$
    \max_{j \leq d} \frac{1}{T} \sum_{t=1}^T (\hat e_{jt} - e_{jt})^2 = O_p\left( \frac{\log(d)}{T} + \frac{1}{s_r^2} \right). 
    $$
    Let $A_{j:}$ denote the $j^{th}$ row of $A$ and $A_{ji}$ denote the $(j,i)$ entry of $A$. Observe that 

    \begin{align*}
    \max_{j \leq d} \frac{1}{T} \sum_{t=1}^T (\hat e_{jt} - e_{jt})^2 &= \max_{j \leq d} \frac{1}{T} \sum_{t=1}^T \left(\hat A_{j:}^\top \htf_t - A_{j:}^\top \tf_t\right)^2 \\ 
    &= \max_{j \leq d} \frac{1}{T} \sum_{t=1}^T \left( \sum_{i=1}^r \hat A_{ji} \htf_{it} - A_{ji} \tf_{it} \right)^2 \\ 
    &\leq r \sum_{i=1}^r \max_{j \leq d} \frac{1}{T} \sum_{t=1}^T \left(\hat A_{ji} \htf_{it} - A_{ji} \tf_{it} \right)^2 
    \end{align*} 

    Since $r = O(1)$, it is sufficient to bound $\max_{j \leq d} \frac{1}{T} \sum_{t=1}^T \left(\hat A_{ji} \htf_{it} - A_{ji} \tf_{it} \right)^2$. 

    \begin{align*}
        \max_{j \leq d} \frac{1}{T} \sum_{t=1}^T \left(\hat A_{ji} \htf_{it} - A_{ji} \tf_{it} \right)^2 &= \max_{j \leq d} \frac{1}{T} \sum_{t=1}^T \left( (\hat A_{ji} - A_{ji}) \tf_{it} + A_{ji} (\htf_{it} - \tf_{it}) + (\hat A_{ji} - A_{ji})(\htf_{it} - \tf_{it}) \right)^2 \\ 
        \begin{split}
            &\leq 3 \frac{1}{T} \sum_{t=1}^T (\htf_{it} - \tf_{it})^2 \ \max_{j \leq d} A_{ji}^2 + 3 \frac{1}{T} \sum_{t=1}^T \tf_{it}^2 \ \max_{j \leq d} (\hat A_{ji} - A_{ji})^2 \\ 
            &\quad + 3 \frac{1}{T} \sum_{t=1}^T (\htf_{it} - \tf_{it})^2 \ \max_{j \leq d} (\hat A_{ji} - A_{ji})^2  
        \end{split} \\
        &= G_1 + G_2 + G_3. 
    \end{align*} 

    For $G_1$, by the proof of Lemma \ref{lemma:samplemeanfitfjt}(vi), $\frac{1}{T} \sum_{t=1}^T (\htf_{it} - \tf_{it})^2 = O_p\left( s_i^2 \psi^2 + 1\right)$, and by Assumption \ref{asmp:threshold}(i), $\max_{j \leq d} A_{ji}^2 = O_p(1 / s_i^2)$. Therefore, $G_1 = O_p(\psi^2 + \frac{1}{s_i^2})$. Note that if we assume $\max_{j \leq d} a_{ik,j} \leq c / \sqrt{d_k}$ where $a_{ik,j}$ is the $j^{th}$ entry of $a_{ik}$, we have $\max_{j \leq d} A_{ji} \lesssim 1 / \sqrt{d}$. In this case, we have $G_1 = O_p\left( \frac{s_i^2}{d} \psi^2 + \frac{1}{d} \right)$. 

    For $G_2$, $\frac{1}{T} \sum_{t=1}^T \tf_{it}^2 = O_p(s_i^2)$. For $\max_{j \leq d} (\hat A_{ji} - A_{ji})^2$, since $\max_{j \leq d} (\hat A_{ji} - A_{ji})^2 = (\max_{j \leq d} |\hat A_{ji} - A_{ji})|^2$, I will bound 
    $\max_{j \leq d} | \hat A_{ji} - A_{ji}|$. Denote the indices for $a_{ik}$ with respect to $j$ by $j_1, \ldots, j_K$ such that $\hat A_{ji} = \prod_{k=1}^K \hat a_{ik,j_k}$ and the counterpart for $A_{ji}$. Observe that
    \begin{align*}
        \max_{j \leq d} | \hat A_{ji} - A_{ji} | &= \max_{j \leq d} | \hat a_{i1,j_1} \hat a_{i2, j_2} \cdots \hat a_{iK, j_K} - a_{i1,j_1} a_{i2, j_2} \cdots a_{iK, j_K} | \\ 
        \begin{split}
            &\leq \binom{K}{1} \max_{j_k \leq d_k} | \hat a_{ik, j_k} - a_{ik,j_k} | \max_{j_l \leq d_l, l \neq k} \prod_{l} | a_{il,j_l} | \\ 
            &\quad + \binom{K}{2} \max_{j_{k_1} \leq d_{k_1}} | \hat a_{ik_1, j_{k_1}} - a_{ik_1, j_{k_1}} | \max_{j_{k_2} \leq d_{k_2}} | \hat a_{ik_2, j_{k_2}} - a_{ik_2,j_{k_2}} | \max_{j_l \leq d_l, l \neq k_1, k_2} \prod_l | a_{il,j_l} | \\ 
            &\quad + \cdots \\ 
            &\quad + \prod_{k=1}^K \max_{j_k \leq d_k} | \hat a_{ik,j_k} - a_{ik,j_k} |, 
        \end{split}
    \end{align*} 
    where the first term is the leading term. 

    For $1 \leq k \leq K$, note that 
    $$
    \begin{aligned}
        \max_{j_k \leq d_k} | \hat a_{ik,j_k} - a_{ik,j_k} | &= \max_{j_k \leq d_k} e_{j_k}^\top |\hat a_{ik} - a_{ik} | \\ 
        &= \max_{j_k \leq d_k} e_{j_k}^\top (I_{d_k} - a_{ik} a_{ik}^\top) \hat a_{ik} + e_{j_k}^\top a_{ik} a_{ik}^\top \hat a_{ik} - e_{j_k}^\top a_{ik} \\ 
        &\leq \max_{j_k \leq d_k} e_{j_k}^\top P_{a_{ik}}^\bot \hat a_{ik} + \max_{j_k \leq d_k} (e_{j_k}^\top a_{ik}) (a_{ik}^\top \hat a_{ik} - 1) \\ 
        &:= \Psi _1 + \Psi_2, 
    \end{aligned}
    $$ 
    where $e_{j_k}$ is the $j_k^{th}$ standard basis vector in $\mathbb{R}^{d_k}$ and $P_{a_{ik}}^\bot = I_{d_k} - a_{ik} a_{ik}^\top$. 

    By the definition of the CC-ISO algorithm by \cite{infCP2024}, $\hat a_{ik}$ is the top eigenvector of $\hat \Sigma_{ik}$, where 
    \begin{align*}
        \hat \Sigma_{ik} &= \hat \Sigma \times_{l \neq k, K+k}^{2K} \hat g_{il}^\top \\ 
        \begin{split}
            &= \left( \left(\prod_{l \neq k, K+k}^{2K} a_{il}^\top \hat g_{il}\right) \frac{1}{T} \sum_t s_i^2 f_{it}^2\right) a_{ik} a_{ik}^\top + \frac{1}{T} \sum_t \tilde f_{it} \left(\prod_{l \neq k}^K a_{il}^\top \hat g_{il}\right) \left(\calE_t \times_{l \neq k}^K \hat g_{il}^\top\right) a_{ik}^\top \\ 
            &\quad + \frac{1}{T} \sum_t \tilde f_{it} \left(\prod_{l \neq k}^K a_{il}^\top \hat g_{il}\right) a_{ik} \left(\calE_t \times_{l \neq k}^K \hat g_{il}^\top\right)^\top + \frac{1}{T} \sum_t \left(\calE_t \otimes \calE_t\right) \times_{l \neq k, K+k}^{2K} \hat g_{il}^\top  \\ 
            &\quad + \sum_{i_1 \neq i} s_{i_1}^2 \frac{1}{T} \sum_t \tf_{i_1,t}^2 \left( \prod_{l \neq k, K+k}^{2K} a_{i_1l}^\top \hat g_{i_1l} \right) a_{i_1k} a_{i_1k}^\top  \\ 
            &\quad + \sum_{i_1 \neq i_2} s_{i_1} s_{i_2} \frac{1}{T} \sum_t \tf_{i_1, t} \tf_{i_2, t} \left(\prod_{l \neq K}^K a_{i_1 l}^\top \hat g_{il}\right) \left(\prod_{l\neq K}^K a_{i_2l}^\top \hat g_{il}\right) a_{i_1k} a_{i_2k}^\top \\ 
            &\quad + \sum_{i_1 \neq i} s_{i_1} \frac{1}{T} \sum_t \tf_{i_1,t} \left(\prod_{l \neq k}^K a_{i_1l}^\top \hat g_{il}\right) \left(\calE_t \times_{l \neq k}^K \hat g_{il}^\top\right) a_{i_1k}^\top \\ 
            &\quad + \sum_{i_1 \neq i} s_{i_1} \frac{1}{T} \sum_t \tf_{i_1,t} \left(\prod_{l \neq k}^K a_{i_1l}^\top \hat g_{il}\right) a_{i_1k} \left(\calE_t \times_{l \neq k}^K \hat g_{il}^\top\right)^\top
        \end{split} \\ 
        &:= \tilde s_i^2 a_{ik} a_{ik}^\top + \phi_1 + \phi_2 + \phi_3 + \phi_4 + \phi_5 + \phi_6 + \phi_7 := \tilde s_i^2 a_{ik} a_{ik}^\top + \Phi, 
    \end{align*} 
    where $\tilde s_i^2 = \left( \left(\prod_{l \neq k, K+k}^{2K} a_{il}^\top \hat g_{il}\right) \frac{1}{T} \sum_t s_i^2 f_{it}^2\right)$ and $\Phi$ is the sum of the rest terms. By the proof of Theorem 4.2 and Theorem 4.3 of \cite{infCP2024}, $ \| \frac{1}{s_i^2} \Phi \|_2 = O_p(\psi^2)$ and $\| \phi_4 + \phi_5 + \phi_6 + \phi_7 \|_2 = O_p(\psi^2)$ when the algorithm coverges. And by equation \eqref{eqn:aikbikhat}, $\tilde s_i^2 = s_i^2 + s_i^2 O_p(\psi) \asymp s_i^2$. 

    Define $P_{a_{ik}} = a_{ik} a_{ik}^\top$, applying Theorem 1 of \cite{xia2021}, we have 
    \begin{align*}
        \hat a_{ik} \hat a_{ik}^\top - a_{ik} a_{ik}^\top &= \frac{1}{\tilde s_i^2} P_{a_{ik}}^\bot \Phi \Paik + \frac{1}{\tilde s_i^2} P_{a_{ik}} \Phi \Paikortho \\ 
        &\quad + \frac{1}{\tilde s_i^4} \left(\Paik \Phi \Paikortho \Phi \Paikortho + \Paikortho \Phi \Paikortho \Phi \Paik + \Paikortho \Phi \Paik \Phi \Paikortho\right) \\ 
        &\quad - \frac{1}{\tilde s_i^4} \left(\Paik \Phi \Paik \Phi \Paikortho + \Paik \Phi \Paikortho \Phi \Paik + \Paikortho \Phi \Paik \Phi \Paik \right) \\ 
        &\quad + R, 
    \end{align*}

    where $\| R \|_2 = O_p(\| \frac{1}{s_i^6} \Phi \|_2^3) = O_p(\psi^3)$ at convergence. 

    Pre and post multiplying by $a_{ik}$: 
    \begin{align*}
        (\hat a_{ik}^\top a_{ik})^2 - 1 &= a_{ik}^\top (\hat a_{ik} \hat a_{ik}^\top - a_{ik} a_{ik}^\top) a_{ik} \\ 
        &\asymp - \frac{1}{s_i^4} a_{ik}^\top \Phi \Paikortho \Phi a_{ik} \\ 
        &= O_p(\psi^2). 
    \end{align*}

    Thus, 
    \begin{align*}
        \Psi_2 &= \max_{j_k \leq d_k} (e_{j_k}^\top a_{ik}) (a_{ik}^\top \hat a_{ik} - 1) \\ 
        &= \max_{j_k \leq d_k} a_{ik,j_k} ( (a_{ik}^\top \hat a_{ik})^2 - 1) (a_{ik}^\top \hat a_{ik} + 1) \\ 
        &= O_p(\psi^2) \max_{j_k \leq d_k} a_{ik,j_k}. 
    \end{align*} 

    For $\Psi_1$, 
    \begin{align*}
        \max_{j_k \leq d_k} e_{j_k}^\top \Paikortho \hat a_{ik} &= \max_{j_k \leq d_k} e_{j_k}^\top (\hat a_{ik} \hat a_{ik}^\top - a_{ik} a_{ik}^\top) (\hat a_{ik} - a_{ik}) + e_{j_k}^\top(\hat a_{ik} \hat a_{ik}^\top - a_{ik} a_{ik}^\top)\\
    \end{align*} 

    The first term is bounded by $\|(\hat a_{ik} \hat a_{ik}^\top - a_{ik} a_{ik}^\top) \|_2 \|\hat a_{ik} - a_{ik}\|_2 = O_p(\psi^2).$ The second term is asymptotically equal to $s_i^{-2} e_{j_k}^\top \Paikortho \Phi a_{ik} + o_p\left(s_i^{-2} e_{j_k}^\top \Paikortho \Phi a_{ik}\right)$. 

    Expanding $\Phi$: 
    \begin{align*}
        \max_{j_k \leq d_k} s_i^{-2} e_{j_k}^\top \Paikortho \Phi a_{ik} &\lesssim s_i^{-2} \left( \prod_{l \neq k}^K a_{il}^\top \hat g_{il}\right) \max_{j_k \leq d_k} \frac{1}{T} \sum_t \tf_{it} e_{j_k}^\top \left(\calE_t \times_{l \neq k}^K \hat g_{il}^\top\right) \\ 
        &\quad + s_i^{-2} \max_{j_k \leq d_k}  \frac{1}{T} \sum_{t} e_{j_k}^\top \Paikortho \left(\calE_t \otimes \calE_t\right) \times_{l \neq k, K+k}^{2K} \hat g_{il}^\top \\ 
        &:= \Upsilon_1 + \Upsilon_2.
    \end{align*} 

    For $\Upsilon_1$, as $\prod_{l \neq k}^K a_{il}^\top \hat g_{il} = O_p(1)$, 
    \begin{align*}
        &s_i^{-1} \left( \prod_{l \neq k}^K a_{il}^\top \hat g_{il}\right) \max_{j_k \leq d_k} \frac{1}{T} \sum_t f_{it} e_{j_k}^\top \left(\calE_t \times_{l \neq k}^K \hat g_{il}^\top\right) \\ 
        &\lesssim s_i^{-1} \sum_{l \neq k, K+k}^{K} \left\| \hat g_{il} - g_{il}\right\|_2 \max_{j_k \leq d_k} \max_{\|u_l\|_2 = 1} \frac{1}{T} \sum_t f_{it} e_{j_k}^\top \Paikortho \calE_t \times_{l \neq k, K+k}^{K} u_l^\top \\ 
        &\quad + s_i^{-1} \max_{j_k \leq d_k} \frac{1}{T} \sum_t f_{it} e_{j_k}^\top \Paikortho \left(\calE_t \times_{l \neq k}^K g_{il}^\top\right)
    \end{align*} 
    The first term is bounded by $s_i^{-1} \sum_{l \neq k, K+k}^{K} \left\| \hat g_{il} - g_{il}\right\|_2 \max_{\|u_l\|_2 = 1} \frac{1}{T} \sum_t f_{it} \calE_t \times_{l=1}^{K} u_l^\top = O_p(\psi^2)$ by Equation \eqref{eqn:D6_bound}. 

    For the second term, denote $e_{t,a_{ik}} = \Paikortho \left(\calE_t \times_{l \neq k}^K g_{il}^\top\right)$. By Assumption \ref{asmp:error}, $e_{t,a_{ik}}$ is a general sub-exponential random vector with mean zero. Then, by Assumption \ref{asmp:threshold}(ii) and by the argument of Lemma A.3 of \cite{Fan2011}, which is by Bernstein's inequality for weak-dependent sub-exponential by \cite{merlevede2011}, we can show that 
    \begin{equation}\label{eqn:maxfe}
        \max_{j_k \leq d_k} \frac{1}{T} \sum_t | f_{it} e_{t,a_{ik}, j_k} | = O_p\left(\sqrt{\frac{\log(d_k)}{T}}\right). 
    \end{equation}

    Putting them together yields $\Upsilon_1 = O_p\left(\psi^2\right)$. 

    The bound of $\Upsilon_2$ can be derived similarly. By Assumption \ref{asmp:error} and \ref{asmp:threshold}(ii), we have $\Upsilon_2 = O_p\left(\psi^2 + \frac{1}{s_i^2}\right)$, which implies that 
    $$
    \Psi_1 = O_p\left(\psi^2 + \sqrt{\frac{\log(d_k)}{s_i^2 T}} + \frac{1}{s_i^2}\right).
    $$
    So we have 
    $$
    \max_{j_k \leq d_k} | \hat a_{ik,j_k} - a_{ik,j_k} | = O_p\left(\psi^2 + \frac{1}{s_i^2}\right). 
    $$
    Therefore, 
    $$
    \max_{j \leq d} | \hat A_{ji} - A_{ji} | = O_p\left(\psi^2 + \frac{1}{s_i^2}\right) \max_{j_l \leq d_l, l \neq k} \prod_{l} | a_{il,j_l} |.
    $$
    Then we have, by Assumption \ref{asmp:threshold}(i),
    $$
    G_2 = \frac{1}{T} \sum_{t=1}^T f_{it}^2 \ s_i^2 \max_{j \leq d} (\hat A_{ji} - A_{ji})^2 = O_p\left(\psi^4 + \frac{1}{s_i^4}\right), 
    $$
    which is dominated by $G_1$. If we assume $a_{ik,j} \leq c / \sqrt{d_k}$, we have $G_2 = O_p\left(\psi^4 + \frac{1}{s_i^4} \right)O_p\left(s_i^2 \frac{d_{\max}}{d}\right)$.
    
    Since $G_3$ is dominated by $G_1$ and $G_2$, we have 
    $$
    \max_{j \leq d} \frac{1}{T} \sum_{t=1}^T (\hat e_{jt} - e_{jt})^2 = O_p\left(\psi^2 + \frac{1}{s_r^2} \right). 
    $$

    By Assumption \ref{asmp:threshold}(iii), $\psi^2 + \frac{1}{s_r^2} = O\left(\log(d) / T + 1/ s_r^2\right)$. If we assume $a_{ik,j} \leq c / \sqrt{d_k}$, udner additional mild rate conditions, we have $\max_{j \leq d} \frac{1}{T} \sum_{t=1}^T (\hat e_{jt} - e_{jt})^2 = O_p\left(\psi^2 + \frac{1}{d}\right) = O_p\left(\frac{\log(d)}{T} + \frac{1}{d}\right)$.

\end{proof} 

\begin{proof}[Proof of Theorem \ref{thm:hd_di}] 
    To show (i) in the theorem, by an analogous argument of Lemma A.7 in \cite{Adamek2023}, one can show that, under the event 
    $$
    E_{\Sigma_V} = \left\{ \maxnorm{\hat \Sigma_{\hat V} - \Sigma_V} \leq C / p_0 \right\}
    $$
    and Assumption \ref{asmp:hd_di}(iii), suppose the tuning parameter $\lambda \geq \frac{C'}{T} \infnorm{\tU^\top \hat V}$, where $\tU = \tilde Y - \hat V \beta_0$, for some constant $C, C'$ that are large enough, then 
    $$
    \onorm{\hat \beta_0 - \beta_0} \lesssim \ p_0 \lambda. 
    $$ 
    By Lemma \ref{lemma:eventEv_bound}, we have the probability of $E_{\Sigma_V}$ approaches to 1 under the assumptions of Theorem \ref{thm:hd_di}. By Lemma \ref{lemma:maxutvhat_bound}, we have $\frac{1}{T}\infnorm{\tU^\top \hat V} = O_p\left( \psi^2 + \frac{1}{s_r^2} + \sqrt{\frac{\log(d)}{T}}\right)$. So the result for (i) follows. 

    For (ii), observe that 
    $$
    \hat \beta_1 - H^{-1} \beta_1 = \left( \hat \beta_1^* - H^{-1} \beta_1^*\right) - \left( \hat \Lambda - \Lambda H^{-1}\right)^\top \hat \beta_0 - H^{-1} \Lambda^\top \left( \hat \beta_0 - \beta_0\right). 
    $$ 
    So we have 
    \begin{align*}
        \tnorm{\hat \beta_1 - H^{-1} \beta_1} &\leq \tnorm{\hat \beta_1^* - H^{-1} \beta_1^*} + \max_{j \leq p} \tnorm{\hat \Lambda_j - H^{-1} \Lambda_j} \onorm{\beta_0} \\ 
        &\quad + \max_{j \leq p} \tnorm{\Lambda_j} \tnorm{H^{-1}} \onorm{ \hat \beta_0 - \beta_0} + \max_{j \leq p} \tnorm{\hat \Lambda_j - H^{-1} \Lambda_j} \onorm{ \hat \beta_0 - \beta_0} \\ 
        &= O_p\left(p_0  \left( \psi + \frac{1}{s_r} + \sqrt{\frac{\log(p)}{T}}\right)\right), 
    \end{align*}
    by Lemma \ref{lemma:beta1star_bound}, \ref{lemma: lambdaj_bound} and the result of (i).
    For (iii), observe that 
    \begin{align*}
        \abs{\hat y_{T+h | T} - y_{T+h | T}} &= \abs{\hat V_T^\top \hat \beta_0 + \hat f_T^\top \hat \beta_1^* - V_T^\top \beta_0 - f_T^\top H H^{-1} \beta_1^*} \\ 
        &\leq \abs{V_T^\top \left( \hat \beta_0 - \beta_0\right)} + \abs{\left(\hat V_T - V_T\right)^\top \beta_0} + \abs{f_T^\top H \left( \hat \beta_1^* - \beta_1^*\right)} \\ 
        &\quad + \abs{\left( \hat f_T - H f_T\right)^\top H^{-1} \beta_1^*} + \abs{ \left(\hat V_T - V_T\right)^\top \left( \hat \beta_0 - \beta_0\right)} \\ 
        &\quad + \abs{\left(\hat f_T - H f_T\right)^\top \left( \hat \beta_1^* - H^{-1} \beta_1^*\right)} \\ 
        &\leq \infnorm{V_T} \onorm{\hat \beta_0 - \beta_0} + \infnorm{\hat V_T - V_T} \onorm{\beta_0} + \tnorm{f_T} \tnorm{H} \tnorm{\hat \beta_1^* - H^{-1} \beta_1^*} \\ 
        &\quad + \tnorm{\hat f_T - H f_T} \tnorm{H^{-1}} \tnorm{\beta_1^*} + \infnorm{\hat V_T - V_T} \onorm{\hat \beta_0 - \beta_0} \\ 
        &\quad + \tnorm{\hat f_T - H f_T} \tnorm{\hat \beta_1^* - H^{-1} \beta_1^*} \\ 
        &:= \calI_1 + \calI_2 + \calI_3 + \calI_4 + \calI_5 + \calI_6.
    \end{align*} 
    For $\calI_1$, by Assumption \ref{asmp:hd_di}(i) and Bonferroni's inequality, we have 
    $$
    \PP\left(\max_{j \leq p} |V_{Tj}| > t\right) \leq p \exp\left(-\frac{t^{\eta_1}}{C}\right). 
    $$
    Let $t = \left( C' \log(p)\right)^{1/\eta_1}$ for some $C' > C$, we have 
    $$
    \PP\left(\max_{j \leq p} |V_{Tj}| > \left( C' \log(p)\right)^{1/\eta_1}\right) \to 0, 
    $$
    which implies $\infnorm{V_T}= O_p\left(\log(p)^{1/\eta_1}\right)$. By the result on (i), we have 
    $$
    \calI_1 = O_p\left( p_0\log(p)^{1/\eta_1} \left(\psi^2 + \frac{1}{s_r^2} + \sqrt{\frac{\log(p)}{T}}\right)\right). 
    $$
    For $\calI_2$, we have 
    \begin{align*}
        \max_{j \leq p} \left| \hat V_{Tj} - V_{Tj} \right| &\leq \max_{j \leq p} \tnorm{\hat \Lambda_j - H^{-1} \Lambda_j} \tnorm{\hat f_T} + \max_{j \leq p} \tnorm{\Lambda_j} \tnorm{H^{-1}} \tnorm{\hat f_T - H f_T} \\ 
        &= O_p\left(\psi + \frac{1}{s_r} + \sqrt{\frac{\log(p)}{T}}\right), 
    \end{align*}
    by Lemma \ref{lemma: lambdaj_bound} and Theorem \ref{thm:factor1}. 
    So we have 
    $$
    \calI_2 = O_p\left(p_0 \left(\psi + \frac{1}{s_r} + \sqrt{\frac{\log(p)}{T}}\right)\right). 
    $$
    And $ \calI_3 = O_p\left(p_0 \left(\psi + \frac{1}{s_r} + \sqrt{\frac{\log(p)}{T}}\right)\right)$ by Lemma \ref{lemma:beta1star_bound}. $\calI_4 = O_p\left(\psi\right)$ by Theorem \ref{thm:factor1}. $\calI_5$ and $\calI_6$ are dominated by $\calI_2$ and $\calI_3$ and $\calI_4$. By the rate condition on Theorem \ref{thm:hd_di}, we have $\log(p)^{1/\eta_1} \left(\psi + 1/s_r\right) = o(1)$. So we have the result for (iii).
    
\end{proof}
\section{Lemmas and Proofs}  

\begin{lemma}\label{lemma:samplemeanfitfjt}
    Under assumptions of Theorem \ref{thm:factor1}, 
    $$
    \hat f_{it} = \hat s_i^{-1} \htf_{it}, 
    $$
    then  
    \begin{itemize}
        \item[(i)]  $\frac{1}{T} \sum_{t=1}^T \left( \hat f_{it} \hat f_{jt} - h_i h_j f_{it} f_{jt}\right) = O_p( \psi)$; 
        \item[(ii)] $\frac{1}{T} \sum_{t=1}^T \left( \hat f_{it} \hat f_{jt} - f_{it} f_{jt}\right) = O_p(\Omega)= O_p( \psi+T^{-1/2}) $;  
        \item[(iii)] $\frac{1}{T} \sum_{t=1}^T \left\| \hat f_{t} - H f_{t} \right\|_2 = O_p(\psi + \frac{1}{s_r})$; 
        \item[(iv)] $\frac{1}{T} \sum_{t=1}^T \left\| \hat f_{t} - f_{t} \right\|_2 = O_p(\Omega) = O_p( \psi+T^{-1/2}) $; 
        \item[(v)] $\frac{1}{T} \sum_{t=1}^T \left\| \hat f_{t} - H f_{t} \right\|_2^2 = O_p(\psi^2 + \frac{1}{s_r^2})$; 
        \item[(vi)] $\frac{1}{T} \sum_{t=1}^T \left\| \hat f_{t} - f_{t} \right\|_2^2 = O_p(\Omega^2) = O_p(\psi^2+ s_r^{-2} + T^{-1})$; 
        \item[(vii)] $\left\| \frac{1}{\sqrt{T}} \sum_{t=1}^{T} (\hat f_t - H f_t) \epsilon_{t+h} \right\|_2 = O_p\left(\psi \left( \frac{\sqrt{d_{\max}}}{s_r} + \frac{d_{\max}^{1/\eta_5}}{s_r \sqrt{T}} + 1\right) + \frac{1}{s_r}\right)$; 
        \item[(viii)] $\left\| \frac{1}{\sqrt{T}} \sum_{t=1}^{T} z_t (\hat f_t - H f_t)^\top \right\|_2 = O_p\left( \sqrt{T} \psi \left( \sqrt{\frac{d_{\max}}{s_r^2T}} + \frac{d_{\max}^{1/\eta_4}}{s_r T} + 1\right) + \frac{1}{s_r}\right)$. 
        
    \end{itemize}
    where $\Omega$ is defined in \eqref{eqn:omega_bound}. 
    
\end{lemma}

\begin{proof}[Proof of Lemma \ref*{lemma:samplemeanfitfjt}] ~\\
    Let $\Omega=\max_{1\le i\le r}\Omega_i$, where $\Omega_i= s_i^{-2} (\hat s_i^2 - s_i^2)$ is given in \eqref{eqn:omega_bound}. Then, by \eqref{eqn:sitaylor}, we have
    \begin{align*}
     \hat s_i^{-1} - s_i^{-1} &= -\frac12 s_i^{-1} \Omega_i + s_i^{-1} O(\Omega_i^2) ,\\
     \Omega&=\max_{1\le i\le r}\Omega_i =O_p(\psi+T^{-1/2})=o_p(1).
    \end{align*}
    
    For $(i)$ and $(ii)$,
    \begin{align*}
        &\frac{1}{T} \sum_{t=1}^T \left( \hat f_{it} \hat f_{jt} - f_{it} f_{jt}\right) \\
        &=\frac{1}{T} \sum_{t=1}^T \hat s_i^{-1} \hat s_j^{-1} \htf_{it} \htf_{jt} - s_i^{-1} s_j^{-1} \tf_{it} \tf_{jt} \\ 
        &= \hat s_i^{-1} \hat s_j^{-1} \frac{1}{T} \sum_{t=1}^T (\htf_{it} \htf_{jt} - \tf_{it} \tf_{jt}) + (\hat s_i^{-1} \hat s_j^{-1} - s_i^{-1} s_j^{-1}) \frac{1}{T} \sum_{t=1}^T \tf_{it} \tf_{jt} \\ 
        &= \frac{1}{T} \sum_{t=1}^T \left( \hat f_{it} \hat f_{jt} - h_i h_j f_{it} f_{jt}\right) + (\hat s_i^{-1} \hat s_j^{-1} - s_i^{-1} s_j^{-1}) \frac{1}{T} \sum_{t=1}^T \tf_{it} \tf_{jt} \\ 
        &:= D_1 + D_2. 
    \end{align*}
    For $D_1$, 
    \begin{align*}
        \hat s_i^{-1} \hat s_j^{-1}&= (\hat s_i^{-1} - s_i^{-1} + s_i^{-1})(\hat s_j^{-1} - s_j^{-1} + s_j^{-1}) \\ 
        &= (\hat s_i^{-1} - s_i^{-1})(\hat s_j^{-1} - s_j^{-1}) + (\hat s_i^{-1} - s_i^{-1})s_j^{-1} + s_i^{-1}(\hat s_j^{-1} - s_j^{-1}) + s_i^{-1}s_j^{-1} \\ 
        &= \frac14 s_i^{-1}s_j^{-1} \big(\Omega_i\Omega_j +O(\Omega_i^2\Omega_j^2) \big) - \frac12 s_i^{-1} s_j^{-1} \big(\Omega_i+\Omega_j +O(\Omega_i^2+\Omega_j^2) \big) +  s_i^{-1}s_j^{-1} \\ 
        &= s_i^{-1}s_j^{-1} (1 + O(\Omega) ). 
    \end{align*}
    By \eqref{eqn:fitfjt_bound}, 
    $$
    \begin{aligned}
       D_1  = O_p(\psi)(1 + O(\Omega) ) =O_p (\psi)
    \end{aligned}
    $$
    For $D_2$, by the argument above
    \begin{align*}
       \hat s_i^{-1} \hat s_j^{-1} - s_i^{-1} s_j^{-1} = s_i^{-1}s_j^{-1}  O(\Omega) .
    \end{align*} 
    By \eqref{eqn:fitfjt_bound}, 
    $$ 
    D_2 = O_p\left(\left(1 + \frac{1}{\sqrt{T}}\right)\Omega \right) = O_p(\Omega). 
    $$ 
    Therefore 
    $$ 
    \frac{1}{T} \sum_{t=1}^T \left( \hat f_{it} \hat f_{jt} - f_{it} f_{jt}\right) = O_p(\Omega)= O_p( \psi+T^{-1/2}).
    $$ 
    For $(iii)$, we have 
    $$
    \frac{1}{T} \sum_{t=1}^T \left\| \hat f_{t} - H f_{t} \right\|_2 = \frac{1}{T} \sum_{t=1}^T \sqrt{\sum_{i=1}^r (\hat f_{it} - h_i f_{it})^2} \leq \sum_{i=1}^r \frac{1}{T} \sum_{t=1}^T \left| \hat f_{it} - h_i f_{it} \right|. 
    $$
    As $r$ is fixed, it is sufficient to show that 
    $$
    \frac{1}{T} \sum_{t=1}^T \left| \hat f_{it} - h_i f_{it} \right| = O_p(\psi + \frac{1}{s_i}).
    $$
    The same argument applies to $(iv)$ $\sim$ $(vi)$. 
    For $(iii)$ and $(iv)$, 
    \begin{align*}
        \frac{1}{T} \sum_{t=1}^T \left( \hat f_{it} - f_{it} \right) &= \frac{1}{T} \sum_{t=1}^T \hat s_i^{-1} \htf_{it} - s_i^{-1} \tf_{it} \\ 
        &= \frac{1}{T} \sum_{t=1}^T \hat s_i^{-1} (\htf_{it} - \tf_{it}) + (\hat s_i^{-1} - s_i^{-1}) \frac{1}{T} \sum_{t=1}^T \tf_{it} \\ 
        &= \frac{1}{T} \sum_{t=1}^T (\hat f_{it} - h_i f_{it}) + (\hat s_i^{-1} - s_i^{-1}) \frac{1}{T} \sum_{t=1}^T \tf_{it} \\
        &= M_1 + M_2. 
    \end{align*} 
    Following similar argument with $D_1$ and the proof of Theorem \ref{thm:factor1}, 
    $$
    M_1 = O_p(\psi + \frac{1}{s_i}).
    $$
    And similar to the argument for $D_2$, 
    $$
    M_2 = O_p(\Omega).
    $$
    So the result follows. ~\\~\\ 
    For $(v)$ and $(vi)$, 
    \begin{align*}
        \frac{1}{T} \sum_{t=1}^T (\hat f_{it} - f_{it})^2 &= \frac{1}{T} \sum_{t=1}^T (\hat s_i^{-1} \htf_{it} - s_i^{-1} \tf_{it})^2 \\
        &= \frac{1}{T} \sum_{t=1}^T (\hat s_i^{-1} \htf_{it} - \hat s_i^{-1} \tf_{it} + \hat s_i^{-1} \tf_{it} - s_i^{-1} \tf_{it})^2 \\ 
        &\leq 2 \hat s_i^{-2} \frac{1}{T} \sum_{t=1}^T (\htf_{it} - \tf_{it})^2 + 2 (\hat s_i^{-1} - s_i^{-1})^2 \frac{1}{T} \sum_{t=1}^T \tf_{it}^2 \\ 
        &= 2 \frac{1}{T} \sum_{t=1}^T \left( \hat f_{it} - h_i f_{it} \right)^2 + 2 (\hat s_i^{-1} - s_i^{-1})^2 \frac{1}{T} \sum_{t=1}^T \tf_{it}^2 \\ 
        &= N_1 + N_2.
    \end{align*}
    For $N_2$, 
    $$
    (\hat s_i^{-1} - s_i^{-1})^2 \frac{1}{T} \sum_{t=1}^T \tf_{it}^2 = s_i^{-2} \Omega^2 O_p(s_i^2) = O_p(\Omega^2).
    $$
    For $N_1$, by Taylor expansion, 
    $$
    \begin{aligned}
    \hat s_i^{-2} &= \hat s_i^{-2} - s_i^{-2} + s_i^{-2} \\ 
    &=s_i^{-2}(\Omega + O(\Omega^2)) + s_i^{-2} \\ 
    &= s_i^{-2} + s_i^{-2}o_p(1). 
    \end{aligned}
    $$

    Expanding the square, 
    \begin{align*}
        &\frac{1}{T} \sum_{t=1}^T (\htf_{it} - \tf_{it})^2 \\ 
        &= \frac{1}{T} \sum_{t=1}^T\left( \tf_{it} \left( \prod_{k=1}^K a_{ik}^\top \hat b_{ik} - 1\right)+ \sum_{j \neq i}^r \tf_{jt} \prod_{k=1}^K a_{jk}^\top\hat b_{ik} + \cE_t \times_{k=1}^K \hat b_{ik}^\top\right)^2 \\ 
        \begin{split}
            &= \left( \prod_{k=1}^K a_{ik}^\top \hat b_{ik} - 1\right)^2 \frac{1}{T} \sum_{t=1}^T \tf_{it}^2 + 2 \sum_{j \neq i}^r \left( \prod_{k=1}^K a_{ik}^\top \hat b_{ik} - 1\right) (\prod_{k=1}^K a_{jk}^\top\hat b_{ik}) \frac{1}{T} \sum_{t=1}^T \tf_{it} \tf_{jt} \\ 
            &\quad + 2 \left( \prod_{k=1}^K a_{ik}^\top \hat b_{ik} - 1\right) \frac{1}{T} \sum_{t=1}^T \tf_{it} \cE_t \times_{k=1}^K \hat b_{ik}^\top + \sum_{j \neq i}^r \sum_{l \neq i}^r (\prod_{k=1}^K a_{jk}^\top\hat b_{ik})(\prod_{k=1}^K a_{lk}^\top\hat b_{ik}) \frac{1}{T} \sum_{t=1}^T \tf_{jt} \tf_{lt} \\ 
            &\quad + 2 \sum_{j\neq i}^r (\prod_{k=1}^K a_{jk}^\top\hat b_{ik}) \frac{1}{T} \sum_{t=1}^T \tf_{jt} \cE_t \times_{k=1}^K \hat b_{ik}^\top + \frac{1}{T} \sum_{t=1}^T (\cE_t \times_{k=1}^K \hat b_{ik}^\top)^2 \\ 
            &:= \Pi_1 + \Pi_2 + \Pi_3 + \Pi_4 + \Pi_5 + \Pi_6 
        \end{split}
    \end{align*}
    By similar argument in \eqref{eqn:fitfjt_bound} and Assumption \ref{asmp:weakfactor}, 
    \begin{align*}
        \Pi_1 &= O_p(\psi^2 s_i^2) \\ 
        \Pi_2 &= O_p(\psi^{K+1} s_1 s_i) = O_p(\psi^2 s_r s_i) \\ 
        \Pi_3 &= O_p\left(\psi s_i\left(\frac{1}{\sqrt{T}} + \sqrt{\frac{d_{\max}}{T}} + \frac{d_{\max}^{1/\nu^*}}{T}\right)\right) = O_p(\psi^2 s_i s_r)\\ 
        \Pi_4 &= O_p(\psi^{3} s_r^2) \\ 
        \Pi_5 &= O_p\left(\psi s_r\left(\frac{1}{\sqrt{T}} + \sqrt{\frac{d_{\max}}{T}} + \frac{d_{\max}^{1/\nu^*}}{T}\right) \right) = O_p\left( s_r^2 \psi^2\right)\\  
        \Pi_6 &= O_p\left(1 + \psi + \psi\left( \sqrt{\frac{d_{\max}}{T}} + \frac{d_{\max}^{1/\nu}}{T}\right)\right) = O_p(1+\psi+ s_r^2\psi^2). 
    \end{align*} 
    So 
    $$
    N_1 = O_p\left(\psi^2 + \psi \frac{1}{s_i}\left(\sqrt{\frac{d_{\max}}{T}} + \frac{d_{\max}^{1/\nu^*}}{T}\right) + \frac{1}{s_i^2}\right) = O_p(\psi^2 + \frac{1}{s_i^2}). 
    $$
    And 
    $$
    N_1 + N_2 = O_p(\Omega^2 + \psi^2 + \frac{1}{s_i^2}) = O_p(\psi^2+ s_i^{-2} + T^{-1}).
    $$ 

    For $(vii)$, Since $r$ is fixed, it is sufficient to show that 
    $$
    \frac{1}{\sqrt{T}} \sum_{t=1}^T (\hat f_{it} - h_i f_{it}) \epsilon_{t+h} = O_p\left(\psi \frac{\sqrt{d_{\max}}}{s_r} + \psi \frac{d_{\max}^{1/\nu^*}}{s_i \sqrt{T}} + \frac{1}{s_r} \right). 
    $$ 
    Note that 
    \begin{align*}
       \frac{1}{\sqrt{T}} \sum_{t=1}^T (\hat f_{it} - h_i f_{it}) \epsilon_{t+h} &=  \frac{1}{\sqrt{T}} \sum_{t=1}^T \hat s_i^{-1} (\htf_{it} - \tf_{it}) \epsilon_{t+h} \\
       \begin{split}
           &= \hat s_i^{-1} \frac{1}{\sqrt{T}} \sum_{t=1}^T \tf_{it} \epsilon_{t+h} \left(\prod_{k=1}^K a_{ik}^\top \hat b_{ik} - 1\right) + \hat s_i^{-1} \sum_{j \neq i} \frac{1}{\sqrt{T}} \sum_{t=1}^T \tf_{jt} \epsilon_{t+h} \left( \prod_{k=1}^K a_{jk}^\top \hat b_{ik}\right) \\
           &\quad + \hat s_i^{-1} \frac{1}{\sqrt{T}} \sum_{t=1}^T \epsilon_{t+h}  \left(\calE_t \times_{k=1}^K \hat b_{ik}^\top\right)
       \end{split}\\ 
       &:= \Phi_1 + \Phi_2 + \Phi_3
    \end{align*} 
    For $\Phi_1$, $\left(\prod_{k=1}^K a_{ik}^\top \hat b_{ik} - 1\right) = O_p(\psi)$, and by Assumption \ref{asmp:factor} and \ref{asmp:di}, 
    \begin{align*}
        \frac{1}{\sqrt{T}} \hat s_i^{-1} \sum_{t=1}^T \tf_{it} \epsilon_{t+h} &= (\hat s_i^{-1} - s_i^{-1}) \frac{1}{\sqrt{T}} \sum_{t=1}^T \tf_{it} \epsilon_{t+h} + s_i^{-1} \frac{1}{\sqrt{T}} \sum_{t=1}^T \tf_{it} \epsilon_{t+h} \\ 
        &= O_p\left(\Omega_i\right) + O_p\left(1\right). 
    \end{align*} 
    So, $\Phi_1 = O_p \left(\psi\right)$. As $\Phi_2$ is similar to $\Phi_1$, we have $\Phi_2 = O_p\left( \frac{s_1 \psi^K}{s_i}\right) = O_p\left( \frac{s_r \psi }{s_i}\right) = O_p(\psi).$  ~\\ 
For $\Delta_6$, following the same argument for $\Delta_6$ but replacing $f_{it}$ with $\epsilon_{t+h}$, we have 
$$
\Phi_3 = \sqrt{T} \hat s_i^{-1} O_p\left(\psi \right) \max_{\|u_{ik}\|=1} \frac{1}{T} \sum_{t=1}^T  \epsilon_{t+h} \left(\calE_t \times_{k=1}^K u_{ik}^\top\right) + \sqrt{T} \hat s_i^{-1} \frac{1}{T} \sum_{t=1}^T\epsilon_{t+h} \left(\calE_t \times_{k=1}^K b_{ik}^\top\right). 
$$

By Lemma \ref{lemma:expprod} together with Assumption \ref{asmp:factor_error} and \ref{asmp:di}, for any unit vector $u_{ik}$, $k=1,\ldots,K$, $\epsilon_{t+h} \calE_t \times_{k=1}^K u_{ik}$ has exponential tail probability bound with coefficient $\nu_1\nu_4 / (\nu_1 + \nu_4)$. By the argument for $\Delta_6$ and CLT for $\alpha$-mixing process, 
$$
\Phi_3 = O_p\left( \sqrt{T} \psi \sqrt{\frac{d_{\max}}{s_i^2 T} } + \frac{d_{\max}^{1/\eta_7}}{s_iT} \right) + O_p\left( \frac{1}{s_i} \right), 
$$
where $1/\eta_7 =(\nu_1 + \nu_4) /  (\nu_1\nu_4) + 1/\gamma$. Result for $(vii)$ follows. Analysis for $(viii)$ is similar. Following the same decomposition and argument of the bound, we can show that 
$\Phi_1' = \Phi_2' = O_p\left( \sqrt{T} \psi \right)$ and $\Phi'_3 = O_p\left( \sqrt{T} \psi \sqrt{\frac{d_{\max}}{s_i^2 T} } + \frac{d_{\max}^{1/\eta_7}}{s_iT} \right) + O_p\left( \frac{1}{s_i} \right)$, where $\Phi'_i$ is the counterpart of the decomposition $\Phi_i$ for $(vii)$. The rate of $\Phi'_1$ and $\Phi'_2$ is different from $(vii)$ because the process $f_{it}z_t$ is not necessarily mean zero. 
    

\end{proof} 


\begin{lemma}\label{lemma:hateit2_bound}
    Denote the $(i,j)^{th}$ element of $\Sigma_e$ as $\sigma_{ij}$ and denote $\hat \sigma_{ij} = \frac{1}{T} \sum_{t=1}^T \hat e_{it} \hat e_{jt}$. Suppose Assumption \ref{asmp:error} and \ref{asmp:threshold}(ii) hold. And assume that 
    $$
    P \left(\max_{i \leq d} \frac{1}{T} \sum_{t=1}^T (\hat e_{it} - e_{it})^2 > C a_T^2 \right) < O(\kappa(d,T)) 
    $$
    for some $a_T = o(1)$ and $\kappa(d,T) = o(1)$. Then we have
    $$
    P\left( \max_{i,j \leq d} \left| \hat \sigma_{ij} - \sigma_{ij} \right| \leq C\left(a_T + \sqrt{\frac{\log(d)}{T}}\right) \right) \geq 1 - O(d^{-2}) - O(\kappa(d,T)), 
    $$
    for some constant $C>0$. 
\end{lemma} 

Lemma \ref{lemma:hateit2_bound} is part of Lemma A.3 in \cite{Fan2011}. The proof is omitted here. 

\begin{lemma}\label{lemma:thresholding_bound}
    Suppose Assumption \ref{asmp:error} and \ref{asmp:threshold}(ii) hold. Assume that $\Sigma_e \in \calU(q,c_0(d), M)$ defined in \eqref{eqn:covclass}. And assume that 
    $$
    P \left(\max_{i \leq d} \frac{1}{T} \sum_{t=1}^T (\hat e_{it} - e_{it})^2 > C a_T^2 \right) < O(\kappa(d,T)). 
    $$
    for some $a_T = o(1)$ and $\kappa(d,T) = o(1)$. Denote $\hat \Sigma_e^\calT = \calT_{\lambda}(\frac{1}{T} \sum_{t=1}^T \hat e_t \hat e_t^\top)$ where the thresholding operator $\calT(\cdot)$ satifies condition (i) to (iii) in section \ref{sec:threshold}. Let $\lambda = C' \sqrt{\frac{\log(d)}{T}} + a_T$ for some constant $C'>0$ that is large enough. Then, 
    $$
    \left\| \hat \Sigma_e^\calT - \Sigma_e \right\|_2 = O_p\left( c_0(d) \left(\sqrt{\frac{\log(d)}{T}} + a_T\right)^{1-q} \right)
    $$
\end{lemma} 

\begin{proof}
    Denote the choice of threshold by $C' \,b_T$, i.e., $b_T := \sqrt{\frac{\log(d)}{T}} + a_T$, where $C' > 0$ is sufficiently large. Define event 
    \begin{align*}
        E &= \left\{ \max_{i,j \leq d} \left| \hat \sigma_{ij} - \sigma_{ij} \right| \leq C' b_T  \right\}
    \end{align*}
    
    By Lemma \ref{lemma:hateit2_bound}, the probability of event $E$ is bounded by $1 - O(d^{-2}) - O(\kappa(d,T))$. Under $E$, 
     $|\hat \sigma_{ij}| \leq C' b_T$ implies $|\sigma_{ij}| \leq (C' + 1) b_T \leq C'' b_T$ and $|\hat \sigma_{ij}| > C' b_T$ implies $|\sigma_{ij} | > (C' - 1) b_T > C b_T$. 
    
    Under $E$, by the inequality for spectral norm: $\| \Sigma_e \|_2 \leq \max_{i \leq d} \sum_{j=1}^d |\sigma_{ij}|$ and the conditions on $\calT(\cdot)$, 
    \begin{align*}
        \left\| \hat \Sigma_e^\calT - \Sigma_e \right\|_2 &\leq \max_{i \leq d} \sum_{j=1}^d |\hat \calT(\sigma_{ij}) - \sigma_{ij}| \\ 
        \begin{split} 
            &\leq \max_{i \leq d} \sum_{j=1}^d | \sigma_{ij} | \one \left\{ |\hat \sigma_{ij} | \leq C' b_T \right\} + \max_{i \leq d} \sum_{j=1}^d \left| \calT(\hat \sigma_{ij}) - \hat \sigma_{ij} \right| \one \left\{ | \hat \sigma_{ij} | > C' b_T \right\}  \\ 
            &\quad + \max_{i \leq d} \sum_{j = 1}^d \left| \hat \sigma_{ij} - \sigma_{ij} \right| \one \left\{ | \hat \sigma_{ij} | > C' b_T \right\}
        \end{split}\\ 
        \begin{split} 
            &\leq \max_{i \leq d} \sum_{j=1}^d | \sigma_{ij} | \one \left\{ |\sigma_{ij} | \leq C'' b_T \right\} + \max_{i \leq d} \sum_{j=1}^d \left| \calT(\hat \sigma_{ij}) - \hat \sigma_{ij} \right| \one \left\{ | \sigma_{ij} |> C b_T \right\}  \\ 
            &\quad + \max_{i \leq d} \sum_{j = 1}^d \left| \hat \sigma_{ij} - \sigma_{ij} \right| \one \left\{ | \sigma_{ij} | > C b_T \right\}
        \end{split}\\ 
        &= D_1 + D_2 + D_3. 
    \end{align*} 

    By the definition of $\calU$ and condition (iii) of $\calT(\cdot)$, 
    $$
    \begin{aligned}
        D_1 &\lesssim \max_{i\leq d} \sum_{j=1}^d | \sigma_{ij} |^q b_T^{1-q} \leq c_0(d) b_T^{1-q}, \\ 
        D_2 &\lesssim b_T \sum_{j=1}^d \one \left\{ | \sigma_{ij} |> C b_T \right\} \leq b_T \sum_{j=1}^d | \sigma_{ij} |^q b_T^{-q} \leq c_0(d) b_T^{1-q} \\
    \end{aligned}
    $$
    For $D_3$, 
    \begin{align*}
        D_3 &\leq \max_{i,j \leq d} \left| \hat \sigma_{ij} - \sigma_{ij} \right| \sum_{j=1}^d \one \left\{ | \sigma_{ij} | > C b_T \right\} \\ 
        &\lesssim b_T \sum_{j=1}^d \one \left\{ | \sigma_{ij} | > C b_T \right\} \\
        &\leq b_T \sum_{j=1}^d | \sigma_{ij} |^q b_T^{-q} \leq c_0(d) b_T^{1-q}.
    \end{align*} 
    Therefore, with probability at least $1 - O(d^{-2}) - O(\kappa(d,T))$, 
    $$
    \left\| \hat \Sigma_e^\calT - \Sigma_e \right\|_2 = O_p\left( c_0(d) b_T^{1-q} \right) = O_p\left( c_0(d) \left(\sqrt{\frac{\log(d)}{T}} + a_T\right)^{1-q} \right).
    $$
\end{proof} 

\begin{lemma}
    Suppose the assumptions of Theorem \ref{thm:factor1} hold, in particular $r = O(1)$ and $K = O(1)$. Denote $A = A_K * A_{K-1} * \cdots * A_1$ and $\hat A = \hat A_K * \hat A_{K-1} * \cdots * \hat A_1$ where $*$ denotes Khatri-Rao product. Then 
    $$
    \|\hat A - A\|_2 = O(\psi), 
    $$
    where $\psi = \max_{i \leq r, k \leq K} \|\hat a_{ik} \hat a_{ik}^\top - a_{ik} a_{ik}^\top \|_2$. 
\end{lemma} 

\begin{proof}
    Let $a_i$ denote the column of $A$ and $\hat a_i$ denote the column of $\hat A$. Then 
    $$
    \| \hat A - A \|_2 \leq r \max_{i \leq r} \|\hat a_i - a_i\|_2 \lesssim \max_{i \leq r} \|\hat a_i - a_i\|_2. 
    $$
    By the definition of Khatri-Rao product, 
    \begin{align*}
    \max_{i \leq r} \| \hat a_i - a_i \|_2 &= \max_{i \leq r} \| \hat a_{i1} \odot \cdots \odot \hat a_{iK} - a_{i1} \odot \cdots \odot a_{iK} \|_2 \\ 
    &= \max_{i \leq r} \| (\hat a_{i1} - a_{i1}) \odot \hat a_{i2} \odot \cdots \odot \hat a_{iK} + a_{i1} \odot \hat a_{i2} \odot \cdots \odot \hat a_{iK} - a_{i1} \odot \cdots \odot a_{iK} \|_2 \\ 
    \begin{split}
        &\leq \max_{i \leq r} \| \hat a_{i1} - a_{i1} \|_2 \\
        &\quad + \max_{i \leq r} \| a_{i1} \odot (\hat a_{i2} - a_{i2}) \odot \cdots \odot \hat a_{iK} + a_{i1} \odot a_{i2} \odot \cdots \odot \hat a_{iK} - a_{i1} \odot \cdots \odot a_{iK} \|_2
    \end{split}\\
    &\leq \cdots \\
    &\leq \sum_{k=1}^{K} \max_{i \leq r} \|\hat a_{ik} - a_{ik} \|_2 + \| a_{i1} \odot \cdots \odot a_{iK} - a_{i1} \odot \cdots \odot a_{iK} \|_2 \\ 
    &\leq K \max_{i \leq r, k \leq K} \|\hat a_{ik} - a_{ik} \|_2 \\  
    &\leq K\sqrt{2}\psi, 
\end{align*}
where the last equality is from \eqref{eqn:ahatminua}. As $K = O(1)$, the results follows.
\end{proof} 

The following three lemmas bound the estimation error of $\beta_1^*$, which is used to bound the estimation error of $\beta_1$ in the proof of Theorem \ref{thm:hd_di}. 

\begin{lemma}\label{lemma:beta1star_bound} 
    Under the assumption of Theorem \ref{thm:hd_di}, 
    $$ 
    \left\|\hat \beta_1^* - H^{-1} \beta_1^* \right\|_2 = O_p\left(p_0  \left( \psi + \frac{1}{s_r} + \sqrt{\frac{\log(p_0)}{T}}\right)\right). 
    $$
\end{lemma} 

\begin{lemma}\label{lemma: lambdaj_bound}
    Under the assumption of Theorem \ref{thm:hd_di}, let $\calS_0$ denote the set of non-zero indices of $\beta_0$, we have
    \begin{align*}
        &\max_{j \in \calS_0} \left\| \hat \Lambda_j - \Lambda_j H^{-1} \right\|_2 = O_p\left( \psi + \frac{1}{s_r} + \sqrt{\frac{\log(p_0)}{T}}\right). 
    \end{align*}
\end{lemma} 

\begin{lemma}\label{lemma: vfhatf_bound} 
    Under the assumption of Theorem \ref{thm:hd_di}, let $\calS_0$ denote the set of non-zero indices of $\beta_0$, we have 
    $$
    \max_{j \in \calS_0} \tnorm{ \frac{1}{T} \sum_{t=1}^T V_{tj} \left(\hat f_{t} - H f_{t}\right)} = O_p\left(\psi^2\right), 
    $$
    where $V_{tj}$ indicates the $j^{th}$ entry of $V_t$. 
\end{lemma} 

The proofs of three lemmas will proceed in reverse order. 

\begin{proof}[Proof of Lemma \ref{lemma: vfhatf_bound}] 
    Decompose the objective:
    \begin{align*}
        \max_{j \in \calS_0} \left\| \frac{1}{T} \sum_{t=1}^T V_{tj} \left(\hat f_t - H f_t\right) \right\|_2 &\lesssim r  \max_{i \leq r} \max_{j \in \calS_0} s_i^{-1} \left| \frac{1}{T} \sum_{t=1}^T V_{tj} \left( \htf_{it} - \tf_{it}\right)  \right|
    \end{align*}
    Since $r = O(1)$, it is sufficient to bound the term inside the outer maximum. Decompose the term: 
    $$ 
    \begin{aligned}
        \max_{j \in \calS_0} \frac{1}{T} \sum_{t=1}^T V_{tj} \left( \htf_{it} - \tf_{it}\right) &= \left( \prod_{k=1}^K a_{ik}^\top \hat b_{ik} - 1 \right) \max_{j \in \calS_0} \frac{1}{T} \sum_t V_{tj} \tf_{it} \\ 
        &\quad + \sum_{i' \neq i}^r \prod_{k=1}^K a_{i'k}^\top \hat b_{ik} \max_{j \in \calS_0} \frac{1}{T} \sum_t V_{tj} \tf_{i't} \\
        &\quad + \left( \hat b_i - b_i\right)^\top \max_{j \in \calS_0} \frac{1}{T} \sum_t V_{tj} e_t + \max_{j \in \calS_0} \frac{1}{T} \sum_t V_{tj} b_i^\top e_t \\
    \end{aligned}
    $$
    By Assumption \ref{asmp:factor}, \ref{asmp:error} and \ref{asmp:hd_di}, with a similar argument with Equation \eqref{eqn:maxfe}, we have 
    $$
    \max_{j \in \calS_0} \frac{1}{T} \sum_t V_{tj} \tilde f_{it} = O_p\left( s_i \sqrt{\frac{\log(p_0)}{T}}\right) \quad \max_{j \in \calS_0} \frac{1}{T} \sum_t V_{tj} b_i^\top e_t = O_p\left( \sqrt{\frac{\log(p_0)}{T}} \right), 
    $$
    and by the similar argument for the first term of $\Upsilon_1$ in the proof of Theorem \ref{thm:coverate}, 
    \begin{align*}
        \left( \hat b_i - b_i\right)^\top \max_{j \in \calS_0} \frac{1}{T} \sum_t V_{tj} e_t &\leq \left\| \hat b_i - b_i \right\|_2 \| \max_{j \in \calS_0} \frac{1}{T} \sum_t V_{tj} e_t \|_2 \\ 
        &= O_p(\psi) O_p\left( \sqrt{\frac{d_{\max} + \log(p_0)}{T}} \right) = O_p\left( \psi \sqrt{\frac{d_{\max}}{T}} \right). 
    \end{align*}
    Putting all together gives the result. 
\end{proof} 

\begin{proof}[Proof of Lemma \ref{lemma: lambdaj_bound}]
    By the construction of $\hat \Lambda$, 
    \begin{align*}
        \hat \Lambda &= \left( \frac{1}{T} \sum_t w_t \hat f_t^\top\right) \left( \frac{1}{T} \sum_t \hat f_t \hat f_t\top\right) \\ 
        &= \left( \Lambda H^{-1} \frac{1}{T} \sum_t H f_t \hat f_t^\top\right) S_f^{-1} + \frac{1}{T} \sum_t V_t \hat f_t^\top S_f^{-1}, 
    \end{align*}
    where $S_f = \sum_t \hat f_t \hat f_t^\top / T$. So, 
    \begin{align*}
        \hat \Lambda - \Lambda H^{-1} &= \left(\Lambda H^{-1} \frac{1}{T} \sum_t \left(H f_t - \hat f_t\right)\hat f_t^\top \right) S_f^{-1} + \left( \frac{1}{T} \sum_t V_t f_t^\top H\right) S_f^{-1} \\ 
        &\quad + \frac{1}{T} \sum_t V_t \left( \hat f_t - H f_t\right)^\top S_f^{-1}. 
    \end{align*} 

    \begin{align*}
        \max_{j \in \calS_0} \left\| \hat \Lambda_j - \Lambda_j H^{-1} \right\|_2 &\leq \max_{j \in \calS_0} \left\| \Lambda_j\right\|_2 \tnorm{H^{-1}} \tnorm{\tmean \left(Hf_t - \hat f_t\right) \hat f_t^\top} \tnorm{S_f^{-1}} \\ 
        &\quad + \max_{j \in \calS_0} \left\| \tmean V_{tj} f_t^\top \right\|_2 \tnorm{H} \tnorm{S_f^{-1}} \\ 
        &\quad + \max_{j \in \calS_0} \tnorm{\tmean V_{tj} \left( \hat f_t - H f_t\right)} \tnorm{S_f^{-1}} \\ 
        &=  \Gamma_1 + \Gamma_2 + \Gamma_3. 
    \end{align*} 
    By Lemma \ref{lemma:samplemeanfitfjt}, one can show that 
    $$
    \begin{aligned}
        &\tnorm{\tmean \left(Hf_t - \hat f_t\right) \hat f_t^\top} = O_p\left(\psi + 1/s_r\right) \\ 
        & \tnorm{H} = \tnorm{H^{-1}} = O_p(1) \\ 
        &\tnorm{S_f^{-1}} = O_p(1).
    \end{aligned}
    $$ 
    Therefore, 
    $$
    \Gamma = O_p\left(\psi + \frac{1}{s_r}\right). 
    $$
    By the proof of Lemma \ref{lemma: vfhatf_bound}, 
    $$
    \Gamma_2 = O_p\left(\sqrt{\frac{\log(p_0)}{T}}\right). 
    $$
    And by Lemma \ref{lemma: vfhatf_bound}, 
    $$
    \Gamma_3 = O_p\left(\psi^2\right). 
    $$ 
    Therefore, 
    $$\max_{j \in \calI_0} \left\| \hat \Lambda_j - \Lambda_j H^{-1} \right\|_2 = O_P\left( \psi + \frac{1}{s_r} + \sqrt{\frac{\log(p_0)}{T}}\right).$$
\end{proof}

\begin{proof}[Proof of Lemma \ref{lemma:beta1star_bound}]
    By the construction of $\hat \beta_1^*$, 
    $$
    \begin{aligned}
        \hat \beta_1^* &= S_f^{-1} \tmean{ \hat f_t \left(V_t^\top \beta_0 + f_t^\top H H^{-1} \beta_1^* + \epsilon_{t+h}\right)} \\ 
        &= S_f^{-1} \tmean \hat f_t V_t^\top \beta_0 + S_f^{-1} \tmean\hat f_t \left(H f_t - \hat f_t\right)^\top H^{-1} \beta_1^* \\
        &\quad + S_f^{-1} \tmean \hat f_t \hat f_t^\top H^{-1} \beta_1^* + S_f^{-1} \tmean{ \hat f_t \epsilon_{t+h}}. \\
    \end{aligned}
    $$ 
    So, 
    \begin{align*}
        \hat \beta_1^* - H^{-1}\beta_1 &= S_f^{-1} \tmean{H f_t \epsilon_{t+h}} + S_f^{-1} \tmean{\left(\hat f_t - H f_t\right)\epsilon_{t+h}} \\ 
        &\quad + S_f^{-1} \tmean{\hat f_t \left(H f_t - \hat f_t\right)^\top H^{-1} \beta_1}  \\ 
        &\quad + \left( \hat \Lambda - \Lambda H^{-1}\right) \beta_0 \\ 
        &:= \calD_1 + \calD_2 + \calD_3 + \calD_4.
    \end{align*} 
    For $\calC_1$, 
    $$
    \tnorm{\calD_1} \leq \tnorm{S_f^{-1}} \tnorm{H} \tnorm{\tmean f_t \epsilon_{t+h}} = O_p\left(\frac{1}{\sqrt{T}}\right). 
    $$
    By Lemma \ref{lemma:samplemeanfitfjt}, 
    $$
    \tnorm{\calD_2} = O_p\left(\psi + \frac{1}{s_r}\right).
    $$
    For $\calD_3$, denote $\hat F = (\hat f_1, \ldots, \hat f_T) \in \R^{r \times T}$ and $F = (f_1, \ldots, f_T) \in \R^{r \times T}$. Then
    $$
    \tnorm{\calD_3} \leq \tnorm{S_f^{-1}} \tnorm{\hat F} \tnorm{\hat F - H F^\top} \tnorm{\beta_1}. 
    $$
    By Lemma \ref{lemma:samplemeanfitfjt}, 
    \begin{align*}
        \| \hat F - H F \|_2^2 \leq \| \hat F - H F \|_F^2 \leq \sum_t \| \hat f_t - H f_t \|_2^2 = O_p\left(T \psi^2 + \frac{T}{s_r^2}\right), 
    \end{align*} 
    and 
    \begin{align*}
        \tnorm{\hat F} \leq \| \hat F - H F \|_2^2 + \| H F \|_2^2 \leq O_p\left(T \psi^2 + \frac{T}{s_r^2} + T\right) = O_p\left(T\right).
    \end{align*} 
    Therefore, 
    $$
    \tnorm{\calD_3} = O_p\left( \psi + \frac{1}{s_r} \right). 
    $$
    For $\calD_4$, by Lemma \ref{lemma: lambdaj_bound}, 
    $$
    \tnorm{\calD_4} \leq \max_{j \in \calS_0} \tnorm{\hat \Lambda_j - \Lambda_j H^{-1}} \| \beta_0\|_1 = O_p\left( p_0\left(\psi + \frac{1}{s_r} + \sqrt{\frac{\log(p_0)}{T}}\right)\right). 
    $$
    $\calD_4$ is the leading term so the result follows. 
\end{proof} 

\begin{lemma}\label{lemma:maxvjhat_bound} 
    Under the assumptions of Theorem \ref{thm:hd_di}, 
    $$
    \frac{1}{T} \max_{j \leq p} \tnorm{\hat V_j - V_j}^2 = O_p\left(\psi^2 + \frac{1}{s_r^2} + \frac{\log(p)}{T}\right). 
    $$
\end{lemma} 

\begin{proof}[Proof of Lemma \ref{lemma:maxvjhat_bound}] 
    \begin{align*}
        \frac{1}{T} \max_{j \leq p} \tnorm{\hat V_j - V_j}^2 &\leq \frac{1}{T} \max_{j \leq p} \tnorm{H^{-1}\Lambda_j \left( H F^\top - \hat F^\top\right)}^2 + \frac{1}{T} \max_{j \leq p} \tnorm{\left( \Lambda_j - H^{-1} \Lambda_j\right) \hat F^\top}^2 \\ 
        &\leq \frac{1}{T} \tnorm{HF^\top - \hat F^\top}^2 \tnorm{H}^2 \max_{j \leq p} \tnorm{\Lambda_j}^2 + \frac{1}{T} \tnorm{\hat F}^2 \max_{j \leq p} \tnorm{\Lambda_j - H^{-1} \Lambda_j}^2
    \end{align*} 
    By Lemma \ref{lemma:samplemeanfitfjt}, \ref{lemma: lambdaj_bound} and Assumption \ref{asmp:hd_di}(v), 
    $$
    \frac{1}{T} \max_{j \leq p} \tnorm{\hat V_j - V_j}^2 = O_p\left(\psi^2 + \frac{1}{s_r^2} \right) + O_p\left(\psi^2 + \frac{1}{s_r^2} + \frac{\log(p)}{T} \right). 
    $$
    The result follows. 
\end{proof}

\begin{lemma}\label{lemma:betas_bound} 
    For an index set $\calS$, define event 
    $$
    E_{\Sigma_V} = \left\{ \maxnorm{\hat \Sigma_{\hat V} - \Sigma_V} \leq \frac{c}{|\calS|} \right\}, \quad \text{for some constant $c>0$}, 
    $$
    where $\hat \Sigma_{\hat V} = \hat V^\top \hat V / T$. Assume that $\onorm{\beta_\calS}^2 \leq C |\calS| \beta^\top \Sigma_V \beta$ for some constant $C>0$ and $\beta \in \calC(S,3)$, then under event $E_{\Sigma_V}$, 
    $$
    \| \beta_S\|_1 \leq C \sqrt{|\calS| \beta^\top \hat \Sigma_{\hat V} \beta}, 
    $$
    for some constant $C>0$ and $\beta \in \calC(S,3)$.
\end{lemma}

This is the Lemma A.5 of \cite{Adamek2023}, which directly follows by Corollary 6.8 in \cite{buhlmann2011}. Proof is omitted here. 

\begin{lemma}\label{lemma:eventEv_bound}
    Under the assumptions of Theorem \ref{thm:hd_di}, 
    $$
    p_0 \maxnorm{\hat \Sigma_{\hat V} - \Sigma_V} = o_p(1), 
    $$
    which implies that the probability of event $E_{\Sigma_V}$ for $\calS_0$ converges to one. 
\end{lemma}

\begin{proof}[Proof of Lemma \ref{lemma:eventEv_bound}]
    Denote $\hat \Sigma_V  = V^\top V / T$. We have 
    $$
    \maxnorm{\hat \Sigma_{\hat V} - \Sigma_V} \leq \maxnorm{\hat \Sigma_{\hat V} - \hat \Sigma_V} + \maxnorm{\hat \Sigma_V - \Sigma_V} := \calG_1 + \calG_2.
    $$ 
    By Assumption \ref{asmp:hd_di}(vi),  $\sqrt{p_0} \leq C \sqrt{T / \log(p)}$ for some constant $C > 0$. Then by the argument in Lemma A.3 of \cite{Fan2011}, since $\log(p)^{\eta_1/2 - 1} = o(T)$, we have 
    $$
    \PP\left( \sqrt{p_0} \maxnorm{\hat \Sigma_V - \Sigma_V} \geq C' / \sqrt{p_0}\right) \leq \PP\left( \sqrt{p_0} \maxnorm{ \hat \Sigma_V - \Sigma_V} \geq C \sqrt{\frac{\log(p)}{T}}\right) = O\left(1 / p^2\right), 
    $$
    which bounds $\sqrt{p_0} \calG_2$. For $\calG_1$, 
    \begin{align*}
        \maxnorm{\hat \Sigma_{\hat V} - \hat \Sigma_V} &= \maxnorm{\frac{1}{T} \hat V^\top \hat V - \frac{1}{T} \hat V^\top \hat V} \\ 
        &\leq \frac{2}{T} \maxnorm{\hat V^\top \left( \hat V - V\right)} + \frac{1}{T} \maxnorm{\left(\hat V - V\right)^\top \left( \hat V - V\right)} \\ 
        &= \calG_{11} + \calG_{12}. 
    \end{align*}
    Observe that 
    $$
    \calG_{11} = \hat V^\top \left( F H H^{-1} \Lambda^\top - \hat F \hat \Lambda^\top\right) = \hat V^\top F H H ^{-1} \Lambda^\top. 
    $$
    So by Assumption \ref{asmp:hd_di}(v) and Lemma \ref{lemma:samplemeanfitfjt}, 
    \begin{align*}
        \frac{2}{T} \maxnorm{\hat V^\top \left( \hat V - V\right)} &= \frac{2}{T} \max_{j \leq p} \infnorm{\hat V^\top \left( \hat V^\top \left( \hat F - FH\right) H^{-1} \Lambda_j\right)} \\ 
        &\leq \frac{2}{T} \max_{l \leq p} \tnorm{\hat V_l^\top \left( \hat F - FH\right)} \tnorm{H^{-1}} \max_{j \leq p} \tnorm{\Lambda_j} \\ 
        &= \frac{2}{T} \max_{l \leq p} \tnorm{\hat V_l^\top \left( \hat F - FH\right)} O_p(1). 
    \end{align*} 

    \begin{align*}
        \frac{2}{T} \max_{l \leq p} \tnorm{\hat V_l^\top \left( \hat F - FH\right)} &\leq \frac{2}{T} \max_{l \leq p} \tnorm{\left( \hat V_l - V_l\right)^\top \left( \hat F - FH\right)} + \frac{2}{T} \max_{l \leq p} \tnorm{V_l^\top \left( \hat F - FH\right)} \\ 
        &:= \calG_{111} + \calG_{112}.
    \end{align*} 

    By Lemma \ref{lemma: vfhatf_bound}, $\calG_{112} = O_p\left(\psi^2\right)$. For $\calG_{111}$, take the square of the term and apply the Cauchy-Schwarz inequality: 
    \begin{align*}
        \frac{2}{T} \max_{l \leq p} \tnorm{\left( \hat V_l - V_l\right)^\top \left( \hat F - FH\right)}^2 &\leq \frac{4}{T} \tnorm{\hat F - FH}^2 \frac{1}{T} \max_{l \leq p} \tnorm{\hat V_l - V_l}^2 \\ 
        &= O_p\left(\psi^2 + \frac{1}{s_r^2} \right) O_p\left( \psi^2 + \frac{1}{s_r^2} + \frac{\log(p)}{T}\right),  
    \end{align*}
    by Lemma \ref{lemma:samplemeanfitfjt} and \ref{lemma:maxvjhat_bound}. Therefore, 
    $$
    \calG_{111} = O_p\left( \psi^2 + \frac{1}{s_r^2} + \sqrt{\frac{\log(p)}{T}} \left( \psi + \frac{1}{s_r}\right)\right). 
    $$ 
    And $\calG_{112} = O_p\left(\psi^2\right)$. Therefore, 
    $$
    \calG_{11} = O_p\left( \psi^2 + \frac{1}{s_r^2} + \sqrt{\frac{\log(p)}{T}} \left( \psi + \frac{1}{s_r}\right)\right).
    $$ 

    For $\calG_{12}$, 
    $$
    \begin{aligned}
        \frac{1}{T} \maxnorm{\left( \hat V - V\right)^\top \left( \hat V - V\right)} &= \max_{j \leq p, l \leq p} \tmean \left( \hat V_{tj} - V_{tj}\right) \left( \hat V_{tl} - V_{tl}\right) \\ 
        &\leq \max_{j \leq p, l \leq p} \frac{1}{T} \tnorm{\hat V_j - V_j} \tnorm{\hat V_l - V_l} \\ 
        &= \max_{j \leq p} \frac{1}{T} \tnorm{\hat V_j - V_j}^2 = O_p\left(\psi^2 + \frac{1}{s_r^2} + \frac{\log(p)}{T}\right) \\ 
        &= O_p\left( \psi^2 + \frac{1}{s_r^2} + \frac{\log(p)}{T}\right), 
    \end{aligned}
    $$
    by Lemma \ref{lemma:maxvjhat_bound}. So we have 
    $$
    \calG_1 = O_p\left( \psi^2 + \frac{1}{s_r^2} + \frac{\log(p)}{T}\right).
    $$
    
    By Assumption \ref{asmp:hd_di}(vi), 
    $$
    p_0\left( \calG_1 + \calG_2\right) = O_p\left( p_0 \left( \psi^2 + \frac{1}{s_r^2}\right) + \sqrt{\frac{p_0\log(p)}{T}} \right) = o_p(1), 
    $$
    which proves the lemma. 

\end{proof} 

\begin{lemma}\label{lemma:maxutvhat_bound} 
    Denote $\tU = \tilde Y - \hat V \beta_0 $. Under the assumptions of Theorem \ref{thm:hd_di}, 
    $$
    \frac{1}{T} \infnorm{\tU^\top \hat V} = O_p\left(\psi^2 + \frac{1}{s_r^2} + \sqrt{\frac{\log(p)}{T}}\right).
    $$
\end{lemma} 

\begin{proof}[Proof of Lemma \ref{lemma:maxutvhat_bound}] 
    Denote $\epsilon = (\epsilon_{1+h}, \ldots, \epsilon_{T+h})$. Whether it ends at $T+h$ or $T+h-1$ does not affect the result. Observe that 
    $$
    \tU^\top \hat V = \beta_1^\top F^\top \hat V + \epsilon^\top \hat V, 
    $$
    which implies that 
    $$
    \frac{1}{T} \infnorm{\tU^\top \hat V} = \frac{1}{T} \infnorm{\beta_1^\top F^\top \hat V} + \frac{1}{T} \infnorm{\epsilon^\top \hat V} := \calH_1 + \calH_2.
    $$ 
    For $\calH_1$, 
    \begin{align*}
        \frac{1}{T} \infnorm{\beta_1^\top F^\top \hat V} &= \frac{1}{T}\infnorm{\hat V^\top \left( \hat F - FH\right)H^{-1}\beta_1} \\ 
        &\leq \frac{1}{T} \max_{j \leq p} \tnorm{\hat V_j^\top \left( \hat F - FH\right)} \tnorm{H^{-1}} \tnorm{\beta_1} \\ 
        &= O_p\left(\psi^2 + \frac{1}{s_r^2} + \sqrt{\frac{\log(p)}{T}} \left( \psi + \frac{1}{s_r}\right)\right), 
    \end{align*}
    by the argument analogous to $\calG_{11}$ in the proof of Lemma \ref{lemma:eventEv_bound}. For $\calH_2$, 
    \begin{align*}
        \frac{1}{T} \infnorm{\epsilon^\top \hat V} &\leq \frac{1}{T} \infnorm{\left( \hat V - V\right)^\top \epsilon} + \frac{1}{T} \infnorm{V^\top \epsilon} \\ 
        &\leq \frac{1}{T} \infnorm{\Lambda H^{-1} \left( FH - \hat F\right)^\top \epsilon} + \frac{1}{T} \infnorm{\left( \hat \Lambda - \Lambda H^{-1}\right)H^\top F^\top \epsilon}  \\
        &\quad + \frac{1}{T} \infnorm{\left( \hat \Lambda - \Lambda H^{-1}\right)\left( \hat F - FH \right)^\top \epsilon}  + \frac{1}{T} \infnorm{V^\top \epsilon} \\ 
        &:= \calH_{21} + \calH_{22} + \calH_{23} + \calH_{24}.
    \end{align*}

    By the same argument of Lemma B.1(ii) in \cite{Fan2011} and the rate assumption of Theorem \ref{thm:hd_di}, 
    $$
    \calH_{24} = O_p\left(\sqrt{\frac{\log(p)}{T}}\right).
    $$ 
    For $\calH_{21}$, 
    \begin{align*}
        \frac{1}{T} \infnorm{\Lambda H^{-1} \left( FH - \hat F\right)^\top \epsilon} &\leq \frac{1}{T} \infnorm{\left( FH - \hat F\right)^\top \epsilon} \tnorm{H^{-1}} \max_{j \leq p} \tnorm{\Lambda_j} \\ 
        &= O_p\left( \psi^2 + \frac{1}{s_r\sqrt{T}}\right), 
    \end{align*}
    by Lemma \ref{lemma:samplemeanfitfjt} and Assumption \ref{asmp:hd_di}(v). For $\calH_{22}$, 
    \begin{align*}
        \frac{1}{T} \infnorm{\left( \hat \Lambda - \Lambda H^{-1}\right)H^\top F^\top \epsilon} &\leq \max_{j \leq p} \tnorm{\hat \Lambda_j - \Lambda_j H^{-1}} \tnorm{H^\top} \frac{1}{T} \tnorm{F^\top \epsilon} \\ 
        &= O_p\left(\frac{\psi}{\sqrt{T}} + \frac{1}{s_r\sqrt{T}} + \frac{\sqrt{\log(p)}}{T}\right), 
    \end{align*}
    by Lemma \ref{lemma: lambdaj_bound}. 

    For $\calH_{23}$, 
    \begin{align*}
        \frac{1}{T} \infnorm{\left( \hat \Lambda - \Lambda H^{-1}\right)\left( \hat F - FH \right)^\top \epsilon} &\leq \max_{j \leq p} \tnorm{\hat \Lambda_j - \Lambda_j H^{-1}} \frac{1}{T} \tnorm{ \left(\hat F - FH\right)^\top \epsilon} \\ 
        &= O_p\left(\psi + \frac{1}{s_r} + \sqrt{\frac{\log(p)}{T}}\right) O_p\left( \psi^2 + \frac{1}{s_r\sqrt{T}}\right), 
    \end{align*}
    which is dominated by $\calH_{21}$ and $\calH_{22}$ as $\sqrt{T} \psi^2 = o(1)$ by assumption. Therefore,
    $$
    \calH_2= O_p\left( \psi^2 + \sqrt{\frac{\log(p)}{T}} \right). 
    $$
    Putting them all together yields 
    $$
    \frac{1}{T} \infnorm{\tU^\top \hat V} = O_p\left(\psi^2 + \frac{1}{s_r^2} + \sqrt{\frac{\log(p)}{T}}\right). 
    $$
    
\end{proof} 

\begin{lemma}\label{lemma:expprod}
    Suppose that the random variables $Z_1$, $Z_2$ such that for any $s > 0$,
    $$
    P\left( | Z_i | > s \right) \leq \exp\left(1 - (s/b_i)^{r_i}\right), \quad i=1,2. 
    $$
    Define $r = r_1r_2/(r_1 + r_2)$ and $b_3 = \left( 1 + \log2 \right)^{1/r}b_1b_2$, then we have 
    $$
    P\left( |Z_1Z_2|>s \right)< \exp \left( 1 - (s/b_3)^{r} \right). 
    $$
\end{lemma}
It is a simple modification of Lemma A.2 and its proof in \cite{Fan2011}, so we omit the proof here.

\section*{Appendix C: An illustrative example}

\begin{figure}[htbp!]
    \label{fig:cove_demo}
    \centering
    \includegraphics[width=0.5\textwidth]{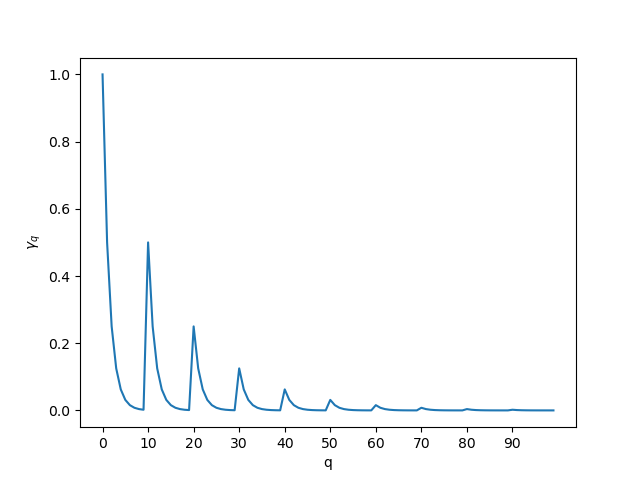}
    \caption{Plot of $\gamma_q$ for $d_k = 10$ and $\tau = 0.5$.}
\end{figure}

To illustrate the performance of HAC-type estimator, consider the strong matrix factor model with one factor where $\tilde A := \sqrt{d} A$ is a $d$-dimensional vector of ones. In this case, $ \Sigma_{Be} = \lim \frac{1}{d} \sum_{j=1}^d \sum_{l=1}^d \E{e_{jt}e_{lt}}$. Suppose the idiosyncratic error matrix is generated by: 
$$
\cE_t = \Sigma_{\cE,1}^{1/2} Z_t \Sigma_{\cE,2}^{1/2}, \quad Z_{t} \sim MN(0, I_{d_1}, I_{d_2}). 
$$ 
Let $d_1 = d_2$ and $\Sigma_{\cE,1} = \Sigma_{\cE,2} = \text{Toeplitz}(\tau,d_1)$ such that the $(i,j)^{th}$ entry of $\Sigma_{\cE,k}$ is equal to $\tau^{|i-j|}$. It can be verified that $\Sigma_e = \Sigma_{\cE,2} \odot \Sigma_{\cE,1}$ and for $q = |i-j|$, $\E{e_{it}e_{jt}} = \gamma_q$, where $\gamma_q = \tau^q$ for $1 \leq q \leq d_1$, $\gamma_q = \tau \gamma_{q-d_1}$ for $d_1 + 1 \leq q \leq 2 d_1$, $\gamma_q = \tau^2 \gamma_{q-2d_1}$ for $2d_1 + 1 \leq q \leq 3 d_1$, and so on. Therefore, $\max_{j} \sum_{l = 1}^d |\E{ e_{jt} e_{lt}|}\leq \left( \frac{1}{1-\tau}\right)^2 = O(1)$ and Assumption \ref{asmp:factor} (iii) is satisfied. The plot of $\gamma_q$ for $d_1 = 10$ and $\tau = 0.5$ is shown in Figure \ref{fig:cove_demo} for illustration. 

However, due to the Kronecker product structure of $\Sigma_e$, $\gamma_q$ does not decay monotonically. If we choose the tuning parameter $n = \sqrt{d} = d_1 \to \infty$$n$ in the CS-HAC estimator as suggested in \cite{BaiNg2006}, then $\lim \frac{1}{n}\sum_{j=1}^n \sum_{l=1}^n \E{ e_{jt} e_{lt}} = 1 + \lim_{d_1 \to \infty} \sum_{q=1}^{d_1-1} 2 \frac{d_1 - q}{d_1} \gamma_q$, which is the Newey-West sum of $\gamma_q$ before the second peak in Figure \ref{fig:cove_demo}. This estimator is not consistent for $\Sigma_{Be}$ as the sum of $\gamma_{q}$ for $q > d_1$ does not converge to zero. Alternatively, if we choose $n = d^{3/4}$, then  $\lim \frac{1}{n}\sum_{j=1}^n \sum_{l=1}^n \E{ e_{jt} e_{lt}}$ is bounded, the CS-HAC estimator would work as long as $n/T\rightarrow0$. 

A simulation study is conducted to evaluate the performance of cross-sectional HAC-type estimator. Consider the following two-way CP factor model: 
$$
\begin{aligned}
    &\calX_t = \sum_{i=1}^r s_i f_{it} a_{i1} a_{i2}^\top + \calE_t, \\ 
    &f_{it} = \rho_i f_{i,t-1} + \sqrt{1 - \rho_i^2} u_{it}, \quad u_{it} \sim N(0,1), \\ 
    &\calE_t = \Sigma_{\cE,1}^{1/2} Z_t \Sigma_{\cE,2}^{1/2}, \quad Z_{t} \sim MN(0, I_{d_1}, I_{d_2}) \\ 
    &A_k = [a_{1k}, \ldots, a_{rk}] = \Sigma_{A_k} \tilde A_k, \\
\end{aligned}
$$
The matrix $\tilde A_k $ is generated by QR decomposition of the matrix of $d_k \times r$ where each entry is generated from $N(0,1)$ so that $\tilde A_k$ is orthonormal. We consider the following specifications: 
\begin{itemize}
    \item $d_1 = d_2$, $r = 3$; 
    \item $s_i = (r-i+1) \ \sqrt{d}$; 
    \item $\rho_1, \rho_2, \rho_3 = 0.6,0.5,0.4$; 
    \item $\Sigma_{\cE_k} = \Sigma_{A_k} = \text{Toeplitz}(0.6,d_k)$. 
\end{itemize}

The model is estimated by the PCA method on $\vect{\calX_t}$ and we consider three covariance matrices for the factor estimator: 
\begin{itemize} 
    \item $\hat \Gamma^{HAC} = \hat V^{-1} \left( \frac{1}{n} \sum_{j = 1}^n \sum_{l = 1}^n \hat A_{j:} \hat A_{l:}^\top \frac{1}{T} \sum_{t=1}^T \hat e_{jt} \hat e_{lt} \right) \hat V^{-1}$; 
    \item $\Gamma^{HAC} = \hat V^{-1} Q \left( \frac{1}{n} \sum_{j = 1}^n \sum_{l = 1}^n A_{j:} A_{l:}^\top \frac{1}{T} \sum_{t=1}^T e_{jt} e_{lt} \right) Q^\top \hat V^{-1}$; 
    \item $\Gamma = \hat V^{-1} Q \left(A^\top \Sigma_e A / d \right) Q^\top \hat V^{-1}$, 
\end{itemize}
where 
\begin{itemize}
    \item $\Sigma_e = \Sigma_{\cE_2} \odot \Sigma_{\cE_1}$ is the true covariance matrix of the idiosyncratic error; 
    \item $Q = \hat F^\top F / T$; 
    \item $\hat V$ is the diagonal matrix of the first $r$ eigenvalues of $\frac{1}{dT} \sum_{t=1}^T \vect{\calX_t} \vect{\calX_t}^\top$; 
    \item $\hat A_{j:}$ is the $j^{th}$ row of $\hat A$, which is the factor loading estimator by PCA; 
    \item $A_{j:}$ is the $j^{th}$ row of $A$, which is the true factor loading; 
\end{itemize}

$\Gamma$ is the infeasible estimator of the factor covariance matrix, which takes $\Sigma_e$, $A$ and $f_t$ as given, and $\Gamma^{HAC}$ as the ``oracle'' HAC estimator, where the true factor loadings and errors are used. 

We consider two settings for $n$ and $T$: 
\begin{itemize}
    \item $n = \sqrt{\min(d,T)}$, $T = 1000$; 
    \item $n = \lceil d^{3/4} \rceil$ and $T = 500 + \lceil d^{4/5} \rceil$. 
\end{itemize}

\begin{figure}[htbp!]
    \centering 
    \includegraphics[width = 0.9\textwidth]{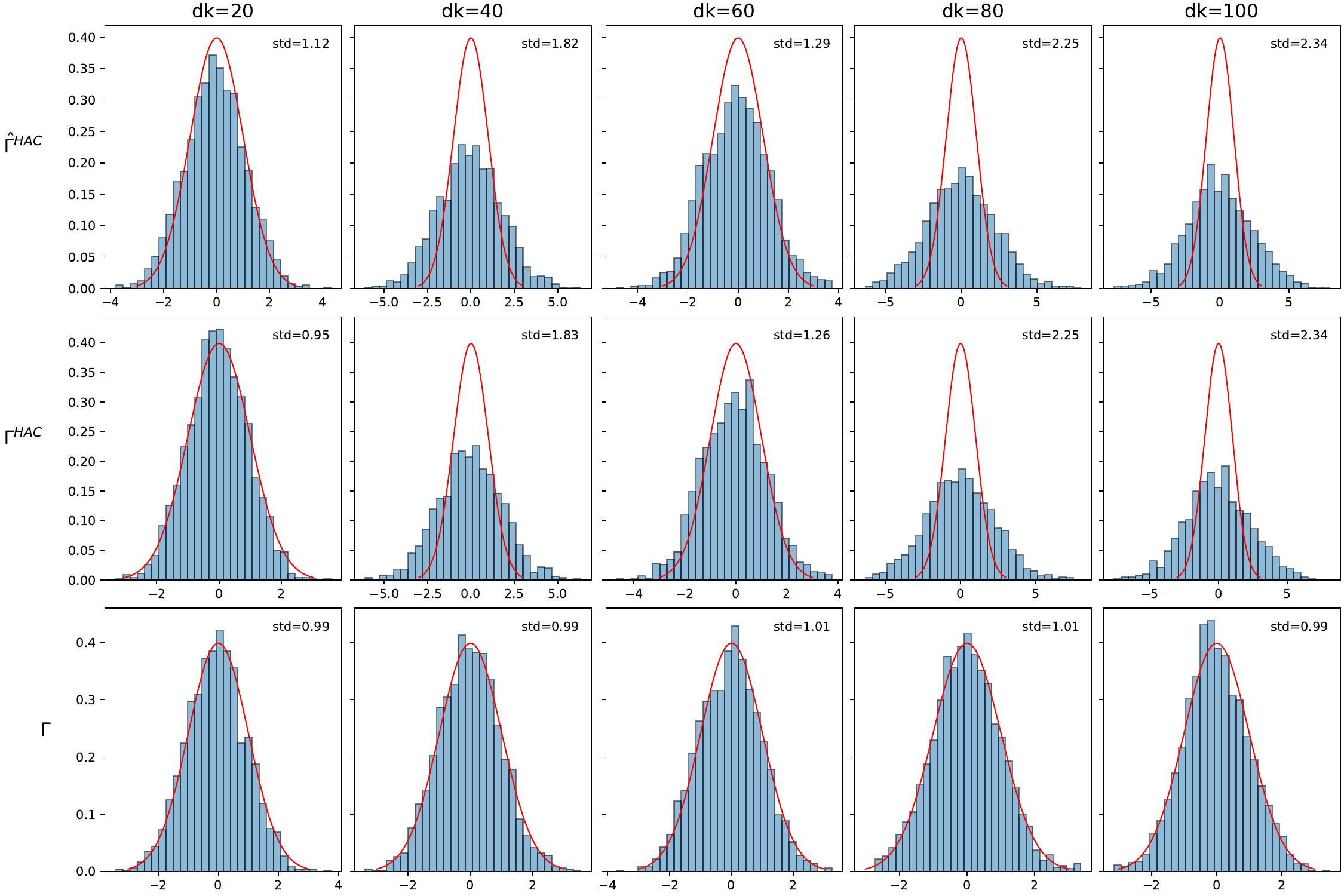} 
    \caption{The histogram of the first entry of $\sqrt{d} \ \Sigma_{Be}^{-1/2} \left(\hat f_{t} - f_{t}\right)$ for $t=0$, under $n = \sqrt{\min(d,T)}$ and $T = 1000$. The first row shows the results for $\hat \Sigma_{Be} = \hat \Gamma^{HAC}$; the second row shows the results for $\hat \Sigma_{Be} = \Gamma^{HAC}$; the third row shows the results for $\hat \Sigma_{Be} = \Gamma$. Columns from left to right show the results for $d = 20,40,60,80,100$. The sample standard deviation of the histograms are shown at the top right corner of each plot.} 
    \label{fig:clt_hac1}
\end{figure}

\begin{figure}[htbp!]
    \centering 
    \includegraphics[width = 0.9\textwidth]{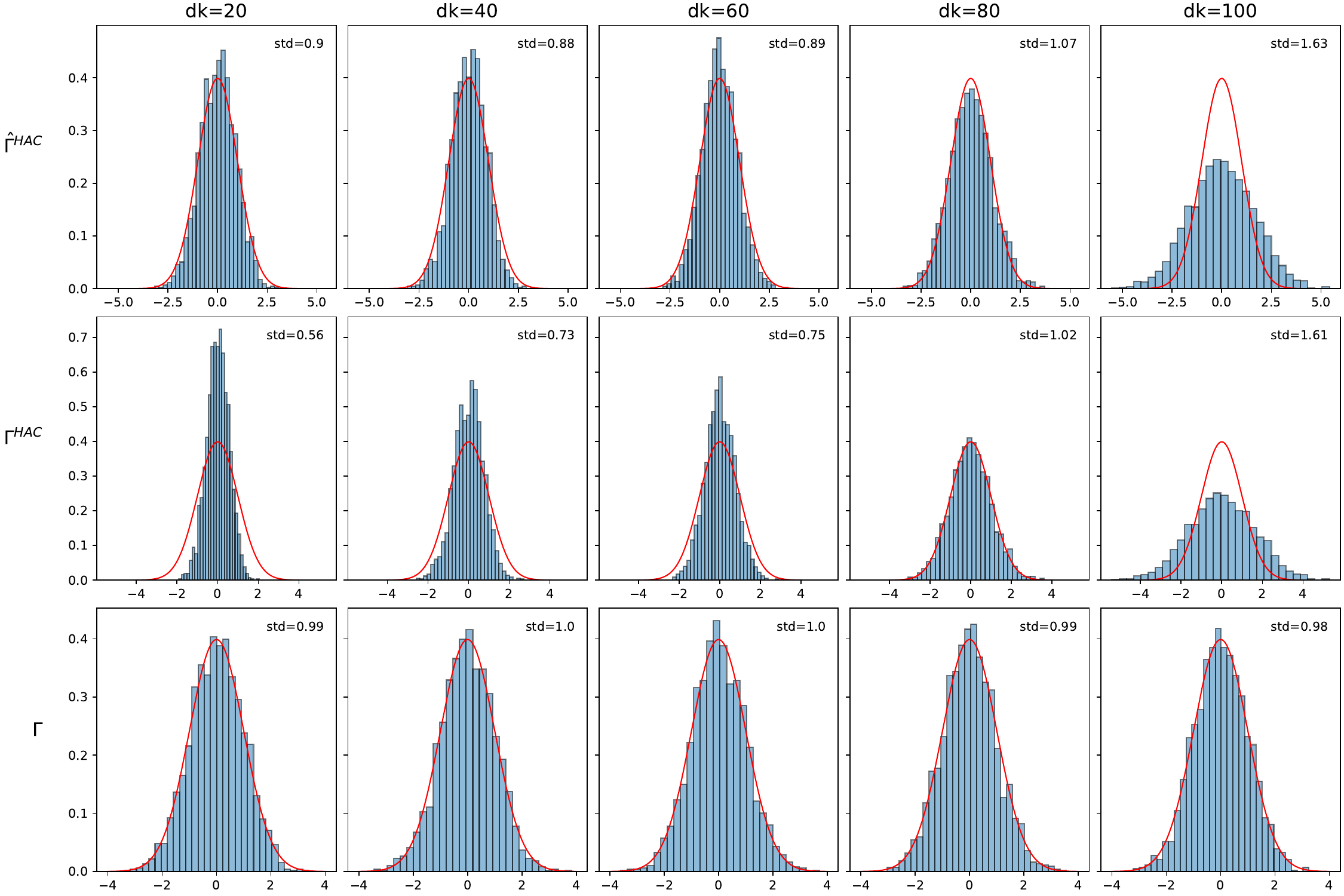} 
    \caption{The histogram of the first entry of $\sqrt{d} \ \Sigma_{Be}^{-1/2} \left(\hat f_{t} - f_{t}\right)$ for $t=0$, under $n = \lceil d^{3/4} \rceil$ and $T = 500 + \lceil d^{4/5} \rceil$. The first row shows the results for $\hat \Sigma_{Be} = \hat \Gamma^{HAC}$; the second row shows the results for $\hat \Sigma_{Be} = \Gamma^{HAC}$; the third row shows the results for $\hat \Sigma_{Be} = \Gamma$. Columns from left to right show the results for $d = 20,40,60,80,100$. The sample standard deviation of the histograms are shown at the top right corner of each plot.} 
    \label{fig:clt_hac2}
\end{figure}

Figure \ref{fig:clt_hac1} and \ref{fig:clt_hac2} show the histogram of the first entry of $\sqrt{d} \ \hat \Sigma_{Be}^{-1/2} \left(\hat f_{t} - f_{t}\right)$ for $t=0$ with choices of $\hat \Sigma_{Be}$ specified above, under two different settings for $n$ and $T$. The sample standard deviation of the histograms are shown at the top right corner of each plot. It can be observed that the HAC estimator as well as the ``oracle'' version does not perform well as $d_k$ grows.

\clearpage 

\section*{Appendix D: Consistency of the factor estimators} 

In this appendix, we show that CC-ISO requires a weaker condition for consistency of the factor estimators than PCA. Consider the model for data generating process in Appendix C, the specifications of the model are as follows:
\begin{itemize} 
    \item $d_1 = d_2$, $r=3$ and $T = 100 + d^{0.3}$;
    \item Factor loadings are generated as in Section \ref{sec:sim}; 
    \item $\rho_1 = 0.6, \rho_2 = 0.5, \rho_3 = 0.4$;
    \item $\Sigma_{\cE_k} = \text{Toeplitz}(0.5,d_k)$; 
    \item $s_i = (r-i+1) \sqrt{d^{\alpha}}$, where $\alpha \in \{0.6,0.5,0.4\}$. 
\end{itemize} 

\begin{figure}[htbp!]
    \centering 
    \includegraphics[width = \textwidth]{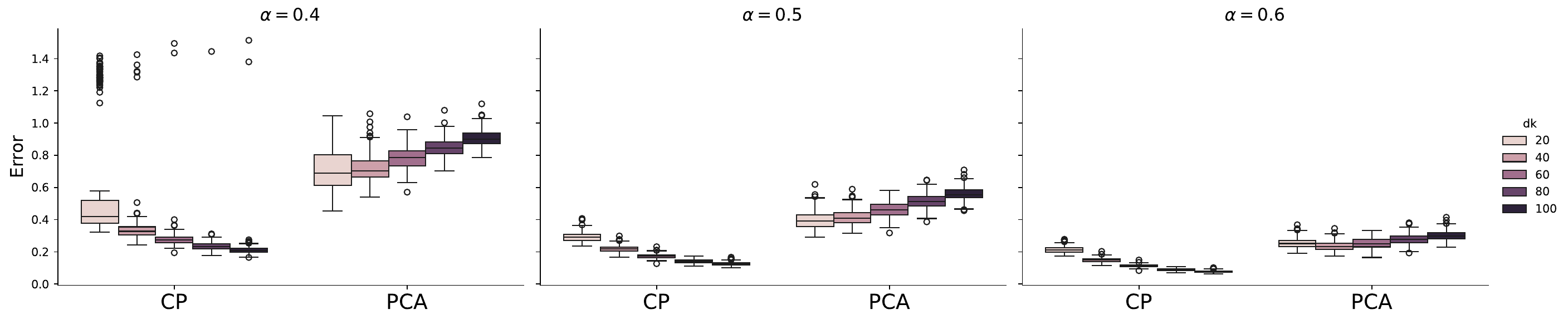} 
    \label{fig:factor_est_iso_vs_pca} 
    \caption{Factor estimation error for PCA and CC-ISO for different choices of $\alpha$ and $d_k$.}
\end{figure}

Under this data generating process, we have $\frac{1}{2} < \alpha + 0.3 < 1 $. As discussed in Remark 3.1, CC-ISO is consistent but PCA is not in theory. Figure \ref{fig:factor_est_iso_vs_pca} shows the estimation error of factors $\frac{1}{\sqrt{T}} \| \hat F - H F\|_2$ for PCA and CC-ISO. It can be observed that PCA is not consistent for the factor estimation, while CC-ISO is consistent, which is in line with the theoretical results. 

\section*{Appendix E: Further discussion of Assumption 3.1(ii)}
The $\alpha$-mixing condition imposed in Assumption 3.1(ii) might not be flexible enough to accommodate some time series models (\cite{Andrews1984}). There are several ways to address this. One possibility is to impose higher-level assumptions, for example assuming directly a CLT and a probability limit, as in \cite{Bai2003} and \cite{StockWatson2002}. Alternatively, we could adopt the more flexible $\tau$-mixing framework (see, \cite{BabiiGhyselsStriaukas2024,han2023probability}) or functional dependence measure \citep{wu2005nonlinear}, which accomodates a broader class of time series processes. However, unlike $\alpha-$mixing, the $\tau-$mixing property is not preserved under measurable transformations, which complicates the analysis of quadratic or product terms. To deal with this, one can use truncation arguments in the proofs. 

For instance, if we replace Assumption 3.1 (ii) with the assumption that $(f_t)$ is $\tau-$mixing with mixing coefficient    \begin{align}\label{cond1tau}
    \tau(m) \le \exp\left( - c_0 m^{\gamma} \right)
    \end{align}
    for some constants $c_0>0$ and $\gamma\ge 0$, then equation (40) in the appendix can still be established, albeit at a slightly slower rate. Specifically, we obtain the following lemma: 
    \begin{lemma} Under the assumptions of Theorem 3.1 with the above $\tau-$mixing condition, we have 
        \[\frac{1}{T}\sum_{t=1}^T\!\big(\tf_{it}^2-s_{it}^2\big)
\;=\;
s_i^2O_p\!\Big(\frac{(\log T)^{2/\nu_2}}{\sqrt{T}}\Big).
\]
    \end{lemma}
    \begin{proof}
        Let
$B_T =\; \Big(C\log T\Big)^{1/\nu_2}$ for some constant $C$. Define the truncated variable and the remainder
\[
h_{it}=\min\{f_{it}^2,B_T^2\}, 
\qquad
r_{it}=f_{it}^2-h_{it}=f_{it}^2\mathbf 1\{|f_{it}|>B_T\}.
\]
Then
\begin{equation}\label{eq:dec-BT}
\frac{1}{T}\sum_{t=1}^T\!\big(\tf_{it}^2-\EE \tf_{it}^2\big)
=
\frac{1}{Ts_i^2}\sum_{t=1}^T\!\big(h_{it}-\EE h_{it}\big)
\;+\;
\frac{1}{Ts_i^2}\sum_{t=1}^T\!\big(r_{it}-\EE r_{it}\big).
\end{equation}

For the first part in equation (\ref{eq:dec-BT}), set $\bar h_{it}=h_{it}-\EE h_{it}$.
We have $|\bar h_{it}|\le B_T^2$ and $h_{it}$ is globally $2B_T$-Lipschitz. 
By the Lipschitz--$\tau$ covariance inequality, we have 
\[
V_T=Var(h_{it})+2\sum_{k\ge1}|Cov(h_{it},h_{i,t+k})|=O(B_T^4).
\]
Then by Bernstein inequality, we have for $\frac{1}{\gamma} = \frac{2}{\gamma_1} + \frac{1}{\gamma_2}$ $c_1,c_2>0$, 
\begin{equation*}
    \begin{split}
    P \left[T s_i^{-2} \left(\frac{1}{T} \sum_{t=1}^T (h_{it} - \EE h_{it}) \geq \varepsilon \right) \right] \leq  (T+1)\exp\!\Big(-c_1 (T\varepsilon)^{\gamma}\Big) + &\exp\!\Big(-\,\frac{T^2\varepsilon^2}{c_2\,[1+T V_T]}\Big).\\
\end{split}
\end{equation*} 
Let $\varepsilon= C\frac{B_T^2}{\sqrt{T}}$. It is easy to check the first exponential term vanishes to 0 and the second term is the dominating term. Note that 
\[\exp\!\Big(-\,\frac{T^2\varepsilon^2}{c_2\,[1+T V_T]}\Big)\asymp \exp\left(-C^2/c_2\right).
\]
Hence, for any $\delta>0$, we can always find $C$, so that 
\begin{equation*}
    \begin{split}
    P \left[T s_i^{-2} \left(\frac{1}{T} \sum_{t=1}^T (h_{it} - \EE h_{it}) \geq \varepsilon \right) \right] \leq \delta,\\
\end{split}
\end{equation*} which implies that 
\[\frac{1}{T} \sum_{t=1}^T (h_{it} - \EE h_{it})=s_i^2 O_p\!\Big(\frac{(\log T)^{2/\nu_2}}{\sqrt{T}}\Big).
\]
For the second part of equation (\ref{eq:dec-BT}), we have 
\[
\EE\,r_{it}
= B_T^2 P(|f_t|>B_T) + \int_{B_T}^\infty 2x\,P(|f_{it}|>x)\,dx\leq C' (\log T)^{(2-\nu_2)/\nu_2}\, T^{-c_2C}
\]
by Assumption 3.1(ii). Therefore, by Markov’s inequality,
\[
P\!\left(\frac{1}{T}\sum_{t=1}^T r_{it} > (\log T)^{2/\nu_2}/\sqrt{T}\right)
\;\le\; 
\frac{\EE r_{it}}{(\log T)^{2/\nu_2}/\sqrt{T}}
\;
=\; O\!\big((\log T)^{-1}T^{-c_2C+1/2}\big)\to 0
\] for large $C$. 
Moreover,
\[
\left|\frac{1}{T}\sum_{t=1}^T\!\big(r_{it}-\EE r_{it}\big)\right|
\;\le\; \frac{1}{T}\sum_{t=1}^T r_{it} + \EE r_{it}
\;=\; o_p\!\Big(\frac{(\log T)^{2/\nu_2}}{\sqrt{T}}\Big).
\] Finally, combining two parts, we have 
\[\frac{1}{T}\sum_{t=1}^T\!\big(\tf_{it}^2-s_{it}^2\big)
\;=\;
s_i^2O_p\!\Big(\frac{(\log T)^{2/\nu_2}}{\sqrt{T}}\Big).
\]
    \end{proof}
    Similar arguments can be used if we replace Assumptions 3.4 and 4.1(ii) with a $\tau$-mixing condition on $R_t = (f_t^\top, z_t^\top, e_t^\top, \varepsilon_t^\top, V_t^\top)^\top$. In that case, truncated products such as $z_t e_t$, $e_t \varepsilon_t$, and $f_t \varepsilon_t$ can be bounded with Bernstein’s inequality, and the corresponding tail terms remain negligible.

We emphasize that these modifications are technically feasible but considerably increase the complexity of the proofs, which are already heavy. For this reason, we chose to work with $\alpha$-mixing in the main text and only provide a discussion of the $\tau$-mixing alternative in the appendix. Importantly, our simulation study confirms that the proposed method is robust to AR(1) dynamics, lending further support to the practical relevance of our assumptions.

\section*{Appendix F: Constructing Prediction Intervals under Post-Selection Debiased LASSO}

This appendix outlines a practical procedure for constructing prediction intervals for 
$\hat y_{T+h|T}$ using the post-selection debiased LASSO (PD-LASSO). 


\paragraph{Step 1. Estimate latent factors.}
Obtain factor estimates $\hat f_t$ and loadings $\hat B$ using the CC-ISO algorithm.

\paragraph{Step 2. Obtain projected residuals.}
Regress $w_t$ on $\hat f_t$ to remove the factor component:
\[
\hat\Lambda = \Big(\sum_{t=1}^T w_t \hat f_t^\top\Big)
              \Big(\sum_{t=1}^T \hat f_t \hat f_t^\top\Big)^{-1}, 
\qquad 
\hat V_t = w_t - \hat\Lambda \hat f_t.
\]
Then regress $y_{t+h}$ on $\hat f_t$ to obtain the projection residuals:
\[
\tilde y_{t+h} = y_{t+h} - \hat\beta_1^{*\top} \hat f_t, 
\qquad 
\hat\beta_1^* = 
  \Big(\sum_{t=1}^{T-h}\hat f_t \hat f_t^\top\Big)^{-1}
  \Big(\sum_{t=1}^{T-h}\hat f_t y_{t+h}\Big).
\]

\paragraph{Step 3. LASSO estimation.}
Estimate the local-predictor coefficients by
\[
\hat\beta_0 = \arg\min_{\beta_0}
  \tfrac{1}{2T}\|\tilde Y - \hat V \beta_0\|_2^2 
  + \lambda \|\beta_0\|_1,
\]
and define the selected support 
$\hat{\mathcal S} = \{ j : \hat\beta_{0,j} \neq 0 \}$.

\paragraph{Step 4. Nodewise precision estimation.}
For each $j \in \hat{\mathcal S}$, estimate the $j$th row of the precision matrix via
\[
\hat\gamma_j = 
  \arg\min_{\gamma_j} 
  \tfrac{1}{T}\|\hat V_j - \hat V_{-j}\gamma_j\|_2^2 
  + \lambda_j \|\gamma_j\|_1,
\qquad
\hat\tau_j^2 = 
  \tfrac{1}{T}\|\hat V_j - \hat V_{-j}\hat\gamma_j\|_2^2 
  + \lambda_j \|\hat\gamma_j\|_1.
\]
Assemble 
$\hat\Gamma_{\hat{\mathcal S}}$ 
and 
$\mathbf T_{\hat{\mathcal S}}$ 
(diagonal with entries $\hat\tau_j^{-2}$), 
and set 
$\hat\Theta_{\hat{\mathcal S}} 
  = \mathbf T_{\hat{\mathcal S}} \hat\Gamma_{\hat{\mathcal S}}$.

\paragraph{Step 5. Post-selection debiasing.}
Compute
\[
\hat\beta_0^{(PL)} 
  = \hat\beta_0 + \tfrac{1}{T}\hat\Theta_{\hat{\mathcal S}} 
    \hat V (\tilde Y - \hat V \hat\beta_0).
\]
Re-estimate $\beta_1$ and residuals:
\[
\hat\beta_1 
  = (\hat F^\top\hat F)^{-1}\hat F(Y - W \hat\beta_0^{(PL)}), 
\qquad 
\hat\varepsilon = Y - W \hat\beta_0^{(PL)} - \hat F \hat\beta_1.
\]

\paragraph{Step 6. Forecast and variance estimation.}
Compute the forecast 
$\hat y_{T+h|T} = w_T^\top\hat\beta_0^{(PL)} + \hat f_T^\top\hat\beta_1$ 
and its variance
\[
\hat\sigma_{\hat y_{T+h|T}} 
  = \hat\sigma_{y,\hat\beta_0} 
  + \hat\sigma_{y,\hat\beta_1} 
  + \hat\sigma_{\hat f_T},
\]
where
\begin{align*}
\hat\sigma_{y,\hat\beta_1}
  &= \tfrac{1}{T}\hat f_T^\top 
     \Big(\tfrac{1}{T}\sum_{t=1}^{T-h}\hat f_t\hat f_t^\top\Big)^{-1}
     \Big(\tfrac{1}{T}\sum_{t=1}^{T-h}\hat f_t\hat f_t^\top\hat\varepsilon_{t+h}^2\Big)
     \hat f_T,\\[1mm]
\hat\sigma_{\hat f_T}
  &= \hat\beta_1^\top \hat S^{-1} \hat B^\top \hat\Sigma_{Be} 
     \hat B \hat S^{-1} \hat\beta_1,\\[1mm]
\hat\sigma_{y,\hat\beta_0}
  &= \tfrac{1}{T}\hat V_T^\top
     \hat\Theta_{\hat{\mathcal S}} \hat\Omega 
     \hat\Theta_{\hat{\mathcal S}} \hat V_T,
\end{align*}
with $\hat\Sigma_{Be}$ and $\hat\Omega$ obtained via the thresholding estimators 
defined in Section~3.3.

\paragraph{Step 7. Construct the prediction interval.}
The $(1-\alpha)\%$ prediction interval is
\[
\Big[
\hat y_{T+h|T} - q_{\alpha/2}\hat\sigma_{\hat y_{T+h|T}},\,
\hat y_{T+h|T} + q_{\alpha/2}\hat\sigma_{\hat y_{T+h|T}}
\Big],
\]
where $q_{\alpha/2}$ is the upper $\alpha/2$ quantile of the standard normal distribution.

\medskip
This PD-LASSO procedure offers a practical approach for constructing approximate 
prediction intervals in high-dimensional forecasting. 
Our simulation study suggests that PD-LASSO intervals achieve 
coverage rates close to nominal levels while remaining narrower than those from 
the fully debiased LASSO, providing a useful balance between validity and efficiency. 

We conduct a simulation study to evaluate its performance. The DGP mainly follows Section 5.4 with minor modifications: 
\begin{itemize}
    \item $p = 100$ and $d_k \in \{40,80,100 \}$. $T = 800 + \lceil d^{3/4} \rceil$.  
    \item $V_t$ is generated independently from $N(0, \Sigma_V)$ where $\Sigma_V = \text{Toeplitz}(0.5,p)$. 
    \item Factor strength $\alpha_i = \alpha \in \{ 1, 0.6 ,0.4\}$. 
    \item $\beta_0 = (3,3,3,0,\ldots, 0)^\top$. 
\end{itemize} 
The tuning parameter $\lambda$ is the Step 3 is selected through BIC\footnote{Here, we use BIC for better control of variable selection consistency.} and $\lambda_j$ in Step 4 are fixed at $\sqrt{\log (d) / T} + \sqrt{1/\hat s_r}$ for all $j$. 

Table \ref{tab:pi_pdlasso} reports the results. Coverage rates of the proposed PD-LASSO are close to the nominal 95\% level. The close coverage rate of PD-LASSO is mainly due to the high variable-selection accuracy achieved by BIC. These findings suggest that PD-LASSO offers a useful compromise, which provides approximately valid coverage with narrower intervals when the selection step is reliable.

\begin{table}[htbp]
  \centering
  \caption{Results of prediction intervals post-selection debiased LASSO }
    \begin{tabular}{ccccccc}
    \toprule
          & \multicolumn{2}{c}{$\alpha = 1$} & \multicolumn{2}{c}{$\alpha = 0.6$} & \multicolumn{2}{c}{$\alpha = 0.4$} \\
    \midrule
    $d_k$    & PI Length & Coverage Rate & PI Length & Coverage Rate & PI Length & Coverage Rate \\
    \midrule
    40    & 0.369 & 0.913 & 0.784 & 0.905 & 1.294 & 0.858 \\
    80    & 0.286 & 0.935 & 0.589 & 0.919 & 1.137 & 0.883 \\
    100   & 0.257 & 0.934 & 0.503 & 0.927 & 1.062 & 0.92 \\
    \bottomrule
    \end{tabular}%
  \label{tab:pi_pdlasso}%
\end{table}%

\section*{Appendix G: More simulation results} 

In this section, we conduct additional simulation studies. Firstly, we check the robustness of our proposed algorithm under (i) more persistent factors, (ii) stronger cross-sectional correlation in errors, and (iii) Student-t errors. For all DGP settings, we evaluate the coverage rate of prediction intervals in the low-dimensional $w_t$ setting and the forecast errors for MS-FASR in the high-dimensional $w_t$ setting. Throughout, we fix $\alpha_i = \alpha = 0.6$ and $d_k = 40$. 

To generate more persistent factors, we specify:  
\begin{align*}
    & g_{i,t+1} = \rho g_{i,t} + \sqrt{1- \rho^2} u_{it}, \quad \rho \in \{0.7,0.8,0.9\} \\ 
    & f_t = \Sigma_{f}^{1/2} g_t
\end{align*}
where $u_{it}$ is generated independently from standard normal and $\Sigma_f = \text{Toeplitz}(0.5,r)$ inducing correlation among factors. Other settings follow Section 5 of the main text. 

To allow stronger error dependence, we vary $\kappa \in \{0.6,0.7,0.8\}$ in $\Sigma_{\calE,1} = \Sigma_{\calE,2} = \text{Toeplitz}(\kappa,d_k),$ thereby varying the level of dependence. The remaining settings are the same as in Section 5. 

For the Student-t errors design, we generate $\calE_t$ as follows: 
$$
\calE_t = \Sigma_{\calE,1}^{1/2} Z_t \Sigma_{\calE,2}^{1/2}, \quad \Sigma_{\calE,1} = \Sigma_{\calE,2} = \text{Toeplitz}(0.5,d_k),
$$
where each entry of $Z_t$ is drawn independently from student-t distribution with degrees of freedom $df \in \{4,5,6\}$. The regression error $\epsilon_{t+h}$ is generated from the same distribution. Other settings again follow Section 5. 

\begin{figure}
    \centering
    \includegraphics[width=0.6\linewidth]{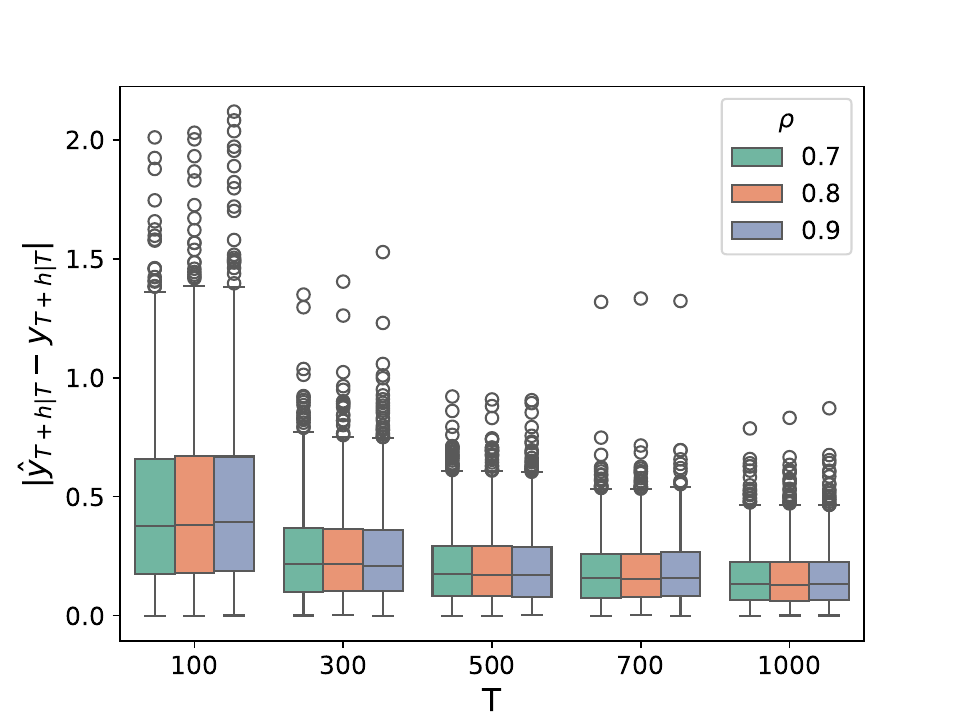}
    \caption{Boxplots of prediction errors $| \hat{y}_{T+h|T} - y_{T+h|T}|$ of MS-FASR under more persistent factors. }
    \label{fig:persistent_factors_hd_pred_err}
\end{figure}

\begin{figure}
    \centering
    \includegraphics[width=0.6\linewidth]{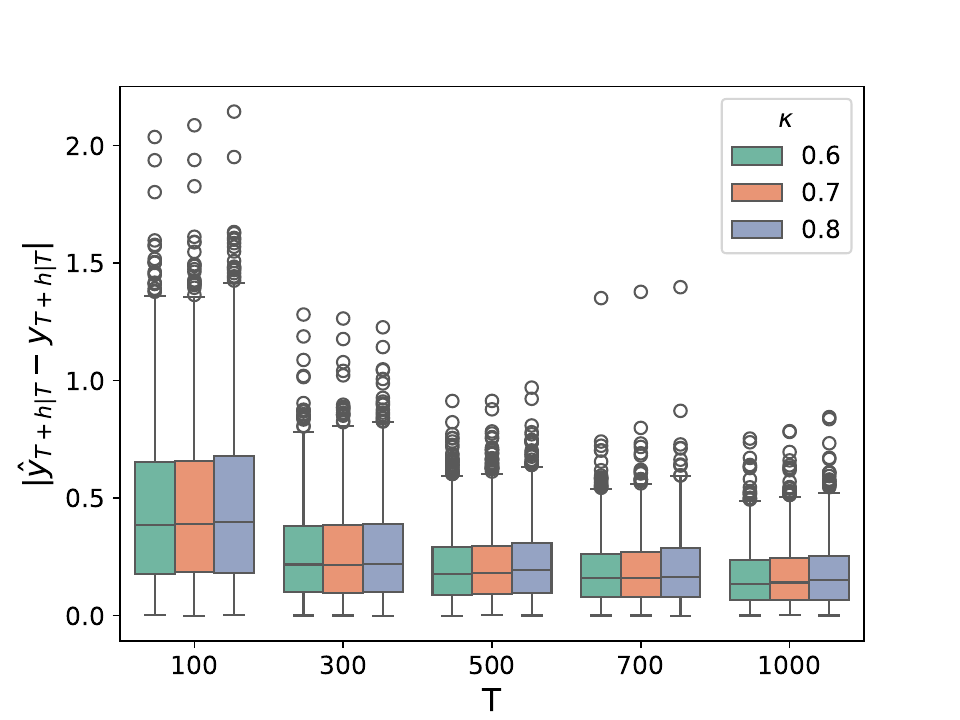}
    \caption{Boxplots of prediction errors $| \hat{y}_{T+h|T} - y_{T+h|T}|$ of MS-FASR under stronger error dependence in CP factor model. }
    \label{fig:high_cs_corr_pred_err}
\end{figure}

\begin{figure}
    \centering
    \includegraphics[width=0.6\linewidth]{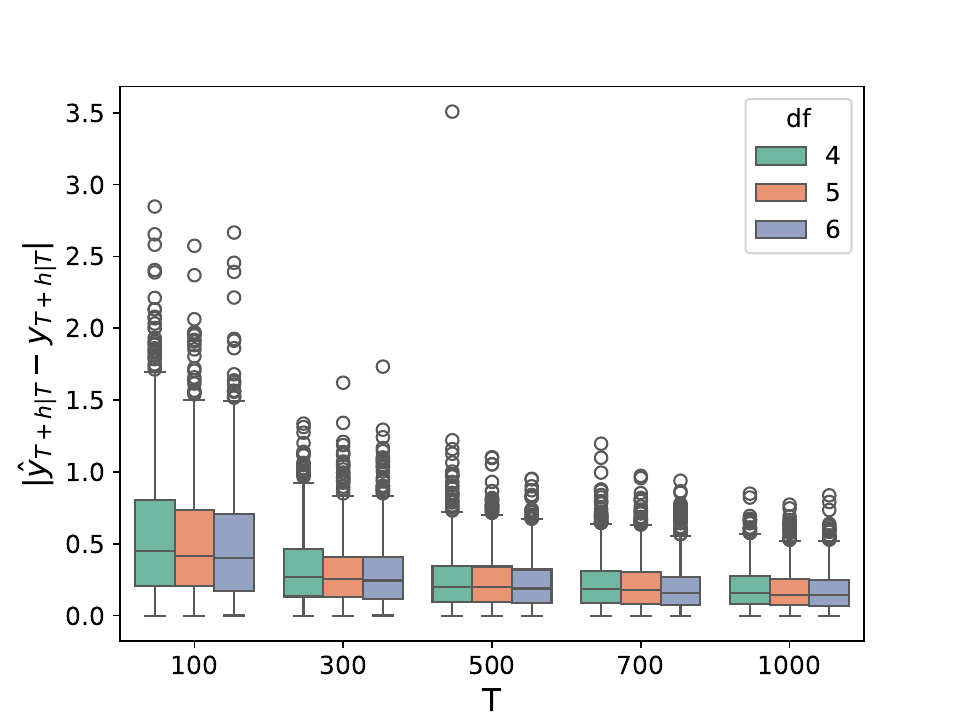}
    \caption{Boxplots of prediction errors $| \hat{y}_{T+h|T} - y_{T+h|T}|$ of MS-FASR under student-t distributions. }
    \label{fig:studentt_pred_err} 
\end{figure}

The prediction errors of MS-FASR are reported in Figure \ref{fig:persistent_factors_hd_pred_err}, \ref{fig:high_cs_corr_pred_err} and \ref{fig:studentt_pred_err}, while the coverage rates of CP and PCA prediction intervals in the low-dimensional regressor setting are presented in Table \ref{tab:rr_ci_tab}. The results show that our method remains robust and performs well across these alternative designs. 

\begin{table}[htbp]
  \centering
  \caption{Coverage rate of CP and PCA prediction intervals}
  \fontsize{10}{15}\selectfont{
    \begin{tabular}{llll}
    \toprule
          & CP    & PCA(H) & PCA(T) \\
    \midrule
    \multicolumn{4}{l}{Persistent factors} \\
    $\rho$   &       &       &  \\
    0.7   & 0.94  & 0.73  & 0.753 \\
    0.8   & 0.942 & 0.724 & 0.753 \\
    0.9   & 0.947 & 0.737 & 0.765 \\
    \midrule
    \multicolumn{4}{l}{Stronger error dependence} \\
    $\kappa$ &       &       &  \\
    0.6   & 0.936 & 0.703 & 0.739 \\
    0.7   & 0.933 & 0.642 & 0.689 \\
    0.8   & 0.915 & 0.47  & 0.527 \\
    \midrule
    \multicolumn{4}{l}{Student t errors} \\
    $df$    &       &       &  \\
    4     & 0.917 & 0.72  & 0.743 \\
    5     & 0.917 & 0.719 & 0.739 \\
    6     & 0.929 & 0.731 & 0.759 \\
    \bottomrule
    \end{tabular}%
    }
    \vspace{0.7em} 
\parbox{0.8\textwidth}{ 
 \small \textit{Note:} (1) The dimension $d_k$ and the factor signal $\alpha$ are fixed at 40 and 0.6, respectively. (2) PCA(T) and PCA(H) refer to the prediction interval constructed using the PCA approach, where the covariance matrix of the factors is estimated via the proposed thresholding covariance estimator and the HAC-type estimator proposed by \cite{BaiNg2006} and \cite{Bai2023}, respectively. (3) The nominal confidence level is $95\%$. 
}
  \label{tab:rr_ci_tab}%
\end{table}%

\begin{figure}
    \centering
    \includegraphics[width=0.5\linewidth]{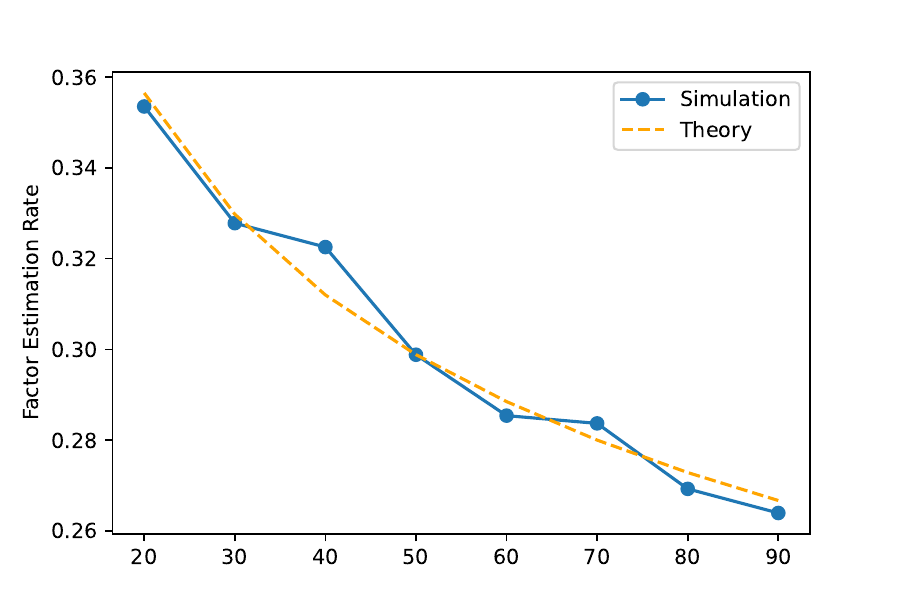}
    \caption{Simulated rates vs. theoretical rates of factor estimation. The blue solid lines show the mean of simulated estimation errors over 200 repetitions. The orange dotted lines show the fitted curve of theoretical rates. The fitted curve is $c_0 \sqrt{\frac{1}{d^{0.1}T}} + c_1 \frac{1}{d^{0.3}}$ where $c_0$ and $c_1$ are calculated by minimizing the distance between the theoretical curve and the simulation curve.}
    \label{fig:rate_fTerr}
\end{figure}

Next, we evaluate the convergence rate in Theorem 3.1 by a simulation study in which the factors are generated independently from standard normal distributions, while all other settings follow Section 5. Under this DGP setting, the theoretical rate for $\| \hat{f_T} - H f_T\|_2$ is $\sqrt{\frac{1}{d^{\alpha-1/2}T}} + \frac{1}{d^{\alpha/2}}$. We fix $T = 500$, $\alpha=0.6$ and let $d_1 = d_2 = \bar{d} $ with increasing $\bar{d}$. Figure \ref{fig:rate_fTerr} shows the comparison between the simulated and the theoretical rates of factor estimation, showing that the simulation curve closely aligns with the theoretical curve.

\begin{figure}
    \centering
    \includegraphics[width=0.6\linewidth]{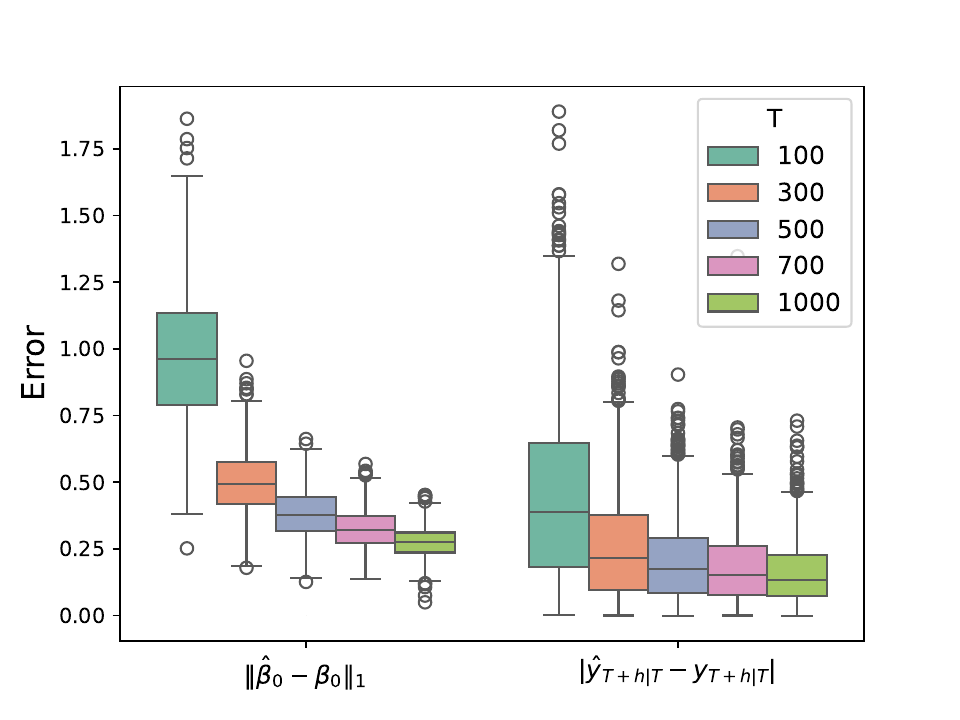}
    \caption{Boxplots of errors for MS-FASR where $W$ is not related to $F$.}
    \label{fig:hd_wnotrelated}
\end{figure}

Finally,  we conducted an additional simulation to evaluate the performance of MS-FASR when $w_t$ is generated independently of $f_t$ (i.e., $\Lambda = 0$). The DGP follows Section 5.4 in the paper with $\alpha = 0.6$, $d_k = 40$ and $T \in \{100,300,500,700,1000\}$. Figure \ref{fig:hd_wnotrelated} shows the boxplots of the estimations error $\| \hat \beta_0 - \beta_0 \|_1$ and forecast error $| \hat{y}_{T+h|T} - y_{T+h|T}|$. The results show that the independence between $w_t$ and $f_t$ does not affect the performance of our algorithm, confirming its robustness to the absence of shared latent structure. See the discussion on Page 21. 

\clearpage

\end{document}